\documentclass[epj]{svjour}
\usepackage{epsfig}
\usepackage{amsmath}
\usepackage{graphicx}
\usepackage{amsfonts}
\usepackage{amssymb}

\begin{document}

\title{Magnetic Field Effects on the 1083 nm Atomic Line of Helium}
\subtitle{Optical Pumping of Helium and Optical Polarisation Measurement 
in High Magnetic Field}
\author{E. Courtade\inst{1}
\and F. Marion\inst{1}
\and P.-J. Nacher\inst{1}\thanks{electronic mail : nacher@lkb.ens.fr} 
\and G. Tastevin\inst{1}
\and K.\ Kiersnowski\inst{2}
\and T.\ Dohnalik\inst{2}}
\institute{Laboratoire Kastler Brossel%
\thanks{Laboratoire de l'Universit\'e Pierre et Marie Curie et de l'Ecole Normale Sup\'erieure, 
associ\'e au Centre National de la Recherche Scientifique (UMR 8552)}, 
24 rue Lhomond, 75231 Paris cedex 05, France.
\and Marian Smoluchowski Institute of Physics, Jagiellonian University, ul. Reymonta 4, 30-059 Krak\'ow, Poland.}

\date{\today}

\abstract{
The structure of the excited $2^{3}\mathrm{S}$ and $2^{3}\mathrm{P}$ triplet states of
$^{3}\mathrm{He}$ and $^{4}\mathrm{He}$ in an applied magnetic field $B$ is
studied using different approximations of the atomic Hamiltonian. All optical
transitions (line positions and intensities) of the 1083~nm $2^{3}\mathrm{S}%
$-$2^{3}\mathrm{P}$ transition are computed as a function of $B$. The effect
of metastability exchange collisions between atoms in the ground state and in the 
$2^{3}\mathrm{S}$ metastable state is studied, and rate equations are derived, for the
populations these states in the general case of an isotopic
mixture in an arbitrary field $B$. It is shown that the usual spin-temperature
description remains valid. A simple optical pumping model based on these rate
equations is used to study the $B$-dependence of the population
couplings which result from the exchange collisions. Simple
spectroscopy measurements are performed using a single-frequency laser diode
on the 1083~nm transition. The accuracy of frequency scans and of
measurements of transition intensities is studied. Systematic experimental
verifications are made for $B$=0 to 1.5~T. Optical pumping effects resulting
from hyperfine decoupling in high field are observed to be in good agreement
with the predictions of the simple model. Based on adequately chosen absorption measurements at
1083~nm, a general optical method to measure
the nuclear polarisation of the atoms in the ground state in an arbitrary field
is described. It is demonstrated at $B\sim$0.1~T, a field for which the usual 
optical methods could not operate.
 }

\titlerunning{The 1083 nm Helium Transition in High Magnetic Field}
\maketitle

\section{Introduction}

Highly polarised $^{3}\mathrm{He}$ is used for several applications in various
domains, for instance to prepare polarised targets for nuclear physics
experiments \cite{Rohe99}, to obtain spin filters for cold neutrons
\cite{Becker98,Jones00}, or to perform magnetic resonance imaging (MRI) of air
spaces in human lungs \cite{Tastevin00,Chupp01}. A very efficient and widely
used polarisation method relies on optical pumping of the $2^{3}\mathrm{S}%
$\ metastable state of helium with 1083~nm resonant light
\cite{Colegrove63,Nacher85}. Transfer of nuclear polarisation to atoms in the
ground state is ensured by metastability exchange collisions. Optical Pumping
(OP) is usually performed in low magnetic field (up to a few~mT), required
only to prevent fast relaxation of the optically prepared orientation. OP can
provide high nuclear polarisation, up to 80\% \cite{Bigelow92}, but
efficiently operates only at low pressure (of order 1~mbar) \cite{Leduc83}.
Efficient production of large amounts of polarised gas is a key issue for most
applications, which often require a dense gas. For instance, polarised gas
must be at atmospheric pressure to be inhaled in order to perform lung MRI.
Adding a neutral buffer gas after completion of OP is a simple method to
increase pressure, but results in a large dilution of the polarised helium.
Polarisation preserving compression of the helium gas after OP using different
compressing devices is now performed by several research groups
\cite{Becker94,Nacher99,Gentile00}, but it is a demanding technique and no
commercial apparatus can currently be used to obtain the large compression
ratio required by most applications.

Improving the efficiency of OP at higher pressure is a direct way to obtain
larger magnetisation densities. Such an improvement was shown to be sufficient
to perform lung MRI in humans \cite{MRIorsay}. It could also facilitate
subsequent mechanical compression by significantly reducing the required
compression ratio and pumping speed. It was achieved by operating OP in a
higher magnetic field (0.1~T) than is commonly used. High field OP in
$^{3}\mathrm{He}$ had been previously reported at 0.1~T \cite{Flowers90} and
0.6~T \cite{Flowers97}, but the worthwhile use of high fields for OP at high
pressures (tens of mbar) had not been reported until recently
\cite{Courtade00}.

An important effect of a high enough magnetic field is to strongly reduce the
influence of hyperfine coupling in the structures of the different excited
levels of helium. In order to populate the $2^{3}\mathrm{S}$\ metastable state
and perform OP, a plasma discharge is sustained in the helium gas. In the
various atomic and molecular excited states which are populated in the plasma,
hyperfine interaction transfers nuclear orientation to electronic spin and
orbital orientations. The electronic angular momentum is in turn converted
into circular polarisation of the emitted light or otherwise lost during
collisions. This process is actually put to use in the standard optical
detection technique \cite{Bigelow92,Pavlovic70} in which the circular
polarisation of a chosen helium spectral line emitted by the plasma is
measured and the nuclear polarisation is inferred. The decoupling effect of an
applied magnetic field unfortunately reduces the sensitivity of the standard
optical detection method above 10~mT, and hence a different measurement
technique must be used in high fields. This transfer of orientation also has
an adverse effect in OP situations by inducing a net loss of nuclear
polarisation in the gas. The decoupling effect of an applied field reduces
this polarisation loss and may thus significantly improve the OP performance
in situations of limited efficiency, such as low temperatures or high
pressures. At low temperature (4.2~K), a reduced metastability exchange cross
section sets a tight bottleneck and strongly limits the efficiency of OP
\cite{Fitzsimmons68,Chapman76,BarbéTh}~; a field increase from 1 to 40~mT was
observed to provide an increase in nuclear polarisation\ from 17~\% to 29~\%
in this situation \cite{NacherTh}. At high gas pressures (above a few mbar)
the proportion of metastable atoms is reduced and the creation of metastable
molecular species is enhanced, two factors which tend to reduce the efficiency
of OP ; it is not surprising that a significant improvement is obtained by
suppressing relaxation channels in high field \cite{Courtade00}.

A systematic investigation of various processes relevant for OP in
non-standard conditions (high field and/or high pressure) has been made, and
results will be reported elsewhere \cite{CourtadeTh,NousMolec}. As mentioned
earlier, an optical measurement method of nuclear polarisation in arbitrary
field must be developed. It is based on absorption measurements of a probe
beam, and requires a detailed knowledge of magnetic field effects on the 1083
nm transition lines. In this article we report on a detailed study of the
effect of an applied magnetic field on the structure of the $2^{3}\mathrm{S}$
and $2^{3}\mathrm{P}$ atomic levels, both in $^{3}\mathrm{He}$ and
$^{4}\mathrm{He}$. In the theoretical section we first present results
obtained using a simple effective Hamiltonian and discuss the accuracy of
computed line positions and intensities at various magnetic fields, then
discuss the effect of an applied field on the metastability exchange
collisions between the $2^{3}\mathrm{S}$ state and the $1^{1}\mathrm{S}_{0}$
ground state level of helium. In the experimental section we present results
of measurements of the line positions and intensities up to 1.5~T, then
describe an optical measurement technique of the nuclear polarisation of
$^{3}\mathrm{He}$.

Let us finally mention that these studies have been principally motivated by
OP developments, but that their results could be of interest to design or
interpret experiments performed on helium atoms in the $2^{3}\mathrm{S}$\ or
$2^{3}\mathrm{P}$\ state in an applied field, as long as metrological accuracy
is not required. This includes laser cooling and Zeeman slowing
\cite{Philips82} of a metastable atom beam, magnetic trapping and evaporative
cooling which recently allowed two groups to obtain Bose Einstein condensation
with metastable $^{4}\mathrm{He}$ atoms \cite{Orsay,Paris}, and similar
experiments which may probe the influence of Fermi statistics with ultracold
metastable $^{3}\mathrm{He}$ atoms.

\section{Theoretical}

\begin{figure}[h]
\centerline{ \psfig{file=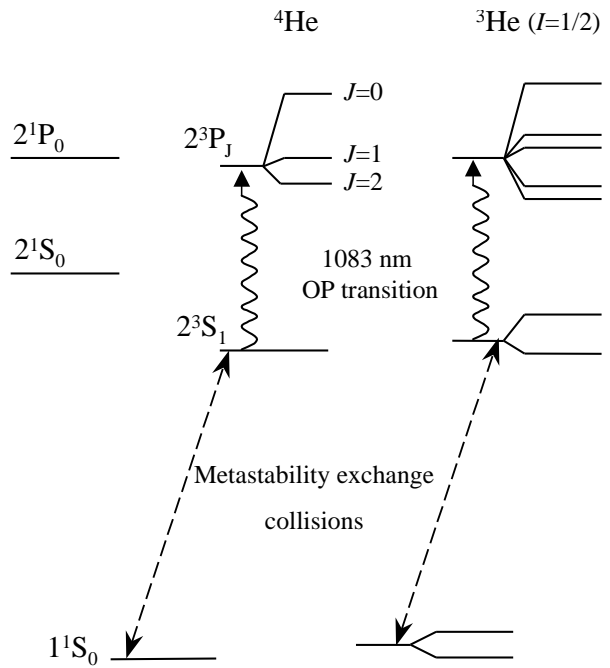, width=8 cm, clip= } } \caption{Ground
state and first excited states of He (not to scale), showing the effects of
fine and hyperfine interactions on excited level structures. The Zeeman effect
is assumed negligible, except in the ground state of $^{3}\mathrm{He}$ in
which it is grossly exaggerated to highlight the existence of two magnetic
sublevels and the possibility of nuclear polarisation. Accurate values of all
splittings are given in the tables in the appendix.}%
\label{figlevels}%
\end{figure}

The low-lying energy states of helium are represented in
figure~\ref{figlevels}. The fine-structure splittings of the $2^{3}\mathrm{P}%
$\ state of $^{4}\mathrm{He}$ and the additional effect of hyperfine
interaction on splittings of the $2^{3}\mathrm{S}$\ and $2^{3}\mathrm{P}%
$\ states of $^{3}\mathrm{He}$ are indicated in null magnetic field.
The $^{4}\mathrm{He}$ $2^{3}\mathrm{P}$\ fine-structure intervals
have been computed and measured with a steadily improved accuracy
\cite{Storry00,Castillega00}, in order to provide a test of QED\ calculations
and to measure the fine-structure constant $\alpha$. High precision isotope
shift measurements of the $2^{3}\mathrm{S}$---$2^{3}\mathrm{P}$\ transition of
helium have also been performed to further test QED calculations and to probe
the $^{3}\mathrm{He}$ nuclear charge radius with atomic physics experiments
\cite{Shiner95}. High precision measurements and calculations of the Zeeman
effect in the $2^{3}\mathrm{P}$ state of helium have also been performed
\cite{Lhuillier76,Yang86,Yan94}. They were required in particular to analyse
the results of fine-structure intervals measured using microwave transitions
in an applied magnetic field in $^{4}\mathrm{He}$ \cite{Kponou81} and
$^{3}\mathrm{He}$ \cite{Prestage85}.

All this work has produced a wealth of data which may now be used to compute
the level structures and transition probabilities with a very high accuracy up
to high magnetic fields (several Tesla), in spite of a small persisting
disagreement between theory and experiments for the $g$-factor of the Zeeman
effect in the $2^{3}\mathrm{P}$ state \cite{Yan94,Gonzalo97}. However, such
accurate computation can be very time-consuming, especially for $^{3}%
\mathrm{He}$ in which fine-structure, hyperfine and Zeeman interaction terms
of comparable importance have to be considered. A simplified approach was
proposed by the Yale group \cite{Hinds85}, based on the use of an effective
Hamiltonian. The hyperfine mixing of the $2^{3}\mathrm{P}$ and singlet
$2^{1}\mathrm{P}$ states is the only one considered in this effective
Hamiltonian, and its parameters have been experimentally determined
\cite{Prestage85}. A theoretical calculation of the $^{3}\mathrm{He}$
$2^{3}\mathrm{P}$ structure with an accuracy of order 1~MHz later confirmed
the validity of this phenomenological approach \cite{Hijikata88}. In the
present article, we propose a further simplification of the effective
Hamiltonian introduced by the Yale group, in which couplings to the singlet
states are not explicitly considered. Instead we implicitly take into account
these coupling terms, at least in part, by using the $2^{3}\mathrm{P}$
splittings measured in zero field to set the eigenvalues of the fine-structure matrices.

Results obtained using this simplified effective Hamiltonian will be compared
in section \ref{discussionth} to those of the one including the couplings to
the $2^{1}\mathrm{P}$ levels, with 3 or 6 additional magnetic sublevels,
depending on the isotope. The effect of additional terms in a more elaborate
form of the Zeeman Hamiltonian \cite{Yan94} than the linear approximation
which we use will also be evaluated.

\subsection{A simple effective Hamiltonian}

\subsubsection{Notations}

\textit{The }$1^{1}S$\textit{ ground level of helium} is a singlet spin state
($S$=0), with no orbital angular momentum ($L$=0), and hence has no total
electronic angular momentum ($J$=0). In $^{3}\mathrm{He}$, the nucleus thus
carries the only angular momentum $I$, giving rise to the two magnetic
sublevels $m_{I}$=$\pm1/2$. Their relative populations, $\left(  1\pm
M\right)  /2,$ define the nuclear polarisation $M$ of the ground state.

\begin{figure}[h]
\centerline{ \psfig{file=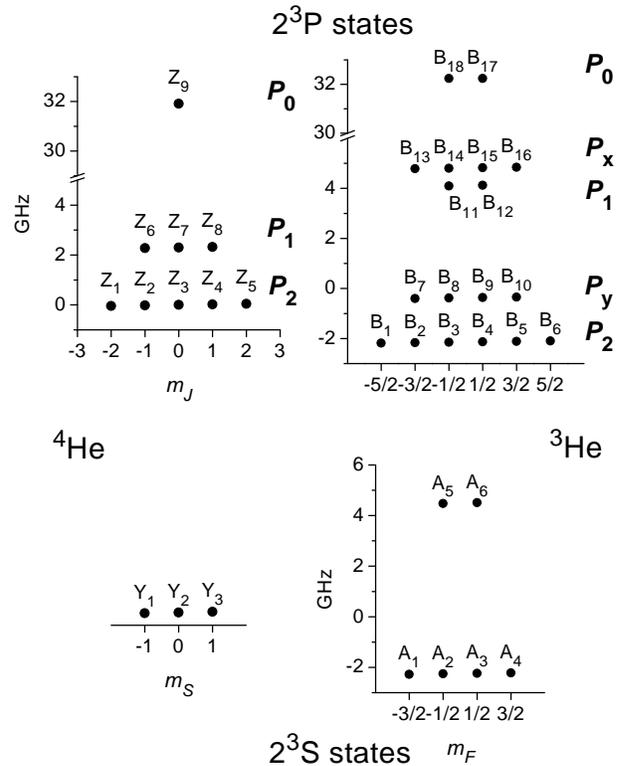, width=8 cm, clip= } }%
\caption{\textbf{Left :} magnetic sublevels of $^{4}\mathrm{He}$ in the
$2^{3}\mathrm{S}$ (bottom) and $2^{3}\mathrm{P}$ (top) levels in very low
magnetic field. $J$ is thus a good quantum number and $m_{J}$ is used to
differentiate the atomic states. When a small field is applied, the sublevel
degeneracies are removed and energies increase for increasing values of
$m_{J}$ (i.e. from $Z_{1}$ to $Z_{9}$). \textbf{Right :} magnetic sublevels of
$^{3}\mathrm{He}$ in very low magnetic field. $F$ and $m_{F}$ are now good
quantum numbers to differentiate the atomic states. Only the $F$=3/2
$\mathrm{P}_{\mathrm{x}}$ and $\mathrm{P}_{\mathrm{y}}$ states involve a
strong mixing of states of different $J$ values \cite{Nacher85}. Energies
increase for increasing values of $m_{F}$ for all sublevels except in the
$2^{3}\mathrm{P}_{\mathrm{0}}$ state (i.e. from $A_{1}$ to $A_{6}$ and $B_{1}$
to $B_{16}$).}%
\label{figniv}%
\end{figure}

\textit{The level structure of }$^{4}He$\textit{ excited states} is determined
from the fine-structure term and the Zeeman term in the Hamiltonian. The
fine-structure term $H_{fs}$ is easily expressed in the total angular momentum
representation $\left(  J\right)  $ using parameters given in
tables\ \ref{sfinemat} and \ref{sfineval} in the Appendix. For the Zeeman term
in the applied magnetic field $B$ we shall use the simple linear form~:
\begin{equation}
H_{Z}^{(4)}=\mu_{B}\left(  g_{L}^{\prime}L+g_{S}^{\prime}S\right)  \cdot B.
\label{Hz4}%
\end{equation}
The values of Bohr magneton $\mu_{B},$ and of the orbital and spin $g$-factors
$g_{L}^{\prime}$ and $g_{S}^{\prime}$ are given in table~\ref{lande} in the
Appendix. The situation is quite simple for the $2^{3}\mathrm{S}$ state, for
which the 3 sublevels $\left|  m_{S}\right\rangle $ (with $m_{S}$=$\pm1,0$)
are eigenstates at any field. For the $2^{3}\mathrm{P}$ state the Zeeman term
couples sublevels of different $J$ (which however have the same $m_{J}$) and
is more easily expressed in the decoupled $\left(  L,S\right)  $
representation. We note $\left|  m_{L},m_{S}\right\rangle $ the vectors of the
decoupled basis used in the $\left(  L,S\right)  $ representation and $\left|
J;m_{J}\right\rangle $ those of the coupled basis, which are the zero-field 9
eigenstates of the $2^{3}\mathrm{P}$ state. We call $\mathrm{Z}_{1}$ to
$\mathrm{Z}_{9}$ (by increasing value of the energy, see figure~\ref{figniv})
the 9 sublevels of the $2^{3}\mathrm{P}$ state in arbitrary field. To simplify
notations used in the following, we similarly call $\mathrm{Y}_{1}$ to
$\mathrm{Y}_{3}$ the 3 sublevels of the $2^{3}\mathrm{S}$ state.

The transition probabilities of all components of the 1083 nm line are
evaluated using the properties of the electric dipole transition operator. For
a monochromatic laser light with frequency $\omega/2\pi,$ polarisation vector
$e_{\lambda},$ and intensity $I_{las}$ (expressed in W/m$^{2}$), the photon
absorption rate for a transition from $\mathrm{Y}_{i}$ to $\mathrm{Z}_{j}$ is
given by~:%
\begin{equation}
\frac{1}{\tau_{ij}^{(4)}}=\frac{4\pi\alpha f}{m_{e}\omega\Gamma^{\prime}%
}~I_{las}~\frac{\left(  \Gamma^{\prime}/2\right)  ^{2}}{\left(  \Gamma
^{\prime}/2\right)  ^{2}+\left(  \omega-\omega_{ij}^{(4)}\right)  ^{2}}%
~T_{ij}^{(4)}(e_{\lambda}), \label{tauij4}%
\end{equation}
where $\alpha$ is the fine-structure constant, $f$ the oscillator strength of
the $2^{3}\mathrm{S}$ -$2^{3}\mathrm{P}$ transition ($f$=0.5391
\cite{reff,DrakeHB}), $m_{e}$ the electron mass, $\Gamma^{\prime}/2$
the\ total damping rate of the optical coherence between the $2^{3}\mathrm{S}$
and the $2^{3}\mathrm{P}$ states, and $T_{ij}^{(4)}(e_{\lambda})$ the
transition matrix element between $\mathrm{Y}_{i}$ and $\mathrm{Z}_{j}$
\cite{Nacher85} for the light polarisation vector $e_{\lambda}$. The
Lorentzian factor in equation~\ref{tauij4} is responsible for resonant
absorption (the absorption rate is decreased when the light frequency differs
from the transition frequency $\omega_{ij}^{(4)}/2\pi,$ $\hbar\omega
_{ij}^{(4)}$ being the energy difference between $\mathrm{Y}_{i}$ and
$\mathrm{Z}_{j}$). At low enough pressure, $\Gamma^{\prime}$ is equal to the
radiative decay rate $\Gamma$ of the $2^{3}\mathrm{P}$ state (the inverse of
its lifetime), related to the oscillator strength $f$ by~:
\begin{equation}
\Gamma=\frac{2\alpha\omega^{2}\hbar}{3mc^{2}}f,
\end{equation}
from which $\Gamma$=1.022$\times$10$^{7}$~s$^{-1}$ is obtained. Atomic
collisions contribute to $\Gamma^{\prime}$ by a pressure-dependent amount of
order 10$^{8}$~s$^{-1}$/mbar~\cite{Bloch85}. The transition matrix elements
$T_{ij}^{(4)}(e_{\lambda})$ are evaluated in the $\left(  L,S\right)  $
representation for each of the three light polarisation states $\lambda=x\pm
iy$ (circularly polarised light $\sigma_{\pm}$ propagating along the $z$-axis
set by the field $B$) and $\lambda=z$ (transverse light with $\pi$
polarisation) using the selection rules given in table~\ref{srule} in the
Appendix. The transformation operator $P_{4}$ given in table~\ref{P4} in the
Appendix is used to change between the $\left(  L,S\right)  $ and the $\left(
J\right)  $ representations in the $2^{3}\mathrm{P}$ state of $^{4}%
\mathrm{He}$ and thus write a simple 9$\times$9 matrix expression for the
Hamiltonian in the $\left(  L,S\right)  $ representation, $H_{9}$~:%
\begin{equation}
H_{9}=P_{4}^{-1}H_{9fs}P_{4}+H_{9Z}, \label{H9}%
\end{equation}
where $H_{9fs}$ is the diagonal matrix of table~\ref{sfinemat} in the Appendix
and $H_{9Z}$ the diagonal matrix written from equation~\ref{Hz4}.

\textit{The level structure of }$^{3}He$\textit{ excited states} is determined
from the total Hamiltonian, including hyperfine interactions. The
fine-structure term $H_{fs}$ is expressed in the total angular momentum
representation $\left(  J\right)  $ using slightly different parameters
(table~\ref{sfineval} in the Appendix) due to the small mass dependence of the
fine-structure intervals. For the Zeeman term we also take into account the
nuclear contribution:
\begin{equation}
H_{Z}^{(3)}=\mu_{B}\left(  g_{L}^{\prime}L+g_{S}^{\prime}S+g_{I}I\right)
\cdot B. \label{Hz3}%
\end{equation}
The values of all $g$-factors for $^{3}\mathrm{He}$ are listed in
table~\ref{lande} in the Appendix. For the hyperfine term, we consider the
simple contact interaction between the nuclear and electronic spins and~a
correction term $H_{hfs}^{cor}$~:%
\begin{equation}
H_{hfs}=A_{\mathrm{L}}I\cdot S+H_{hfs}^{cor}. \label{Hhfs}%
\end{equation}
The matrix elements of the main contact interaction term in the decoupled
$\left(  S,I\right)  $ representation and the values of the constants
$A_{\mathrm{S}}$ and $A_{\mathrm{P}}$ for the $2^{3}\mathrm{S}$ and
$2^{3}\mathrm{P}$ states are given in table~\ref{shfine} in the Appendix. The
correction term only exists for the $2\mathrm{P}$
state~\cite{Hinds85,Hijikata88}. When restricted to the triplet levels, it
only depends on 2 parameters which are given with the matrix elements in table
\ref{shfinecor} in the Appendix.

As in the case of $^{4}\mathrm{He}$, we compute level structure in the
decoupled representation, now $\left(  L,S,I\right)  .$ The vectors of the
decoupled bases are noted $\left|  m_{S}:m_{I}\right\rangle $ for the
$2^{3}\mathrm{S}$ state, and $\left|  m_{L},m_{S}:m_{I}\right\rangle $ for the
$2^{3}\mathrm{P}$ state. For the $2^{3}\mathrm{S}$ state, the 6$\times$6
matrix expression $H_{6}$ of the Hamiltonian is~:%
\begin{equation}
H_{6}=H_{6Z}+F_{6}, \label{H6}%
\end{equation}
where $H_{6Z}$ is the diagonal matrix written from equation~\ref{Hz3} and
$F_{6}$ the matrix representation of $H_{hfs}$\ (table~\ref{shfine} in the
Appendix). For the $2^{3}\mathrm{P}$ state, the 18$\times$18 matrix $H_{18}$
is constructed~ from $H_{9fs},$ $F_{6},$ $H_{hfs}^{cor}$ and $H_{Z}^{(3)}$:%
\begin{multline}
\left\langle m_{L}^{\prime},m_{S}^{\prime}:m_{I}^{\prime}\left|
H_{18}\right|  m_{L},m_{S}:m_{I}\right\rangle =\nonumber\\
\delta\left(  m_{I}^{\prime},m_{I}\right)  \left\langle m_{L}^{\prime}%
,m_{S}^{\prime}\left|  H_{9fs}\right|  m_{L},m_{S}\right\rangle \nonumber\\
+\delta\left(  m_{L}^{\prime},m_{L}\right)  \left\langle m_{S}^{\prime}%
:m_{I}^{\prime}\left|  F_{6}\right|  m_{S}:m_{I}\right\rangle \nonumber\\
+\left\langle m_{L}^{\prime},m_{S}^{\prime}:m_{I}^{\prime}\left|
H_{hfs}^{cor}+H_{Z}^{(3)}\right|  m_{L},m_{S}:m_{I}\right\rangle . \label{H18}%
\end{multline}
The mass-dependent constants in $H_{9fs}$ and $F_{6}$ are set to the values
given for $^{3}\mathrm{He}$ in the Appendix.

Notations similar to those of the zero-field calculation of
reference~\cite{Nacher85} are used. The 6 sublevels of the $2^{3}\mathrm{S}$
state are called $\mathrm{A}_{1}$ to $\mathrm{A}_{6},$ the 18 levels of the
$2^{3}\mathrm{P}$ state $\mathrm{B}_{1}$ to $\mathrm{B}_{18}$ by increasing
values of the energy (see figure~\ref{figniv})\footnote{There is a difference
with notations used in reference~\cite{Nacher85}~: within each level of given
$F,$ labelling of state names was increased for convenience from largest to
lower values of $m_{F}$. In an applied field, this would correspond to an
energy decrease inside each of the $F$ levels (except the $P_{0}$, $F$=1/2
level). This is the reason for the new labelling convention, which is more
convenient in an applied field.}. An alternative notation system, given in
table~\ref{tabnomsAi} in the Appendix, is also used for the 6 sublevels of the
$2^{3}\mathrm{S}$ state when an explicit reference to the total angular
momentum projection $m_{F}$ is desired. Transition probabilities from
$\mathrm{A}_{i}$ to $\mathrm{B}_{j}$ due to monochromatic light are given by a
formula similar to equation~\ref{tauij4}~:
\begin{equation}
\frac{1}{\tau_{ij}}=\frac{4\pi\alpha f}{m_{e}\omega\Gamma^{\prime}}%
~I_{las}~\frac{\left(  \Gamma^{\prime}/2\right)  ^{2}}{\left(  \Gamma^{\prime
}/2\right)  ^{2}+\left(  \omega-\omega_{ij}\right)  ^{2}}~T_{ij}(e_{\lambda
}),\label{tauij}%
\end{equation}
in which the level energy differences $\hbar\omega_{ij}$ and the transition
matrix elements $T_{ij}(e_{\lambda})$ are generalisations of the $B$=0 values
computed in reference~\cite{Nacher85}. To simplify notations, no upper index
(3) is attached to $\tau_{ij},$ $\omega_{ij}$ and $T_{ij}$ in the case of
$^{3}\mathrm{He}$ atoms. The small mass effect on the oscillator strength $f$
(of order 10$^{-4}$~\cite{DrakeHB}) is neglected. Sum rules on the transition
matrix elements for each polarisation $e_{\lambda}$ give the relations:%
\begin{equation}
\sum\nolimits_{i,j}T_{ij}^{(4)}\left(  e_{\lambda}\right)  =3,\hspace
{0.5cm}\sum\nolimits_{i,j}T_{ij}\left(  e_{\lambda}\right)  =6.\label{sumrule}%
\end{equation}
In the following, the dependence of the transition matrix elements on the
polarisation vector $e_{\lambda}$ will not be explicitly written. Given the
selection rules described in the Appendix, the polarisation vector
corresponding to a non-zero transition matrix element can be unambiguously determined.

All optical transition energies\ $\hbar\omega_{ij}^{(4)}$ and $\hbar
\omega_{ij}$ will be referenced to $\hbar\omega_{57}(0),$ the energy of the
C$_{1}$ transition in null magnetic field, which connects levels A$_{5}$ and
A$_{6}$ to levels B$_{7}$ to B$_{10}$. Energy differences will be noted
$\epsilon,$ for instance$~$:%
\begin{equation}
\epsilon_{ij}^{(4)}/\hbar=\omega_{ij}^{(4)}-\omega_{57}(0),~~\epsilon
_{ij}/\hbar=\omega_{ij}-\omega_{57}(0).
\end{equation}
They are determined from computed energy splittings in the $2^{3}\mathrm{S}$
and $2^{3}\mathrm{P}$ states of each isotope. The isotope shift contribution
is adjusted to obtain the precisely measured C$_{9}$-D$_{2}$ interval (810.599
MHz, \cite{Shiner95}).

\subsubsection{\label{SECnumresults}Numerical Results}

Numerical computation of level structure and absorption spectra in an applied
field $B$ is performed in 3 steps. Matrices $H_{9}$ (for $^{4}\mathrm{He}$),
$H_{6}$ and $H_{18}$ (for $^{3}\mathrm{He}$) are computed first. A standard
matrix diagonalisation package (the double-precision versions of the jacobi
and eigsrt routines \cite{Numrec}) is then used to compute and sort all
eigenvalues (energies) and eigenvectors (components of the atomic states in
the decoupled bases). All transition matrix elements $T_{ij}^{(4)}$ and
$T_{ij}$ are finally evaluated, and results are output to files for further
use or graphic display. Actual computation time is insignificant (e.g. 20~ms
per value of $B$\ on a PC using a compiled Fortran program \cite{medemander}),
so that we chose not to use the matrix symmetries to reduce the matrix sizes
and the computational load, in contrast with references
\cite{Hinds85,Hijikata88}. We thus directly obtain all transitions of the 1083
nm line: 19 transitions for $^{4}\mathrm{He}$ (6 for $\sigma_{+}$ and for
$\sigma_{-},$ 7 for $\pi$ light polarisation), and 70 transitions for
$^{3}\mathrm{He}$ (22 for $\sigma_{+}$ and for $\sigma_{-},$ 26 for $\pi$
light polarisation). These numbers are reduced for $B$=0 to 18 for
$^{4}\mathrm{He}$ (the $\mathrm{Y}_{2}\rightarrow\mathrm{Z}_{7},$ i.e.
$\left|  0\right\rangle \rightarrow\left|  1;0\right\rangle $ $\pi$ transition
has a null probability) and to 64 for $^{3}\mathrm{He}$ (the $F$%
=1/2$\rightarrow$ $F$=5/2 transitions, from $\mathrm{A}_{5}$ or $\mathrm{A}%
_{6}$ to any of $\mathrm{B}_{1}$ to $\mathrm{B}_{6}$ are forbidden by
selection rules). Due to level degeneracy, the usual D$_{0}$, D$_{1}$and
D$_{2}$ lines of\ $^{4}\mathrm{He,}$ and C$_{1}$ to C$_{9}$ lines
of\ $^{3}\mathrm{He}$ \cite{Nacher85} are obtained for $B$=0
(figure~\ref{figraies}, and table~\ref{tabB0} in the Appendix).
\begin{figure}[h]
\centerline{ \psfig{file=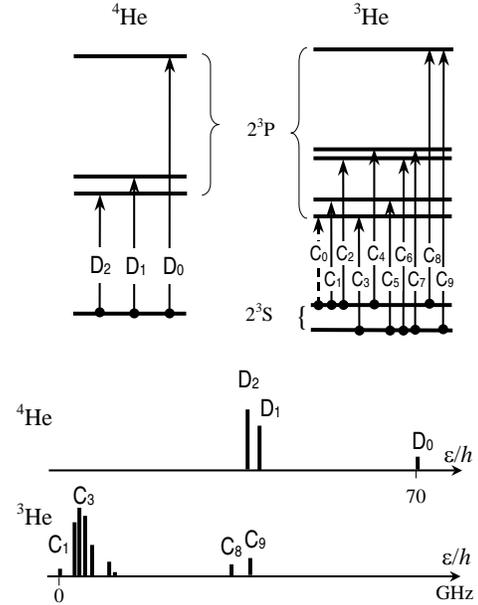, height=8 cm, clip= } } \caption{Top :
Components of the 1083 nm line in $^{4}\mathrm{He}$ (left) and $^{3}%
\mathrm{He}$ (right) for $B$=0. D$_{J}$ is the usual name of the transition to
the $2^{3}\mathrm{P}_{J}$ level. C$_{1}$ to C$_{9}$ refer to the transitions
by increasing energy. C$_{0}$ is the additional lowest energy component
corresponding to a forbidden transition in zero field (see text). Bottom :
positions of all the spectral lines resulting from level splittings and
isotope shift. The total frequency span from C$_{1}$ to D$_{0}$ is
70.442~GHz.}%
\label{figraies}%
\end{figure}

\begin{figure}[h]
\centerline{ \psfig{file=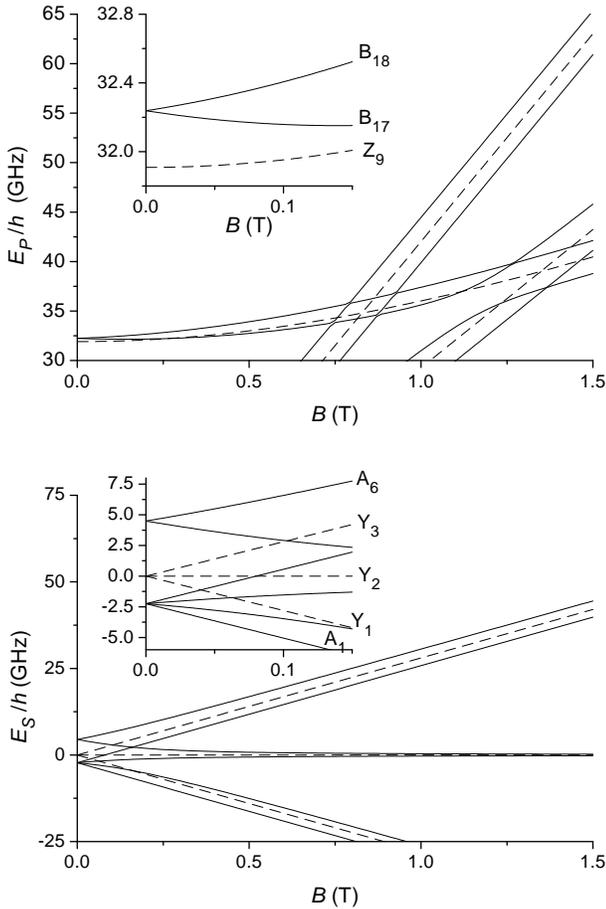, width=8 cm, clip= } } \caption{Level
energies $E_{S}$ and $E_{P}$\ of the $2^{3}\mathrm{S}$ and $2^{3}\mathrm{P}$
states are computed as a function of the applied field $B.$ They result from
the fine-structure and Zeeman energy contributions for $^{4}\mathrm{He}$
(dashed lines),with the additional hyperfine contributions for $^{3}%
\mathrm{He}$ (solid lines). The inserts are close-ups of the low-field
regions. In the $2^{3}\mathrm{P}$ state (top), only the highest lying levels
are shown for simplicity. }%
\label{figSP}%
\end{figure}

An applied magnetic field removes level degeneracies (Zeeman splittings
appear) and modifies the atomic states and hence the optical transition
probabilities. Examples of level Zeeman energy shifts are shown in
figure~\ref{figSP} for the $2^{3}\mathrm{S}$ state and the highest-lying
$2^{3}\mathrm{P}$ levels (originating from the $2^{3}\mathrm{P}_{\mathrm{0}}$
state at low field). In the $2^{3}\mathrm{S}$ state, the two Zeeman splittings are
simply proportional to $B$ (28 GHz/T) for $^{4}\mathrm{He}$, as inferred from
equation~\ref{Hz4}. For $^{3}\mathrm{He,}$ significant deviations from
linearity occur for $\mathrm{A}_{2}$, $\mathrm{A}_{3}$ and $\mathrm{A}_{5}$
above a field of order 0.1~T, for which Zeeman shifts are significant compared
to the hyperfine splitting. Level crossing of $\mathrm{A}_{4}$ and
$\mathrm{A}_{5}$ (and hence interchange of names of the two states) occurs at
0.1619~T. A high-field decoupled regime is almost reached for 1.5~T
(figure~\ref{figSP}, bottom). Analytical expression for state energies and
components on the decoupled $\left\{  \left|  m_{S}:m_{I}\right\rangle
\right\}  $ basis are given in the Appendix (equations \ref{Wplushl} to
\ref{Almoins}). In the $2^{3}\mathrm{P}$ state, the situation is more complex
due to fine-structure interactions and to the larger number of levels. In
particular no level remains unaffected by $B$ in contrast with the situation
for $\mathrm{Y}_{2}.$ As a result, $\pi$ transitions in $^{4}\mathrm{He}$ from
$\mathrm{Y}_{2}$ to $\mathrm{Z}_{3}$ or $\mathrm{Z}_{9}$ have non-zero Zeeman
frequency shifts, which are proportional to $B^{2}$ at low and moderate field
although they originate from the linear Zeeman term in the Hamiltonian
(equation \ref{Hz4}). Even the $2^{3}\mathrm{P}_{\mathrm{0}}$ states, for
which level mixing is weakest due to the large fine-structure gap, experience
significant field effects (figure~\ref{figSP}, top and top insert). The
splitting between $\mathrm{B}_{17}$ and $\mathrm{B}_{18}$ (2.468 GHz/T at low
field) is only weakly affected by the nuclear term in equation~\ref{Hz3}
($\pm$16.2~MHz/T), and mostly results from level mixing and electronic Zeeman effects.

\begin{figure}[h]
\centerline{ \psfig{file=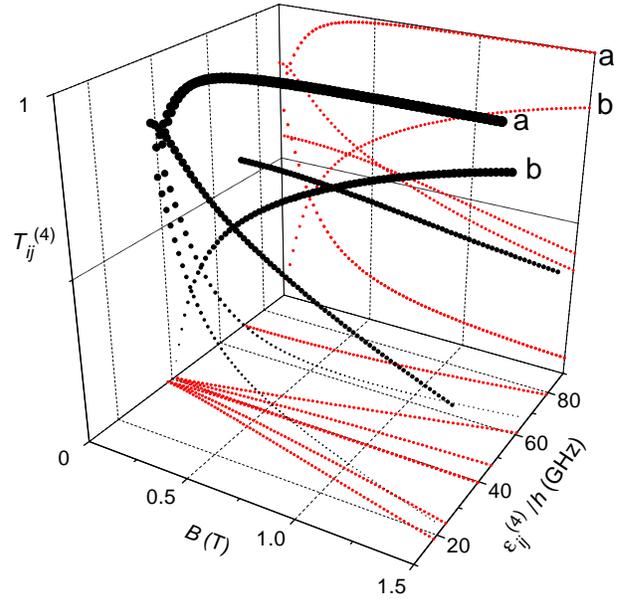, width=8 cm, clip= } } \caption{3D plot of
the transition matrix elements for $\pi$ light polarisation in $^{4}%
\mathrm{He,}$ $T_{ij}^{(4)}\left(  \pi\right)  ,$ (vertical axis) and of
transition energy difference $\epsilon$ (with respect to C$_{1}$) as a
function of $B$. To highlight the changes of the transition intensities, the
symbol sizes (symbol areas) are proportional to $T_{ij}^{(4)}.$ a and b are
the strongest lines in high field (see text). The projection onto the base
plane is the usual Zeeman diagram.}%
\label{figpi4}%
\end{figure}

Examples of optical transition intensities and Zeeman frequency shifts are
shown in figure~\ref{figpi4} for $^{4}\mathrm{He}$ ($\pi$ light polarisation).
Shifts of transition frequencies are clearly visible on the projection onto
the base plane, the usual Zeeman diagram (in which no line crossing occurs).
Large changes are also induced by the applied field on the transition
probabilities. In particular, the forbidden D$_{1}$transition $(\mathrm{Y}%
_{2}\rightarrow\mathrm{Z}_{7})$ has a matrix element which linearly increases
at low field ($T_{27}^{(4)}$=3.3$\times\left|  B\right|  $), and approaches 1
in high field (curve labelled b in figure~\ref{figpi4}). The two other strong
lines at high field (2 curves labelled a superimposed in figure~\ref{figpi4})
originate from the low-field $\mathrm{Y}_{1}\rightarrow\mathrm{Z}_{2}$ and
$\mathrm{Y}_{3}\rightarrow\mathrm{Z}_{8}$ transitions of D$_{2}$ and D$_{1}$
lines. All other line intensities tend to 0 at high field, in a way consistent
with the sum rules of equation~\ref{sumrule}.

\begin{figure}[h]
\centerline{ \psfig{file=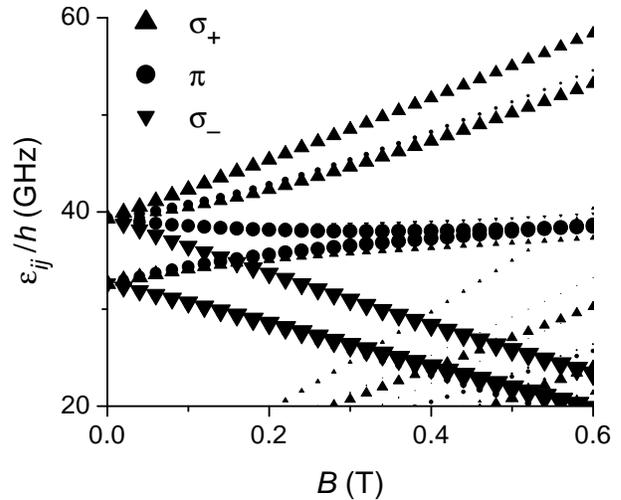, width=8 cm, clip= } } \caption{Plot of
the Zeeman line shifts for the C$_{8}$ and C$_{9}$ lines of $^{3}\mathrm{He.}$
The symbol types refer to different light polarisation states. The symbol
sizes (areas) are proportional to the corresponding transition matrix elements
$T_{ij}$.}%
\label{figC89}%
\end{figure}

The splittings of the C$_{8}$ and C$_{9}$ lines of $^{3}\mathrm{He}$ at
moderate field are displayed in figure \ref{figC89}. Circularly polarised
light, which is used in OP experiments to deposit angular momentum in the gas,
is more efficiently absorbed for $\sigma_{-}$ polarisation when a field is
applied. This can tentatively be related to results of our OP experiments~: at
$B$=0.1~T, the most efficient pumping line is actually found to be C$_{9}$,
$\sigma_{-}.$ OP results at $B$=0.6~T reported in reference~\cite{Flowers97}
also indicate that $\sigma_{-}$ polarisation is more efficient, in spite of
fluorescence measurements suggesting an imperfect polarisation of the light.
One must however note that OP efficiency does not only depend on light
absorption, and that level structures and metastability exchange collisions
(which will be discussed below) also play a key role. Let us finally mention
that the forbidden C$_{0}$ transition becomes allowed in an applied field, but
that the transition probabilities remain very weak and vanish again at high
field. 

\subsubsection{\label{discussionth}Discussion}

To assess the consequences of the main approximation in our calculations,
namely the restriction of the Hamiltonian to the triplet $2^{3}\mathrm{P}$
configuration, we have made a systematic comparison to the results of the full
calculation including singlet-triplet mixing.

In the case of $^{4}\mathrm{He,}$ exact energies are of course obtained for
$B$=0, and level energy differences do not exceed 1~kHz for $B$=2~T. This very
good agreement results from the fact that the lowest order contribution of the
singlet admixture in the $2^{3}\mathrm{P}$ state to the Zeeman effect is a
third order perturbation~\cite{Lhuillier76,Lewis70}. For similar reasons,
computed transition probabilities are also almost identical, and our simple
calculation is perfectly adequate for most purposes if the proper parameters
of table~\ref{sfineval} in the Appendix are used.

For $^{3}\mathrm{He}$ we have compared the results of the full calculation
involving all 24 sublevels in the $2\mathrm{P}$ states as described by the
Yale group~\cite{Hinds85} and two forms of the simplified calculation
restricted to the 18 sublevels of the $2^{3}\mathrm{P}$ state. The simpler
form only contains the contact interaction term $A_{\mathrm{P}}I\cdot S$ in
the hyperfine Hamiltonian of equation~\ref{Hhfs}. The more complete
calculations makes use of the full hyperfine interaction, with the correction
term $H_{hfs}^{cor}$ introducing two additional parameters, $d$ and $e$.

\begin{figure}[h]
\centerline{ \psfig{file=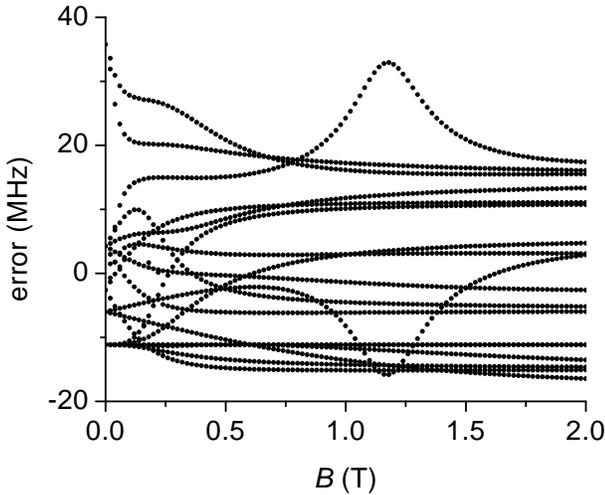, width=8 cm, clip= } } \caption{Errors
in calculated energies for the 18 sublevels of the $^{3}\mathrm{He}$
$2^{3}\mathrm{P}$ state resulting from the sole use of the contact term
$A_{\mathrm{P}}I\cdot S$ in the hyperfine Hamiltonian are plotted for
different magnetic field intensities. Large changes of the errors occur close
to level anticrossings (e.g. $m_{F}$=1/2 $\mathrm{B}_{15}$ and $\mathrm{B}%
_{17}$ levels close to 1.18 T, see figure~\ref{figSP}).}%
\label{figer1par}%
\end{figure}

The simpler 1-parameter calculation is similar to the zero-field calculation
of reference~\cite{Nacher85} (with updated values of all energy parameters),
and provides a limited accuracy as shown in figure~\ref{figer1par}.
Errors on the computed energies of the levels of several tens of
MHz cannot be significantly reduced by adjusting $A_{\mathrm{P}}$ to a more
appropriate value. This is a direct evidence that the correction term
$H_{hfs}^{cor}$ unambiguously modifies the energy level diagram, with clear
field-dependent signatures at moderate field and around 1~T. Errors on the
transition matrix elements up to a few 10$^{-3}$ also result from neclecting
$H_{hfs}^{cor}.$ Using this simple 1-parameter form of the hyperfine
Hamiltonian should thus be avoided when a better accuracy is desired.

\begin{figure}[h]
\centerline{ \psfig{file=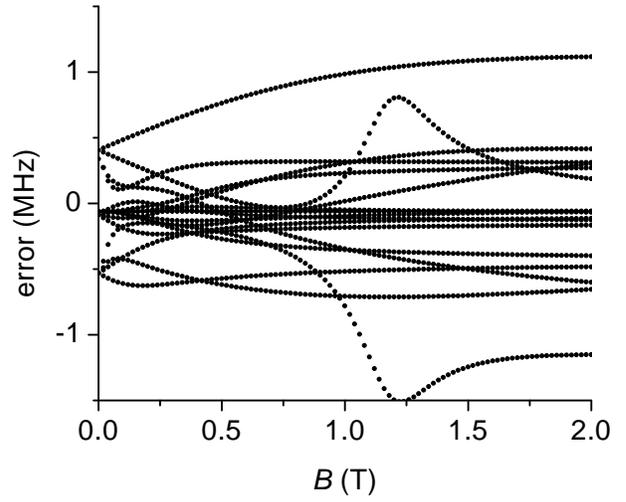, width=8 cm, clip= } } \caption{Errors
in calculated energies for the 18 sublevels of the $^{3}\mathrm{He}$
$2^{3}\mathrm{P}$ state resulting from the use of the full 3-parameter
hyperfine Hamiltonian are plotted for different magnetic field intensities. As
in figure \ref{figer1par}, large changes of the errors occur close to the
$\mathrm{B}_{15}$ - $\mathrm{B}_{17}$ level anticrossing around 1.18 T.}%
\label{figer3par}%
\end{figure}

The 3-parameter calculation has also been compared to the full calculation,
and a slight adjustment of parameters $A_{\mathrm{P}},$ $d$ and $e$ with
respect to the corresponding parameters $C,$ $D/2$ and $E/5$ of the full
calculation (which also includes the off-diagonal parameter $C^{\prime}%
$~\cite{Hinds85}), now provides a more satisfactory agreement for the level
energies, shown in figure~\ref{figer3par}. Transitions matrix elements $T_{ij}$ are also obtained with good
accuracy (1.3$\times$10$^{-4}$ r.m.s. difference on average for all
transitions and fields, with a maximum difference of 5$\times$10$^{-4}$). The
values of the 3 parameters $A_{\mathrm{P}},$ $d$ and $e$ differ by -0.019\%,
3.4\% and 4.6\% from the corresponding parameters in the full calculation.
This adjustment may effectively represent part of the off-diagonal effects in
the singlet admixture, and allows to decrease the energy errors, especially
below 1~T (by a factor 2-3).

The other important approximation in our calculations is related to the
simplifications in the Zeeman Hamiltonian written in equations \ref{Hz4} and
\ref{Hz3}. The first neglected term is the non-diagonal contribution
proportional to the small parameter $g_{x}$~in equation~\ref{Hzlin}. The
non-zero elements of this additional contribution are of order $\mu_{B}%
Bg_{x}/5,$ i.e. 1-2~MHz/T in frequency units. The other neglected term results
from the quadratic Zeeman effect~\cite{Lhuillier76,Yan94,Lewis70}, which
introduces matrix elements of order $\left(  \mu_{B}B\right)  ^{2}/R_{\infty}$
($R_{\infty}=3.29\times10^{6}$~GHz stands for the Rydberg constant in
frequency units), i.e. 60 kHz/T$^{2}$. For most applications, and in
particular for the experimental situations considered in the following,
corrections introduced by these additional terms in the Zeeman Hamiltonian
would indeed be negligible, not larger for instance than the errors in
figure~\ref{figer3par}.

\subsection{\label{ME}Metastability exchange}

So far we have only addressed the effects of an applied magnetic field on the
level energies and structures of the $2^{3}\mathrm{S}$ and $2^{3}\mathrm{P}$
states of helium, and the consequences for the 1083 nm optical transition. In
this section we now consider the effect of the applied field on the angular
momentum transfer between atoms during the so-called metastability exchange
(ME) collisions, which are spin-exchange binary collisions between an atom in
the ground state and one in the metastable $2^{3}\mathrm{S}$ state. They play
an important role in all experimental situations where the atoms in the
$2^{3}\mathrm{S}$ state are a minority species in a plasma (e.g. with a number
density 10$^{10}$--10$^{11}$ cm$^{-3}$ compared to 2.6$\times$10$^{16}$ for
the total density of atoms at 1 mbar in typical OP conditions). These
collisions may be neglected only for experiments on atomic beams or trapped
atoms for which radiative forces are used to separate the metastable atoms
from the much denser gas of atoms in the ground state.

To describe the statistical properties of a mixture of ground state and
$2^{3}\mathrm{S}$ state atoms we shall use a standard density operator
formalism and extend the treatment introduced by Partridge and
Series~\cite{Partridge66} for ME collisions, later improved
\cite{JDR73,Pinard80} and used for an OP model \cite{Nacher85}, to the case of
an arbitrary magnetic field\textit{.} Our first goal is to show that the
``spin temperature'' concept \cite{Anderson59,Happer72,Nacher85} remains
valid, and that the population distribution in the $2^{3}\mathrm{S}$ state is
fully determined (in the absence of OP and of relaxation) by the nuclear
polarisation $M$ of the ground state. This is the result on which the optical
detection method of section~\ref{SECoptmeas} relies. The second goal is to
provide a formalism allowing to study the consequences of hyperfine decoupling
on spin transfer during ME collisions.

\subsubsection{\label{SECrateeq}Derivation of rate equations}

We shall consider situations in which no resonance is driven between the two
magnetic sublevels of the ground state of $^{3}\mathrm{He}$ (the ground state
of $^{4}\mathrm{He}$ has no structure) nor between sublevels of the
$2^{3}\mathrm{S}$ state. We can thus \emph{a priori} assume that the density
operators $\rho_{g}$ (in the ground state of $^{3}\mathrm{He}$), $\rho_{3}$
(in the $2^{3}\mathrm{S}$ state of \ $^{3}\mathrm{He}$) and $\rho_{4}$ (in the
$2^{3}\mathrm{S}$ state of \ $^{4}\mathrm{He}$) are statistical mixtures of
eigenstates of the Hamiltonian. The corresponding density matrices are thus
diagonal and contain only populations (no coherences). With obvious
notations~:%
\begin{align}
\rho_{g}  &  =\frac{1+M}{2}\left|  +\right\rangle \left\langle +\right|
+\frac{1-M}{2}\left|  -\right\rangle \left\langle -\right| \label{defrhos}\\
\rho_{3}  &  =\sum\nolimits_{1}^{6}a_{i}\left|  \mathrm{A}_{i}\right\rangle
\left\langle \mathrm{A}_{i}\right| \\
\rho_{4}  &  =\sum\nolimits_{1}^{3}y_{i}\left|  \mathrm{Y}_{i}\right\rangle
\left\langle \mathrm{Y}_{i}\right|  , \label{defrhos3}%
\end{align}
where $a_{i}$ and $y_{i}$ are relative populations ($%
{\textstyle\sum}
a_{i}$=$%
{\textstyle\sum}
y_{i}$=1).

An important feature of ME in helium is that in practice no depolarisation
occurs during the collisions, due to the fact that all involved angular
momenta are spins \cite{Pinard80}.\ Three kinds of ME collisions occur in an
isotopic mixture, depending on the nature of the colliding atoms
($^{3}\mathrm{He}$ or $^{4}\mathrm{He}$) and on their state (ground state:
$^{3,4}\mathrm{He}$, $2^{3}\mathrm{S}$ state: $^{3,4}\mathrm{He}%
^{\mathrm{\ast}}$).

\begin{enumerate}
\item  Following a ME collision between $^{3}\mathrm{He}$ atoms:%
\[%
\begin{array}
[c]{ccccccc}%
^{3}\mathrm{He} & + & ^{3}\mathrm{He}^{\mathrm{\ast}} & \rightarrow &
^{3}\mathrm{He}^{\mathrm{\ast}} & + & ^{3}\mathrm{He}\\
\rho_{g} &  & \rho_{3} &  & \rho_{3}^{\prime} &  & \rho_{g}^{\prime}%
\end{array}
\]

the nuclear and electronic angular momenta are recombined in such a way that
the density operators just after collision $\rho_{g}^{\prime}$ and $\rho
_{3}^{\prime}$ are given by:%
\begin{align}
\rho_{g}^{\prime} &  =\operatorname{Tr}_{e}\rho_{3}\label{Tre33prime}\\
\rho_{3}^{\prime} &  =\rho_{g}\otimes\operatorname{Tr}_{n}\rho_{3}%
\label{rhoprime}%
\end{align}
where $\operatorname{Tr}_{e},$ $\operatorname{Tr}_{n}$ are trace operators
over the electronic and nuclear variables respectively \cite{JDR73}.

\item  Following the collision between a$\ $ground state $^{4}$\textrm{He
}atom and a $2^{3}\mathrm{S}$ state $^{3}\mathrm{He}$ atom:%
\[%
\begin{array}
[c]{ccccccc}%
^{4}\mathrm{He} & + & ^{3}\mathrm{He}^{\mathrm{\ast}} & \rightarrow &
^{4}\mathrm{He}^{\mathrm{\ast}} & + & ^{3}\mathrm{He}\\
&  & \rho_{3} &  & \rho_{4}^{\prime\prime} &  & \rho_{g}^{\prime\prime}%
\end{array}
\]
the density operators just after collision are:%
\begin{align}
\rho_{g}^{\prime\prime} &  =\operatorname{Tr}_{e}\rho_{3}\label{Tre33second}\\
\rho_{4}^{\prime\prime} &  =\operatorname{Tr}_{n}\rho_{3}.\label{Trnrho3}%
\end{align}

\item  Following the collision between a$\ 2^{3}\mathrm{S}$ state
$^{4}\mathrm{He}$ atom and a ground state $^{3}\mathrm{He}$ atom:%
\[%
\begin{array}
[c]{ccccccc}%
^{3}\mathrm{He} & + & ^{4}\mathrm{He}^{\mathrm{\ast}} & \rightarrow &
^{3}\mathrm{He}^{\mathrm{\ast}} & + & ^{4}\mathrm{He}\\
\rho_{g} &  & \rho_{4} &  & \rho_{3}^{\prime\prime\prime} &  &
\end{array}
\]
the density operator of the outgoing $2^{3}\mathrm{S}$ state $^{3}\mathrm{He}$
atom is:%
\begin{equation}
\rho_{3}^{\prime\prime\prime}=\rho_{g}\otimes\rho_{4}.\label{rhothird}%
\end{equation}
\end{enumerate}

The partial trace operations in equations~\ref{Tre33prime} to \ref{Trnrho3} do
not introduce any coherence term in density matrices. In contrast, the tensor
products in equations \ref{rhoprime} and \ref{rhothird} introduce several
off-diagonal terms. These are driving terms which may lead to the development
of coherences in the density operators, but we shall show in the following
that they can usually be neglected. Since $m_{F}$ is a good quantum number,
off-diagonal terms in the tensor products only appear between 2 states of
equal $m_{F}.$ With the state names defined in the Appendix
(table~\ref{tabnomsAi}), these off-diagonal terms are proportional to the
operators $\left|  \mathrm{A}_{+}^{h}\right\rangle \left\langle \mathrm{A}%
_{+}^{l}\right|  $ and $\left|  \mathrm{A}_{+}^{l}\right\rangle \left\langle
\mathrm{A}_{+}^{h}\right|  $ (for $m_{F}$=1/2), $\left|  \mathrm{A}_{-}%
^{h}\right\rangle \left\langle \mathrm{A}_{-}^{l}\right|  $ and $\left|
\mathrm{A}_{-}^{l}\right\rangle \left\langle \mathrm{A}_{-}^{h}\right|  $ (for
$m_{F}$=-1/2). These four operators contain a similar $\sin\theta_{\pm}%
\cos\theta_{\pm}$ factor, where $\theta_{+}$ and $\theta_{-}$ are
field-dependent level mixing parameters (see equations \ref{Ahplus} to
\ref{Almoins} in the Appendix). They have a time evolution characterised by a
fast precessing phase, $\exp\left(  i\Delta E_{\pm}t/\hbar\right)  ,$
depending on the frequency splittings $\Delta E_{\pm}/\hbar$ of the
eigenstates of equal $m_{F}.$

The time evolution of the density operators due to ME collisions is obtained
from a detailed balance of the departure and arrival processes for the 3 kinds
of collisions. This is similar to the method introduced by Dupont-Roc
\textit{et al}.~\cite{JDR73} to derive a rate equation for $\rho_{g}$ and
$\rho_{3}$ in pure $^{3}\mathrm{He}$ at low magnetic field. It\ is extended
here to isotopic mixtures, and the effect of an arbitrary magnetic field is discussed.

The contribution of ME collisions to a rate equation for the density operator
$\rho_{3}$ of the $2^{3}\mathrm{S}$ state $^{3}\mathrm{He}$ is obtained
computing an ensemble average in the gas of the terms arising from collisions
with a $^{3}\mathrm{He}$ atom (collision type 1, first line of the rate
equation) and from collisions involving different isotopes (types 2 and 3,
second line):
\begin{multline}
\left(  \frac{d}{dt}n_{3}\rho_{3}\right)  _{\mathrm{ME}}=\left\langle
n_{3}N_{3}\sigma v_{33}\left(  -\rho_{3}+\rho_{3}^{\prime}\right)
\right\rangle \\
+\left\langle \sigma v_{34}\left(  -n_{3}N_{4}\rho_{3}+n_{4}N_{3}\rho
_{3}^{\prime\prime\prime}\right)  \right\rangle \label{balance}%
\end{multline}
in which the brackets $\left\langle ~\right\rangle $ correspond to the
ensemble average, $N_{3,4}$ and $n_{3,4}$ are the ground state and
$2^{3}\mathrm{S}$ state atom number densities of $^{3}\mathrm{He}$ and
$^{4}\mathrm{He}$, $\sigma$ is the ME cross section, $v_{33}$ and $v_{34}$ are
the relative velocities of colliding atoms. The ensemble averaging procedure
must take into account the time evolution between collisions. For non-resonant
phenomena, one is interested in steady-state situations, or in using rate
equations to compute slow evolutions of the density operators (compared to the
ME collision rate 1/$\tau$). For the constant or slowly varying populations
($\rho_{3},$ diagonal terms in $\rho_{3}^{\prime}$ and $\rho_{3}^{\prime
\prime\prime}$), the ensemble averages simply introduce the usual thermally
averaged quantities $\overline{\sigma v_{33,34}}$. For each time-dependent
off-diagonal coherence, the average of the fast precessing phase involves a
weighting factor $\exp(-\tilde{t}/\tau)$ to account for the distribution of
times $\tilde{t}$ elapsed after a collision:%
\begin{align}
\left\langle \exp\left(  i\Delta E_{\pm}\tilde{t}/\hbar\right)  \right\rangle
&  =\frac{1}{\tau}\int_{0}^{\infty}\exp\left(  i\Delta E_{\pm}\tilde{t}%
/\hbar\right)  \exp(-\tilde{t}/\tau)d\tilde{t}\nonumber\\
&  =(1-i{}\Delta E_{\pm}\tau/\hbar)^{-1}.
\end{align}
The ME collision rate 1/$\tau$ is proportional to the helium pressure (e.g.
3.75$\times$10$^{6}$~s$^{-1}$/mbar in pure $^{3}\mathrm{He}$ at room
temperature~\cite{JDR71}). The splittings $\Delta E_{\pm}/\hbar$ are equal to
the hyperfine splitting (6.74 GHz) in zero field, and their variation with $B$
can be derived from equations \ref{Wplushl} and \ref{Wmoinshl}: $\Delta
E_{+}/\hbar$ increases with $B$ while $\Delta E_{-}/\hbar$ initially decreases
slightly (down to 6.35 GHz at 0.08 T), and both splittings increase linearly
with $B$ at high field (with a slope of order 27.7 GHz/T). Under such
conditions, $\Delta E_{\pm}\tau/\hbar$%
$>$%
1000 in a gas at 1~mbar. All coherences can thus be safely neglected in
equation~\ref{balance} which, when restricted to its diagonal terms, can be
written as:
\begin{multline}
\left(  \frac{d}{dt}n_{3}\rho_{3}\right)  _{\mathrm{ME}}=n_{3}N_{3}%
\overline{\sigma v_{33}}\left(  -\rho_{3}+\sum\Pi_{i}\rho_{g}\otimes
\operatorname{Tr}_{n}\rho_{3}\Pi_{i}\right) \\
+\overline{\sigma v_{34}}\left(  -n_{3}N_{4}\rho_{3}+n_{4}N_{3}\sum\Pi_{i}%
\rho_{g}\otimes\rho_{4}\Pi_{i}\right)  \label{echmet3}%
\end{multline}
in which we note $\Pi_{i}$=$\left|  \mathrm{A}_{i}\right\rangle \left\langle
\mathrm{A}_{i}\right|  $ the projector on the eigenstate $\mathrm{A}_{i}$.

The main differences with the result derived in~\cite{JDR73} are the
additional term (second line) in equation~\ref{echmet3}, which reflects the
effect of $^{3}\mathrm{He-}^{4}\mathrm{He}$ collisions, and the replacement of
the projectors on the $F$ substates by the more general projectors $\Pi_{i}$
on the eigenstates in the applied field $B$. Considering the field dependence
of the frequency splittings $\Delta E_{\pm}/\hbar,$ the condition on the ME
collision rate 1/$\tau$ is only slightly more stringent at 0.08 T that for
$B$=0, but can be considerably relaxed at high magnetic field. The linear
increase of $\Delta E_{\pm}$ and the 1/$B$ decrease of $\sin\theta_{\pm}$ (see
figure~\ref{figthetas}) in the common factor of all coherences provide a
$B^{2}$ increase of the acceptable collision rate, and hence of the operating
pressure, for which coherences can be neglected in all density matrices, and
equations \ref{defrhos} to \ref{defrhos3} are valid.

The other rate equations are directly derived from equations \ref{Tre33prime},
\ref{Tre33second} and \ref{Trnrho3}. Since there is no tensor product, hence
no coherence source term in the relevant arrival terms, explicit ensemble
averages are not required and the contribution of ME is:%
\begin{multline}
\left(  \frac{d}{dt}N_{3}\rho_{g}\right)  _{\mathrm{ME}}=n_{3}N_{3}%
\overline{\sigma v_{33}}\left(  -\rho_{g}+\operatorname{Tr}_{e}\rho_{3}\right)
\\
+\overline{\sigma v_{34}}\left(  -n_{4}N_{3}\rho_{g}+n_{3}N_{4}%
\operatorname{Tr}_{e}\rho_{3}\right)  \label{echfond}%
\end{multline}%
\begin{equation}
\left(  \frac{d}{dt}n_{4}\rho_{4}\right)  _{\mathrm{ME}}=-n_{4}N_{3}%
\overline{\sigma v_{34}}\rho_{4}+n_{3}N_{4}\overline{\sigma v_{34}%
}\operatorname{Tr}_{n}\rho_{3}. \label{echmet4}%
\end{equation}
A trace operation performed on equations \ref{echmet3} and \ref{echmet4} shows
that
\begin{equation}
\left(  \frac{d}{dt}n_{4}\right)  _{\mathrm{ME}}=-\left(  \frac{d}{dt}%
n_{3}\right)  _{\mathrm{ME}},
\end{equation}
as expected since the three ME processes preserve the total number of atoms in
the $2^{3}\mathrm{S}$ excited state.

\subsubsection{Steady-state solutions}

Total rate equations can be obtained by adding relaxation and OP terms to the
ME terms of equations \ref{echmet3} and \ref{echmet4}. They can be used to
compute steady-state density matrices for the $2^{3}\mathrm{S}$ state, while
the nuclear polarisation $M$ (and hence $\rho_{g}$) may have a very slow rate
of change (compared to the ME collision rate 1/$\tau$). In these steady-state
situations, the contribution of ME collisions to the total rate equations for
the $2^{3}\mathrm{S}$ state just compensates the OP and relaxation
contributions. Since the latter are traceless terms (OP and relaxation only
operate population transfers between sublevels of a given isotope), by taking
the trace of equation \ref{echmet3} or \ref{echmet4} one finds that the
$2^{3}\mathrm{S}$ state and ground state number densities have the same
isotopic ratio $R$:%
\begin{equation}
R=n_{4}/n_{3}=N_{4}/N_{3}.\label{rapmetfond}%
\end{equation}
This results from having implicitly assumed in section~\ref{SECrateeq} that
all kinds of collisions have the same ME cross section\footnote{The assumption
that ME cross sections have the same value $\sigma$ for any isotopic
combination of colliding atoms is valid at room temperature. This would be
untrue at low temperature due to a 10~K isotopic energy difference in the 20
eV excitation energy of the $2^{3}\mathrm{S}$ state atoms (the energy is lower
for $^{3}\mathrm{He}$).}. In fact destruction and excitation of of the
$2^{3}\mathrm{S}$ state in the gas may differently affect $^{3}\mathrm{He}$
and $^{4}\mathrm{He}$ atoms, e.g. due to the different diffusion times to the
cell walls. However in practice this has little effect compared to the very
frequent ME collisions, which impose the condition of
equation~\ref{rapmetfond} for number densities.

To obtain simpler expressions for the rate equations, we introduce the
parameters:
\begin{align}
1/\tau_{e}  &  =N_{3}\overline{\sigma v_{33}}+N_{4}\overline{\sigma v_{34}%
}\label{unsurtaue}\\
1/T_{e}  &  =n_{3}\overline{\sigma v_{33}}+n_{4}\overline{\sigma v_{34}}\\
\mu &  =\overline{\sigma v_{33}}/\overline{\sigma v_{34}}%
\end{align}
The rates $1/T_{e}$ and $1/\tau_{e}$ correspond to the total ME collision rate
for a $^{3}\mathrm{He}$ atom in the ground state and in the $2^{3}\mathrm{S}$
state respectively; in steady state, equation~\ref{rapmetfond} can be used to
show that $\tau_{e}/T_{e}=n_{3}/N_{3}=n_{4}/N_{4}$. Since the thermal velocity
distributions simply scale with the reduced masses of colliding atoms, the
value of the dimensionless parameter $\mu$ can be evaluated from the energy
dependence of $\sigma.$ From reference~\cite{Fitzsimmons68} one can estimate
$\mu\simeq$1.07 at room temperature. Rewriting equation~\ref{unsurtaue} as:
\begin{equation}
1/\tau_{e}=\left(  N_{3}+N_{4}\right)  \overline{\sigma v_{33}}\frac{1+R/\mu
}{1+R},
\end{equation}
the ME rate in an isotopic mixture thus differs from that in pure
$^{3}\mathrm{He}$ by at most 7~\% (at large $R$) for a given total density.

Using these notations and the isotopic ratio $R$ (equation \ref{rapmetfond}),
the contributions of ME to rate equations become:%
\begin{align}
\left(  \frac{d}{dt}\rho_{g}\right)  _{\mathrm{ME}}  &  =\frac{1}{T_{e}%
}\left(  -\rho_{g}+\operatorname{Tr}_{e}\rho_{3}\right) \label{raterhog}\\
\left(  \frac{d}{dt}\rho_{4}\right)  _{\mathrm{ME}}  &  =\frac{1}{\tau_{e}%
}\frac{1}{R+\mu}\left(  -\rho_{4}+\operatorname{Tr}_{n}\rho_{3}\right)
\label{raterho4}\\
\left(  \frac{d}{dt}\rho_{3}\right)  _{\mathrm{ME}}  &  =\frac{1}{\tau_{e}%
}\left\{  -\rho_{3}\right.  +\frac{1}{R+\mu}\sum\Pi_{i}\times\nonumber\\
&  \left.  \times\left(  \mu\rho_{g}\otimes\operatorname{Tr}_{n}\rho_{3}%
+R\rho_{g}\otimes\rho_{4}\right)  \Pi_{i}\right\}  . \label{raterho3}%
\end{align}

In a way similar to that of reference~\cite{Nacher85} we can transform
equations \ref{raterhog} to \ref{raterho3} into an equivalent set of coupled
rate equations for the nuclear polarisation $M$ and for the relative
populations $a_{i}$ and $y_{i}$ of the $2^{3}\mathrm{S}$ states $\mathrm{A}%
_{i}$ and $\mathrm{Y}_{i}$:%
\begin{align}
\left(  \frac{d}{dt}M\right)  _{\mathrm{ME}}  &  =\frac{1}{T_{e}}\left(
-M+\sum_{k=1}^{6}L_{k}a_{k}\right) \label{rateM}\\
\left(  \frac{d}{dt}y_{i}\right)  _{\mathrm{ME}}  &  =\frac{1}{\tau_{e}}%
\frac{1}{R+\mu}\left(  -y_{i}+\sum_{k=1}^{6}G_{ik}^{4}a_{k}\right)
\label{ratey}\\
\left(  \frac{d}{dt}a_{i}\right)  _{\mathrm{ME}}  &  =\frac{1}{\tau_{e}%
}\left\{  -a_{i}+\frac{1}{R+\mu}\right.  \times\nonumber\\
\times\lbrack\mu\sum_{k=1}^{6}(E_{ik}^{3}  &  +MF_{ik}^{3})a_{k}+R\left.
\sum_{k=1}^{3}(E_{ik}^{4}+MF_{ik}^{4})y_{k}]\right\}  . \label{ratea}%
\end{align}
The values of the $M$-independent but $B$-dependent matrices $L,$ $G^{4},$
$E^{3},$ $F^{3},$ $E^{4}$ and $F^{4}$ are provided in the Appendix. Equation
\ref{rateM} directly results from computing $\operatorname{Tr}_{n}\rho
_{g}I_{z}$ using equation \ref{raterhog}. The rate equations on the relative
populations are obtained by computing $\left\langle \mathrm{Y}_{i}\right|
d\rho_{4}/dt\left|  \mathrm{Y}_{i}\right\rangle $ using
equation~\ref{raterho4} and $\left\langle \mathrm{A}_{i}\right|  d\rho
_{3}/dt\left|  \mathrm{A}_{i}\right\rangle $ using equation~\ref{raterho3}.
The linear $M$-dependence in equation \ref{ratea} directly results from that
of $\rho_{g}$ (equation \ref{defrhos}).

\subsubsection{\label{SECspinT}Spin temperature distributions}

Anderson et al.~\cite{Anderson59} have proposed that, under some conditions
including fast spin exchange, the relative populations of sublevels should
follow a Boltzmann-like distribution in angular momentum. This was explicitly
verified for pure $^{3}\mathrm{He}$ in low field~\cite{Nacher85}, and will now
be shown for isotopic mixtures and arbitrary magnetic field. In situations
where OP and relaxation processes have negligible effect on populations, the
steady-state density operators are easily derived from the rate equations
\ref{raterhog} to \ref{raterho3}:%
\begin{align}
\rho_{g} &  =\operatorname{Tr}_{e}\rho_{3}\\
\rho_{4} &  =\operatorname{Tr}_{n}\rho_{3}\label{equilrho4}\\
\rho_{3} &  =\sum\Pi_{i}\rho_{g}\otimes\operatorname{Tr}_{n}\rho_{3}\Pi
_{i}.\label{equilrho3}%
\end{align}
Simply assuming that the populations $a_{i}$ in the $2^{3}\mathrm{S}$ state of
$^{3}\mathrm{He}$ only depend on the $m_{F}$ values in the states
$\mathrm{A}_{i},$ one can directly check using equations \ref{Ahplus} to
\ref{Almoins} that both $\operatorname{Tr}_{e}\rho_{3}$ (and hence equations
\ref{raterhog} and \ref{rateM}) and $\operatorname{Tr}_{n}\rho_{3}$ do not
depend on $B.$ Noting $e^{\beta}$ the ratio (1+$M$)/(1-$M$) of the populations
in the ground state (1/$\beta$ plays the role of a spin temperature), one
derives from equation~\ref{equilrho3} that the ratios of populations in the
$2^{3}\mathrm{S}$ state of $^{3}\mathrm{He}$ are field-independent, and given
by $e^{\beta\Delta m_{F}}$. These populations thus have the same distribution
at all magnetic fields:%
\begin{equation}
a_{i}=e^{\beta m_{F}}/(e^{3\beta/2}+2e^{\beta/2}+2e^{-\beta/2}+e^{-3\beta
/2}).\label{aivsbeta}%
\end{equation}
Using equations \ref{equilrho4} and \ref{Ahplus} to \ref{Almoins}, the
populations in the $2^{3}\mathrm{S}$ state of $^{4}\mathrm{He}$ are in turn
found to obey a similar distribution:%
\begin{equation}
y_{i}=e^{\beta m_{S}}/(e^{\beta}+1+e^{-\beta}).
\end{equation}

\subsubsection{\label{SECOPtheo}OP effects on populations}

The general problem of computing all the atomic populations in arbitrary
conditions is not considered in this work. It has been addressed in the low
field limit using specific models in pure $^{3}\mathrm{He}$ \cite{Nacher85}
and in isotopic mixtures \cite{Larat91}. Here we extend it to arbitrary
magnetic fields, but only consider the particular situation where there is no
ground state nuclear polarisation ($M$=0). It can be met in absorption
spectroscopy experiments, in which a weak probing light beam may induce a
deviation from the uniform population distribution imposed by $M$=0 (infinite
spin temperature). We shall assume in the following that the OP process
results from a depopulation mechanism in which population changes are created
only by excitation of atoms from selected $2^{3}\mathrm{S}$ sublevels and not
by spontaneous emission from the $2^{3}\mathrm{P}$ state (where populations
are randomised by fast relaxation processes). In $^{3}\mathrm{He}$, this
depopulation mechanism has been checked to dominate for pressures above $\sim
$1~mbar~\cite{Colegrove63,Nacher85}.

Under such conditions, and further assuming that Zeeman splittings are large
enough for a single sublevel in the $2^{3}\mathrm{S}$ state to be pumped
($\mathrm{A}_{p}$ or $\mathrm{Y}_{p},$ depending on the probed isotope), the
OP contribution to the time evolution of populations has a very simple form.
Given the ME contribution of equation \ref{ratey} or \ref{ratea}, the total
rate equation for the pumped isotope becomes:%
\begin{align}
\frac{d}{dt}y_{i}  &  =\left(  \frac{d}{dt}y_{i}\right)  _{\mathrm{ME}}%
+\frac{y_{p}}{\tau_{p}}\left(  \frac{1}{3}-\delta_{\mathrm{Y}}(i,p)\right)
\label{rateyOP}\\
\text{or }~\frac{d}{dt}a_{i}  &  =\left(  \frac{d}{dt}a_{i}\right)
_{\mathrm{ME}}+\frac{a_{p}}{\tau_{p}}\left(  \frac{1}{6}-\delta_{\mathrm{A}%
}(i,p)\right)  , \label{rateaOP}%
\end{align}
where 1/$\tau_{p}$ is the pumping rate (equation \ref{tauij4} or \ref{tauij})
and $\delta_{\mathrm{A}}(i,p)$=1 (resp. $\delta_{\mathrm{Y}}(i,p)$=1) if the
level $\mathrm{A}_{i}$ (resp. $\mathrm{Y}_{i}$) is the pumped level, 0
otherwise. In steady state, the relative populations can be obtained from the
kernel of the 9$\times$9 matrix representation of the set of rate equations
\ref{ratey} and \ref{rateaOP}, or \ref{ratea} and \ref{rateyOP}.

\begin{figure}[tbh]
\centerline{ \psfig{file=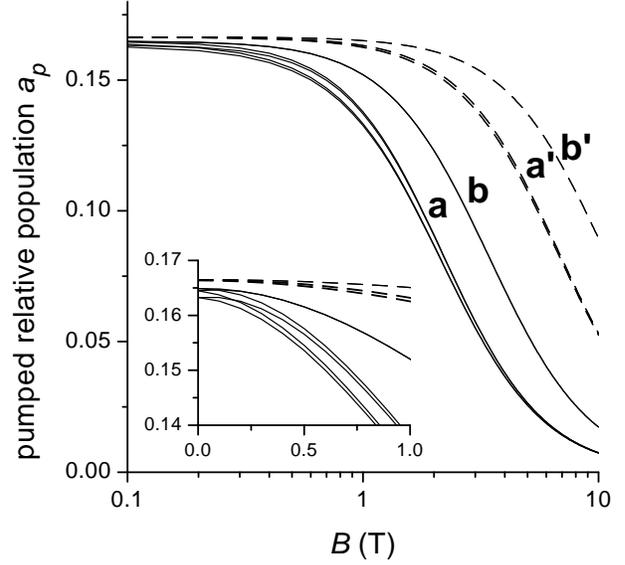, width=8 cm, clip= } } \caption{OP on a
$^{3}\mathrm{He}$ level : the computed decrease of the relative population
$a_{p}$ of the pumped level of $^{3}\mathrm{He}$ below its equilibrium value
(1/6) is plotted as a function of $B$ \ for two values of the reduced pumping
rate $\tau_{e}/\tau_{p}$ : 10$^{-2}$ (solid lines) and 10$^{-3}$ (dashed
lines). For the 2 populations $a_{p}$=$a_{+}^{l}$ or $a_{-}^{h},$ the results
fall on the same curve (labelled b or b$^{\prime}$ depending on $\tau_{e}%
/\tau_{p}$). For the 4 other populations, a significantly larger decrease is
found (groups of curves a and a$^{\prime}$). The insert is an expanded view of
the low-field region (with a linear $B$-scale).}%
\label{figPO3}%
\end{figure}

When a $^{3}\mathrm{He}$ line is probed, the $^{4}\mathrm{He}$ relative
populations $y_{i}$ are fully imposed by ME processes only
(equation~\ref{ratey}), and can be subsituted in equation~\ref{rateaOP}. The
relative populations $a_{i}$ can then be obtained as the kernel of a 6$\times
$6 matrix which depends on the pumped level index $p$, on the reduced pumping
rate $\tau_{e}/\tau_{p}$ and on the magnetic field $B,$ but not on the
isotopic ratio $R$ (as discussed at the end of the Appendix). Results showing
the decrease of the pumped population $a_{p},$ and hence of the absorption of
the incident light, are displayed in figure \ref{figPO3} as a function of $B$.
A strong absorption decrease is found at high $B,$ due to larger
population changes induces by OP when ME less efficiently transfers the
angular momentum to the ground state (let us recall that we assume $M$=0, so
that the ground state is a reservoir of infinite spin temperature). Population
changes are found to be weaker for the two states $\mathrm{A}_{+}^{l}$ and
$\mathrm{A}_{-}^{h}$, which both have $\left|  0:\pm\right\rangle $ as the
main component in high field (curves b and b$^{\prime}$ in figure \ref{figPO3}).

\begin{figure}[tbh]
\centerline{ \psfig{file=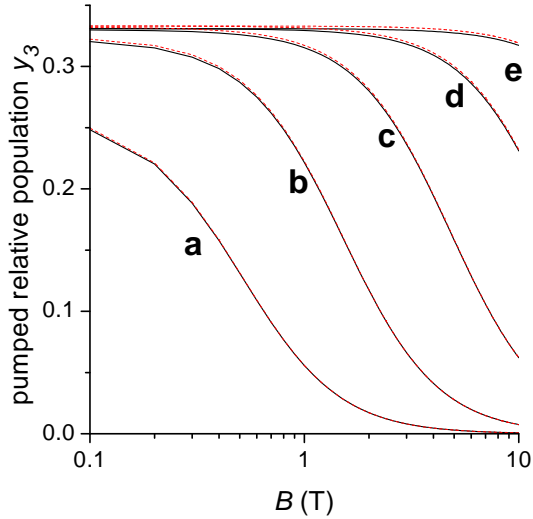, width=7 cm, clip= } } \caption{OP on a
$^{4}\mathrm{He}$ level : the decrease of the $^{4}\mathrm{He}$ population
$y_{3}$ below its equilibrium value (1/3) is plotted as a function of $B.$ The
population has been computed for two values of the reduced pumping rate
$\tau_{e}/\tau_{p}$ : 10$^{-2}$ (solid lines) and 10$^{-3}$ (dashed lines),
and different values of $R$ (ranging from 100 to 10$^{-3}$). The population
decrease mostly depends on the product $R\tau_{e}/\tau_{p}$ (10$^{-1}$%
,$\ $10$^{-2}$, ...10$^{-5}$ for curves a, b, ... e).}%
\label{figPO4}%
\end{figure}

\begin{figure}[tbhtbh]
\centerline{ \psfig{file=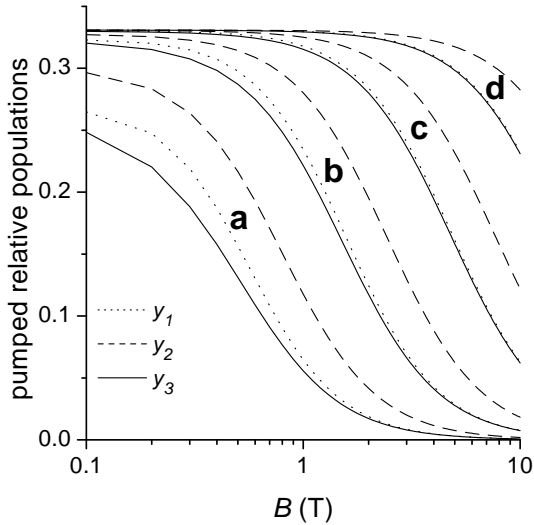, width=7 cm, clip= } } \caption{ OP
on a $^{4}\mathrm{He}$ level : the computed decrease of $^{4}\mathrm{He}$
populations $y_{1},$ $y_{2}$ and $y_{3}$ due to OP effects is plotted as a
function of $B.$ The reduced pumping rate $\tau_{e}/\tau_{p}$ =10$^{-2}$ is
that of the solid lines in figure \ref{figPO4}. Four values of the isotopic
ratio $R$ are used ($R\tau_{e}/\tau_{p}$ =10$^{-2}\ \ $to 10$^{-4}$ for curves
a to d).}%
\label{figPO4states}%
\end{figure}

When a $^{4}\mathrm{He}$ line is probed, the substitution and reduction
performed above cannot be made. The full set of populations turns out to vary
with the isotopic ratio $R.$ However, the OP effects are found to depend
mostly on the product $R\tau_{e}/\tau_{p}$, as shown in figure \ref{figPO4} in
the case of the state Y$_{3}$ ($m_{S}$=1). This influence of the isotopic
composition extends over a wide field range, which justifies the use of
isotopic mixtures to reduce the bias resulting from OP effects in our
systematic quantitative measurements of $^{4}\mathrm{He}$ line intensities.
Figure~\ref{figPO4states} displays a comparison of computed OP
effects for the 3 sublevels of the $^{4}\mathrm{He}$ $2^{3}\mathrm{S}$ states.
Results for the populations $y_{1}$ and $y_{3}$ are quite similar, except at
large $R.$ Weaker OP effects are found for the population $y_{2},$ for which
$m_{S}$=0. This feature is similar to the weaker OP effects found in
$^{3}\mathrm{He}$ for the two states which have $m_{S}$=0, $m_{I}$=$\pm1/2$ as
main angular momenta components in high field (see figure \ref{figPO3}%
).

\section{Experimental}

In this part we present experimental results of laser absorption experiments
probing the fine, hyperfine and Zeeman splittings of the 1083~nm transition in
$^{3}\mathrm{He}$ and $^{4}\mathrm{He}$. Accurate microwave
measurements~\cite{Storry00,Kponou81,Prestage85} and high precision laser
spectroscopy experiments~\cite{Castillega00,Shiner95,Minardi99,Shiner94} are
usually performed on helium atomic beams. In contrast, we have made
unsophisticated absorption measurements using a single free-running laser
diode on helium gas in cells at room temperature, under the usual operating
conditions for nuclear polarisation of $^{3}\mathrm{He}$ by OP. An important
consequence of these conditions is the Doppler broadening of all absorption
lines. Each transition frequency $\omega_{ij}/2\pi$ (in the rest frames of the
atoms) is replaced by a Gaussian distribution of width $\Delta:$%
\begin{equation}
\Delta=\left(  \omega_{ij}/2\pi\right)  \sqrt{2k_{B}T/M_{at}c^{2}},
\label{eqdoppler}%
\end{equation}
depending on the atomic mass $M_{at}$. At room temperature ($T$=300~K), the
Doppler widths for $^{3}\mathrm{He}$ and $^{4}\mathrm{He}$ are $\Delta_{3}%
$=1.19~GHz and $\Delta_{4}$=1.03~GHz (the full widths at half maximum, FWHM,
given by $2\Delta\sqrt{\ln2},$ are respectively 1.98 and 1.72~GHz). The narrow
Lorentzian resonant absorption profiles (with widths $\Gamma^{\prime}/2)$ in
equations \ref{tauij4} and \ref{tauij} are thus strongly modified, and broader
Voigt profiles are obtained for isolated lines in standard absorption
experiments. For $\Gamma^{\prime}\ll\Delta,$ these are almost Gaussian
profiles of width $\Delta$. In the following, we first report on Doppler-free
absorption measurements from which a good determination of frequency
splittings is obtained, then on integrated absorption intensities in different
magnetic fields. We finally demonstrate that absorption measurements may
provide a convenient determination of the nuclear polarisation $M$.

\subsection{\label{SECabsat}Line positions: saturated absorption measurements}

\subsubsection{\label{SECexpsetup}Experimental setup}

\begin{figure}[tbh]
\centerline{ \psfig{file=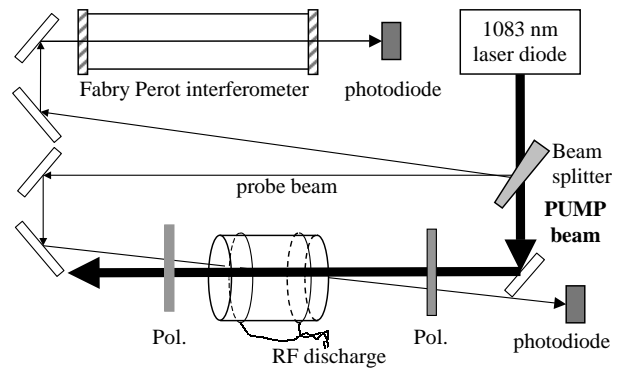, width=8 cm, clip= } } \caption{Main
elements of a saturated absorption experiment. A magnetic field $B$ is applied
either along the pump and probe beams, or perpendicular to the beams.
Polarising elements (Pol.) consist of linear polarisers and/or quarter-wave
plates depending on the desired polarisations of the pump and probe beams.}%
\label{figabsat}%
\end{figure}

The experiment arrangement is sketched in figure~\ref{figabsat}.
The helium gas samples are enclosed in sealed cylindrical Pyrex
glass cells, 5~cm in diameter and 5~cm in length. Cells are filled after a
careful cleaning procedure: bakeout at 700~K under high vacuum for several
days, followed by 100~W RF (27~MHz) or microwave (2.4~GHz) discharges in
helium with frequent gas changes until only helium lines are detected in the
plasma fluorescence. Most results presented here have been recorded in cells
filled with 0.53 mbar of $^{3}\mathrm{He}$ or helium mixture (25\%
$^{3}\mathrm{He}$, 75\% $^{4}\mathrm{He}$). Different gas samples (from 0.2 to
30 mbar, from 10\% to 100\% $^{3}\mathrm{He}$) have also been used to study
the effect of pressure and isotopic composition on the observed signals. A
weak RF\ discharge (%
$<$%
1~W at 3~MHz) is used to populate the $2^{3}\mathrm{S}$ state in the cell
during absorption measurements. The RF excitation is obtained using a pair of
external electrodes. Aligning the RF electric field with $B$ provides a higher
density of $2^{3}\mathrm{S}$ states and better OP results in high magnetic field.

Most data have been acquired using a specially designed air-core resistive
magnet (100~mm bore diameter, with transverse optical access for 20~mm beams).
In spite of a reduced footprint (30$\times$20~cm$^{2}$, axis height 15~cm)
which conveniently permits installation on an optical table, this magnet
provides a fair homogeneity. The computed relative inhomogeneity is
$<$%
10$^{-3}$ over a 5~cm long cylindrical volume 1~cm in diameter (typical volume
probed by the light beams in spectroscopy measurements), and
$<$%
3$\times$10$^{-3}$ over the total cell volume. This usually induces low enough
magnetic relaxation of nuclear polarisation in OP experiments. The
field-to-current ratio was calibrated using optically detected NMR resonance
of $^{3}\mathrm{He}$. Experiments for $B$ up to 0.12~T were performed with
this coil. Several saturated absorption experiments have been repeated and
extended up to 0.22~T using a standard electromagnet in the Institute of
Physics in Krakow.

The laser source is a 50 mW laser diode (model 6702-H1 formerly manufactured
by Spectra Diode Laboratories). Its output is collimated into a quasi-parallel
beam (typically 2$\times$6 mm in size) using an anti-reflection coated lens
($f$=8~mm). A wedge-shaped plate is used to split the beam into a main pump
beam (92\% of the total intensity), a reference beam sent through a confocal
Fabry Perot interferometer, and a probe beam (figure~\ref{figabsat}). A small
aperture limiting the probe beam diameter is used to only probe atoms lying in
the central region of the pump beam. A small angle is set between the
counterpropagating beams. This angle and sufficient optical isolation from the
interferometer are required to avoid any feedback of light onto the laser
diode, so that the emitted laser frequency only depends on internal parameters
of the laser. The polarisations of pump and probe beams are adjusted using
combinations of \ polarising cubes, 1/2-wave, 1/4-wave retarding plates. For
all the measurements of the present work, the same polarisation was used for
pump and probe beams. However valuable information on collisional processes
can be obtained using different polarisation and/or different frequencies for
the two beams.

The probe beam absorption is measured using a modulation technique. The RF
discharge intensity is modulated at a frequency $f_{\mathrm{RF}}$ low enough
for the density of the absorbing atoms $2^{3}\mathrm{S}$ to synchronously vary
($f_{\mathrm{RF}}\sim$100Hz). The signal from the photodiode monitoring the
transmitted probe beam is analysed using a lock-in amplifier. The amplitude of
the probe modulation thus measured, and the average value of the transmitted
probe intensity are both sampled (at 20 Hz), digitised, and stored. Absorption
spectra are obtained from the ratios of the probe modulation by the probe
average intensity. This procedure strongly reduces effects of laser intensity
changes and of optical thickness of the gas on the measured
absorptions~\cite{CourtadeTh}.

With its DBR (Distributed Bragg Reflector) technology, this laser diode
combines a good monochromaticity and easy frequency tuning. It could be
further narrowed and stabilised using external selective elements, and thus
used for high-resolution spectroscopy of the 1083 nm line of
helium~\cite{Minardi99}. Here we have simply used a free running diode, with a
linewidth dominated by the Schalow-Townes broadening factor of order 2-3~MHz
FWHM~\cite{Prevedelli}. The laser frequency depends on temperature and current
with typical sensitivities 20~GHz/K and 0.5~GHz/mA respectively. To reduce
temperature drifts, the diode is enclosed in a temperature regulated
1-dm$^{3}$ shield. Using a custom made diode temperature and current
controller, we obtain an overall temperature stability better than a mK, and a
current stability of order a $\mu$A. Frequency drifts over several minutes and
low frequency jitter are thus limited to 10~MHz at most.

\begin{figure}[h]
\centerline{ \psfig{file=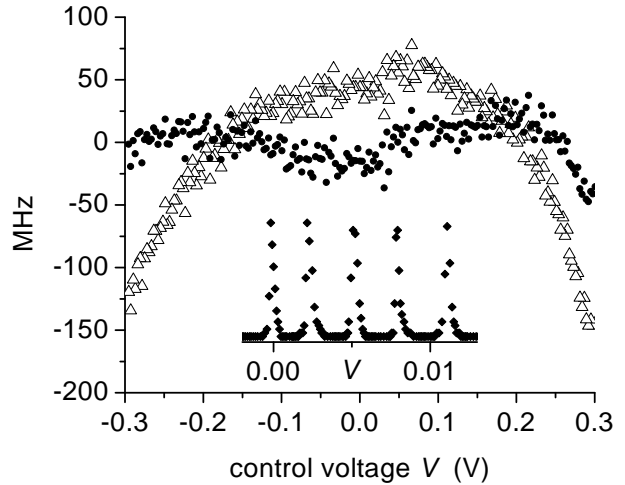, width=8 cm, clip= } }
\caption{Insert : fraction of recording of the light intensity transmitted by
the Fabry Perot interferometer during a temperature controlled laser frequency
scan (1~V of control voltage approximately induces changes of 2.5~K in\ laser
temperature and 50~GHz in emitted frequency). Main plot : residues of a linear
fit (open triangles) or parabolic fit (solid circles) of all transmitted peak
positions.}%
\label{figscanpics}%
\end{figure}

The laser frequency is adjusted, locked or swept by temperature control using
the on-chip Peltier cell and temperature sensor. The small deviation from
linearity of the frequency response was systematically measured during the
frequency sweeps by recording the peaks transmitted by the 150 MHz FSR
interferometer. Figure~\ref{figscanpics} displays an example of such a
measurement: for a control voltage $V$ (in Volts) the frequency offset (in
GHz) is 52.5$\times$(1+0.048$V$)$V$. The linear term in this conversion factor
is usually taken from more accurate zero-field line splitting measurements
(see below), but this small non-linear correction is used to improve the
determination of frequency scales for extended sweeps.

\subsubsection{\label{SECzerofield}Zero-field measurements}

\begin{figure}[h]
\centerline{ \psfig{file=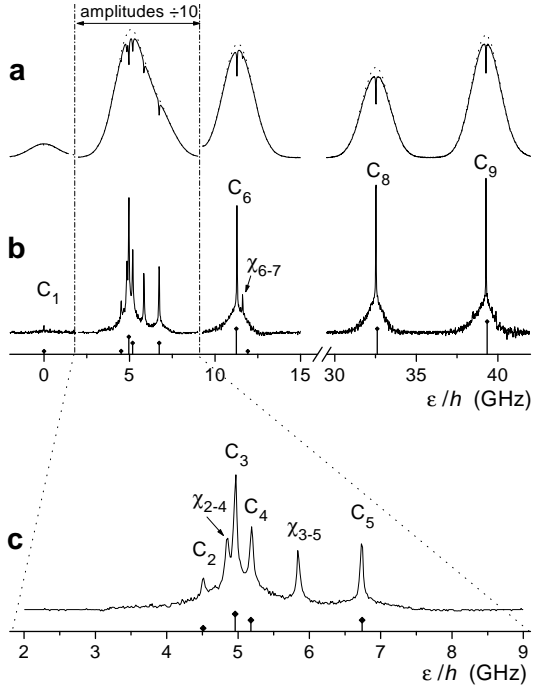, width=7 cm, clip= } }%
\caption{Zero-field saturated absorption spectrum of $^{3}\mathrm{He.}$ The
upper traces (a) represent the probe absorption signals when the pump beam is
applied (solid line) or blocked (dotted line). Traces b and c (an expanded
section of trace b) are the difference of probe absorptions with and without
the pump beam, with a $\times$4 vertical scale factor. The sections of the
spectrum comprising the lines C$_{2}$ to C$_{5}$ provide a much larger signal
and are displayed with a reduced amplitude. Frequency sweep rate is 160~MHz/s,
a value ensuring sufficient signal quality and manageable laser frequency
drifts. Transition lines are labelled C$_{1}$ to C$_{9}$, crossover resonances
between C$_{i}$ and C$_{j}$ are labelled $\chi_{i-j}$. The vertical bars on
the axes represent the computed line positions and intensities.}%
\label{figB0He3}%
\end{figure}

Results presented in this section have in fact been obtained in the earth
field in which Zeeman energy changes are of order 1~MHz or less and thus too
small to affect the results given the accuracy of our measurements. A small
permanent magnet is placed near the cell to strongly reduce the field
homogeneity and thus prevent any nuclear polarisation to build up when the
pump beam is absorbed. An example of frequency sweep over the whole
$^{3}\mathrm{He}$ absorption spectrum is given in figure~\ref{figB0He3}
($^{3}\mathrm{He}$ pressure: 0.53 mbar). The frequency scale is determined using the accurately known value
3$A_{\mathrm{S}}$/2$=$6.7397~GHz of the C$_{8}$-C$_{9}$ splitting. The
saturated absorption spectrum (solid line in figure~\ref{figB0He3}a) combines
the usual broad absorption lines and several narrow dips. The dips reveal a
reduced absorption of the probe by atoms interacting with the pump beam. These
are atoms with negligible velocity along the beam direction (Doppler-free
resonances) or atoms with a velocity such that the Doppler shift is just half
of the splitting of two transitions from or to a common level (crossover
resonances). An absorption signal recorded without the pump beam (dotted line
in figure~\ref{figB0He3}a) only shows the broad features, which can be fit by
a Doppler Gaussian profile for isolated lines (e.g. C$_{8}$ or C$_{9}$). It is
used to obtain the narrow resonances in figure~\ref{figB0He3}b by signal
substraction. The width $\delta$ of the narrow resonances is 50 MHz FWHM in
the conditions of figure~\ref{figB0He3}. It can be attributed to several
combined broadening processes, including pump saturation effects and signal
acquisition filtering.

\begin{figure}[h]
\centerline{ \psfig{file=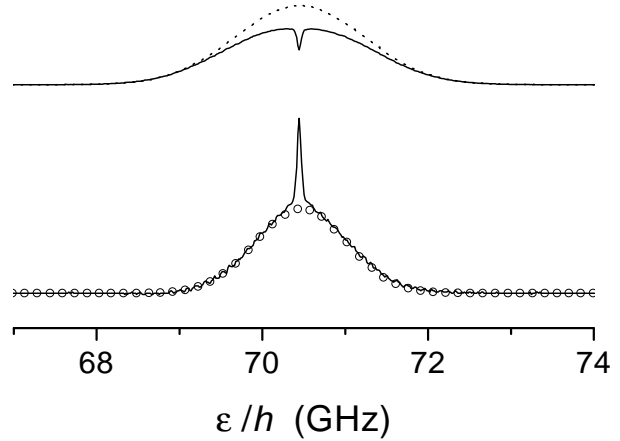, width=8 cm, clip= } }\caption{Zero-field
saturated absorption spectrum of the D$_{0}$ line in a 25\% $^{3}\mathrm{He}$
- 75\% $^{4}\mathrm{He}$ mixture (same total pressure, 0.53~mbar, as for
figure~\ref{figB0He3}). The lower trace also displays the squared absorption
amplitude (open circles), with an arbitrary weight adjusted to fit the
saturated absorption pedestal (see text).}%
\label{figD0Bnul}%
\end{figure}

Broader line features, the ``pedestals'' on which narrow resonances stand,
result from the combined effects of collisions in the gas. Fully elastic
velocity changing collisions tend to impose the same population distribution
in all velocity classes, and thus to identically decrease the probe absorption
over the whole velocity profile. In contrast, ME collisions involving a
$^{3}\mathrm{He}$ atom in a gas where the nuclear polarisation is $M$=0 induce
a net loss of angular momentum\footnote{The orientation is not totally lost at
each ME collision since the electronic part of the angular momentum is
conserved, and recoupled after collision. ME collisions thus contribute both
to orientation transfer between velocity classes and to orientation loss. The
relative importance of these contributions depends on the angular momentum
loss, which is reduced in high field by hyperfine decoupling. A specific
feature of ME collisions is that they involve small impact parameters and
large collision energies due to a centrifugal energy barrier
\cite{Fitzsimmons68,BarbéTh}, and thus usually produce large velocity
changes.} (a uniform population distribution tends to be imposed by the
infinite spin temperature, see section \ref{SECspinT}). In steady-state, the
un-pumped velocity classes will thus acquire only part of the population
changes imposed in the pumped velocity class. The relative amplitude of the
resulting broad pedestal results from the competition of collisional transfer
and collisional loss of orientation. One consequence is that it is almost
pressure-independent, as was experimentally checked. Another consequence is
that it strongly depends on $^{4}\mathrm{He}$ concentration in isotopic
mixtures, as illustrated in figure~\ref{figD0Bnul}. The much larger fractional amplitude of the pedestal results from
the weaker depolarising effect of the ME collisions: only 25\% of the exchange
collisions occur in this case with a depolarised $^{3}\mathrm{He}$ atom and
thus contribute to bleach the pump-induced population changes, while all
collisions induce velocity changes. In figure~\ref{figD0Bnul} is also
displayed the squared absorption profile, which closely matches the shape of
the pedestal. This is a general feature also observed in $^{3}\mathrm{He}$
spectra, which results from the linear response to the rather low pump power.
Assuming that the relative population changes are proportional to the absorbed
pump power, frequency detuning reduces by the same Doppler profile factor both
these population changes and the probed total population. The resulting
squared Doppler profile is still Gaussian, but with reduced width
$\Delta/\sqrt{2}.$

From this detailed analysis of the signal shapes we infer two main results.
First, very little systematic frequency shift is expected to result from the
distortion of the narrow Doppler-free resonances by the broad line pedestals
of neighbouring transitions (no shift at all for isolated lines). It can be
estimated to be of order $\delta^{2}/\Delta,$ i.e. of order 1~MHz and thus
negligible in these experiments. Second, population changes in the whole
velocity profile remain limited in pure $^{3}\mathrm{He}$ or in $^{3}%
\mathrm{He}$-rich isotopic mixtures. They amount for instance to a 7~\% change
for the C$_{9}$ transition when the pump beam is applied, as illustrated in
figure~\ref{figB0He3} (difference between dotted and solid line). They are
significantly larger in a helium mixture, e.g 30~\% for the conditions of
figure~\ref{figD0Bnul}. These OP effects on the populations would be reduced
for a reduced pump power, and the direct effect of the attenuated ($\div$25)
probe itself can be expected to be correspondingly reduced, and hence
negligible under similar conditions. However, this would be untrue in pure
$^{4}\mathrm{He}$ in which strong population relaxation is difficult to
impose, and precise absorption or saturated absorption experiments are less
conveniently performed. A strong magnetic field would also enhance the
perturbing effects of the probe beam, as was discussed in
section~\ref{SECOPtheo} and will be demonstrated in section~\ref{SECOP}.

\begin{figure}[h]
\centerline{ \psfig{file=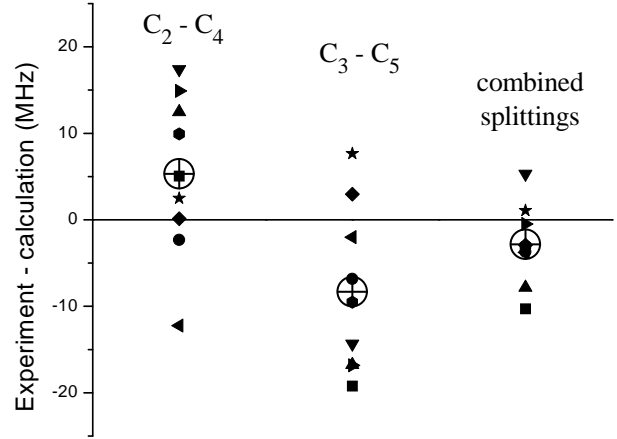, width=8 cm, clip= } } \caption{Difference
between measured and computed hyperfine splittings in the $2^{3}\mathrm{P}$
state of $^{3}\mathrm{He}$ in zero field for selected pairs of individual or
combined line splittings (see text). Each closed symbol corresponds to a
different experiment. The large open symbols are the statistical averages.}%
\label{figBnul}%
\end{figure}

A series of identical sweeps including the C$_{1}$-C$_{7}$ lines, similar to
that in figure~\ref{figB0He3}c, was performed to check for reproducibility.
The frequency scales were determined using the C$_{2}$-C$_{6}$ splittings
(these transitions connect the two hyperfine levels of the $2^{3}\mathrm{S}$
state to the same excited level, see figure~\ref{figraies}, and are thus
separated by 6.7397~GHz just as C$_{8}$-C$_{9}$). To directly probe hyperfine
level splittings in the $2^{3}\mathrm{P}$ state we consider the line
separations C$_{2}$-C$_{4}$ and C$_{3}$-C$_{5},$ which directly measure the
level energy differences between the lower pairs of $2^{3}\mathrm{P}$ levels
(see the zero field level diagram in figure \ref{figniv}). We also consider
the separation of the crossover resonances $\chi_{2-4}$-$\chi_{3-5}$, actually
computed from differences of the four line positions, which measures an
independent energy difference between these levels. Comparison to the values
computed using the 3-parameter hyperfine term in the Hamiltonian
(section~\ref{discussionth}) is shown in figure \ref{figBnul}.
For the two simple splittings, data scatter is $\pm$15 MHz about
the average. It is smaller for the third combined splitting, which involves
averaging pairs of line position measurements and thus a statistically reduced
uncertainty. The three averages have a 6~MHz r.m.s. difference with the
computed values, which is consistent with statistical uncertainty. This
analysis shows that our unsophisticated spectroscopic measurements are quite
accurate and have very little systematic error in the frequency determinations
(at most a few MHz over several GHz). It also justifies the use of the
correction term beyond the contact interaction term in the hyperfine
contribution (equation~\ref{Hhfs}) since our experimental errors are lower
than the effect of this correction term (figure~\ref{figer1par}).

\subsubsection{Field effect on line positions}

\begin{figure}[h]
\centerline{ \psfig{file=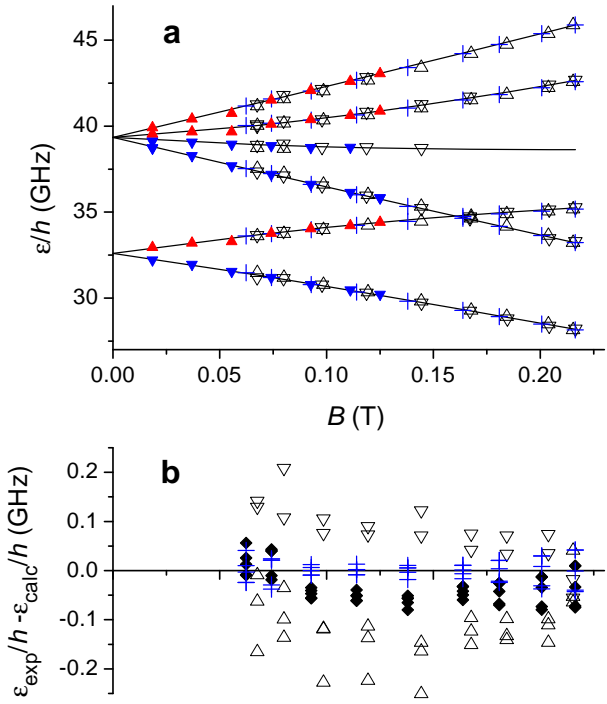, width=8 cm, clip= } }
\caption{\textbf{a~: }Plot of frequency shifts on the $^{3}\mathrm{He}$
$\mathrm{C}_{8}$ and $\mathrm{C}_{9}$ lines in an applied magnetic field
measured by a saturated absorption technique. Solid lines~: computed Zeeman
shifts (same as in figure~\ref{figC89}). Up and down triangles~: data for
$\sigma_{+}$ and $\sigma_{-}$ polarisations of light beams. Solid and open
symbols~: measurement using the air-core and iron yoke magnets respectively.
Crosses~: result of corrections applied the open symbol data set (see text).
\textbf{b~: }Differences between experimental and computed values for the
measurements in the iron yoke magnet are plotted as a function of $B.$ Open
symbols : raw data (no correction) ; filled symbols : data with field
corrections only ; crosses : data with field and frequency corrections (see
text).}%
\label{figexp2kG}%
\end{figure}

The most obvious effect of an applied magnetic field is to considerably
increase the number of observed narrow lines (Doppler-free lines and
intercombination lines) in the saturated absorption recordings. For
simplicity, we only present in figure~\ref{figexp2kG} selected results for the
set of $^{3}\mathrm{He}$ lines originating at low field from C$_{8}$ and
C$_{9}.$ These are the most efficient optical transitions for OP of
$^{3}\mathrm{He}$, and thus are of special interest. Furthermore, the analyses
of transition intensity measurements described in sections \ref{SECisotopic}
and \ref{SECoptmeas} request an accurate knowledge of all line positions for
these transitions. To reduce the number of observed lines and thus facilitate data
processing, the same circular polarisation ($\sigma_{+}$ or $\sigma_{-}$) was
used for the pump and the probe beam, and two measurements were thus performed
for each field value. For the higher field data (iron yoke magnet, open
symbols in figure~\ref{figexp2kG}a) the use of an imperfect circular
polarisation made it possible to usually observe all intense lines in each
measurement. Zero field scans were performed between measurements to evaluate
the slow drift of the laser frequency and provide reference measurements for
the Zeeman shifts.

For the measurements in the air-core magnet (up to 0.125 T), the agreement
between computed and measured line positions $\epsilon/h$ is fair, but
affected by fluctuating frequency offsets. The r.m.s. differences are quite
large (50 MHz and 110 MHz for $\sigma_{+}$ and $\sigma_{-}$ probes
respectively), but the largest contribution to these differences can be
attributed to a global frequency offset from scan to scan. Estimating the
offset of each scan from the average of the 3 recorded line positions, the
remaining differences are significantly reduced (9 MHz r.m.s. for $\sigma_{+}$
and $\sigma_{-}$). The accuracy of these line position measurements is thus
comparable to that of the zero-field measurements (figure~\ref{figBnul}).

For the measurements in the iron yoke electromagnet (up to 0.25 T), a similar
frequency offset adjustment is not sufficient to significantly reduce the
differences between measured and computed line positions
(figure~\ref{figexp2kG}b). For instance, they amount to 130 MHz r.m.s. for the
$\sigma_{+}$ measurements (open up triangles). Moreover, splittings within a
given scan consistently differ from the computed ones. Assuming that the
magnetic field is not exactly proportional to the applied current due to the
presence of a soft iron yoke in the magnet, we use the largest measured
splitting on the C$_{9}$ line (the transitions originating from the sublevels
$m_{F}$=$\pm$3/2) to determine the actual value of the magnetic field during
each scan. Solid symbols in figure~\ref{figexp2kG}b are computed differences
for these actual values of the magnetic field. Crosses (figures
\ref{figexp2kG}a and \ref{figexp2kG}b) correspond to fully corrected data,
allowing also for frequency offset adjusments. The final remaining differences
(22 MHz r.m.s.) are still larger than in the air-core magnet, yet this
agreement is sufficient to support our field correction and offset adjustment
procedures in view of the line intensity measurements described in the
following section.

\subsection{Line intensities}

\subsubsection{\label{SECexpsetup2}Experimental setup}

The experimental arrangement is either the one sketched in
figure~\ref{figabsat}, in which the pump beam is blocked and only a weak probe
beam (usually 0.1-1 mW/cm$^{2}$) is transmitted through the cell, or a
simplified version of the setup when no saturated absorption measurement is performed.

The magnetic field is obtained by different means depending on its intensity.
Data at moderate field (up to 0.22~T) data have been acquired using the
resistive magnets described in section~\ref{SECexpsetup}. Higher field
measurements have been performed in the bore and fringe field of a 1.5 T MRI
superconducting magnet~\cite{CIERM}. With the laser source and all the
electronics remaining in a low-field region several meters away from the
magnet bore, cells were successively placed at five locations on the magnet
axis with field values ranging from $B$=6~mT to 1.5~T. The 1.5~T value (in the
bore) is accurately known from the routinely measured $^{1}\mathrm{H}$ NMR
frequency (63.830~MHz, hence 1.4992~T). The lower field intensities, 6~mT and
0.4~T, have been measured using a Hall probe with a nominal accuracy of 1\%.
The intermediate values, 0.95 T and 1.33 T, are deduced from the measured
Zeeman splittings with a similar accuracy of 1\%. Due to the steep magnetic
field decrease near the edge of the magnet bore, large gradients cause a
significant inhomogeneous broadening of the absorption lines in some
situations discussed in section~\ref{SECgradients}. For these high field
measurements, the Fabry Perot interferometer was not implemented and no
accurate on-site check of the laser frequency scale was performed. Instead,
the usual corrections determined during the saturated absorption measurements
(see section~\ref{SECexpsetup} and figure~\ref{figscanpics}) are used for the
analysis of the recorded absorption spectra.

A systematic study of the effect of experimental conditions such as gas
pressure, RF discharge intensity, and probe beam power and intensity, has been
performed both in low field (the earth field) and at 0.08~T. The main
characteristic features of the recorded spectra have been found to be quite
insensitive to these parameters. Absorption spectra have been analysed in
detail to accurately deduce the relative populations of different sublevels
(from the ratios of line intensities) and the average atomic density of atoms
in the $2^{3}\mathrm{S}$ state in the probe beam (from its absorption). A
careful analysis of lineshapes and linewidths reveals three different
systematic effects which are discussed in the next three sections.

\subsubsection{\label{SECisotopic}Effect of imperfect $^{3}\mathrm{He}$
isotopic purity}

\begin{figure}[h]
\centerline{ \psfig{file=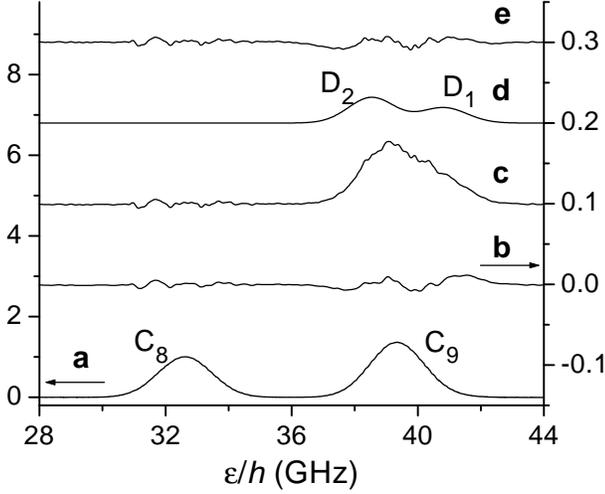, width=8 cm, clip= } }
\caption{Zero-field absorption measurement on the C$_{8}$ and C$_{9}$ lines in
nominally pure $^{3}\mathrm{He}$ at 0.53 mbar. \textbf{a}~: absorption signal
(the vertical scale is 20 times less sensitive than for the other traces).
\textbf{b}~: residue plot of independent Gaussian fits to the two lines.
\textbf{c}~: residue plot of the same fit as in plot~\textbf{b} for C$_{8}$
and the corresponding computed C$_{9}$ profile. \textbf{d}~: computed D$_{2}%
$-D$_{1}$ profile with a $\div$200 amplitude scaling corresponding to the
measured traces of $^{4}\mathrm{He}$ (see text). \textbf{e}~: residue plot of
independent Gaussian fits to the two lines after correction for the
$^{4}\mathrm{He}$ contribution. }%
\label{figfitC8C9}%
\end{figure}

Evidence of a systematic effect on absorption profiles was observed in
experiments with nominally pure $^{3}\mathrm{He}$ gas. A typical recording of
the C$_{8}$ and C$_{9}$ absorption lines for $B$=0\ and residuals to different
Gaussian fits are shown in figure~\ref{figfitC8C9}. A straightforward fit by two independent Gaussian profiles to the
recorded lines (trace a) systematically provides a 1.5-2~\% larger width and
an unsatisfactory fit for the C$_{9}$ transition (trace b, residue plot).
Moreover, the ratio of fitted line strengths (areas) is 7.8\% larger than the
value computed using table~\ref{tabB0} in the Appendix (1.374 instead of
1.274). Conversely, fitting on the C$_{8}$ component and assigning the
expected linewidth, amplitude and frequency shift to the C$_{9}$ component
provides the residue plot of trace c, which suggests the existence of an
additional contribution to absorption. All this is explained by the presence
of traces of $^{4}\mathrm{He}$, which are independently observed by absorption
measurements on the isolated D$_{0}$ line. Indeed this small amount of
$^{4}\mathrm{He}$ (isotopic ratio $R$=0.2 to 0.5 \%) affects the absorption
measurement performed on $^{3}\mathrm{He}$ due to the intense lines which lie
within one Doppler width of the C$_{9}$ line (trace d: computed absorption
lines of \ $^{4}\mathrm{He}$ for $R$=0.2 \%). The reduction of the
$^{3}\mathrm{He}$ gas purity is attributed to the cell cleaning process, which
involves several strong discharges performed in $^{4}\mathrm{He}$ prior to
final filling with pure $^{3}\mathrm{He}$. Reversible penetration of helium
into the glass walls is believed to be responsible for the subsequent presence
of a small (and surprisingly unsteady) proportion of $^{4}\mathrm{He}$ in the
gas. When the appropriate correction is introduced to substract the
contribution of $^{4}\mathrm{He}$ lines from the absorption spectra, the fit
by two independent Gaussian profiles is significantly improved (trace e) and
the linewidths are found to be equal within $\pm$0.2~\%. For the recording of
figure~\ref{figfitC8C9}a one obtains a Doppler width $\Delta$=1.198~GHz,
corresponding to a temperature of 304~K in the gas (consistent with the
ambient temperature and the power deposition of the RF discharge in the plasma).

The ratio of the line amplitudes after this correction is closer to the ratio
of computed transition intensities, still 3.5$\pm$0.2\% larger in this
example. A series of measurements performed in a row provides very close
results, with a scatter consistent with the quoted statistical uncertainty of
each analysis. In contrast, a measurement performed in the same cell after
several days may lead to a different proportion of $^{4}\mathrm{He}$ and a
different ratio of transitions intensities (for instance $R$=0.35\% and a
ratio of intensities 2.9$\pm$0.1\% \emph{smaller} than the computed one).

\begin{figure}[tbh]
\centerline{ \psfig{file=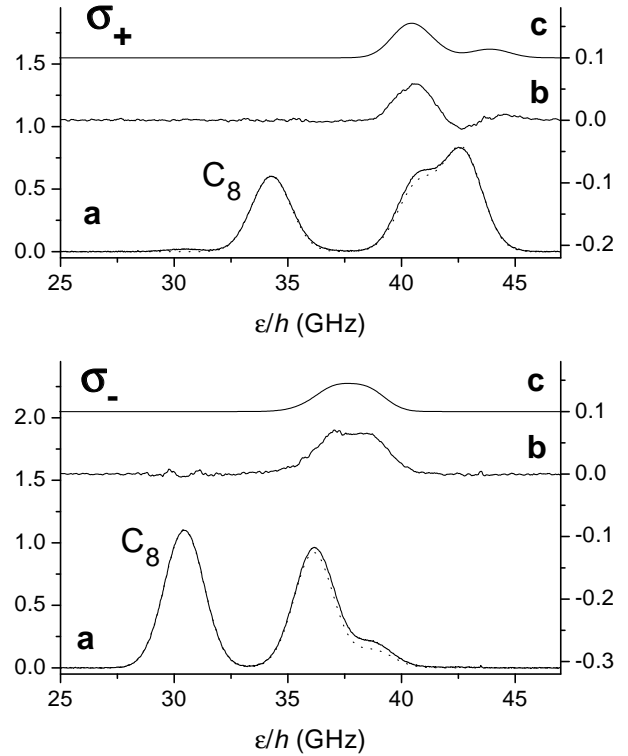, width=8 cm, clip= } }
\caption{Absorption measurements on the C$_{8}$ and C$_{9}$ lines performed in
a field $B$=0.111~T in nominally pure $^{3}\mathrm{He}$ at 0.53 mbar for
circular polarisations (upper graph : $\sigma_{+}$, lower graph : $\sigma_{-}%
$). Traces \textbf{a} correspond to the raw absorption signals (solid lines)
and computed profiles (dotted lines, see text). Traces \textbf{b} are
differences (residue plots) between absorption signals, corrected for light
polarisation defects, and computed profiles. Traces \textbf{c} are the
computed profiles for pure $^{4}\mathrm{He}$, scaled down in order to
correspond to an isotopic ratio $R$=0.4\%. Expanded vertical scales ($\times
$5, right hand side axes) are used for traces \textbf{b} and \textbf{c}.}%
\label{figC8C9kG}%
\end{figure}

Consequences of the presence of a small proportion of $^{4}\mathrm{He}$ on the
lineshapes and intensities of recorded spectra are indeed also observed in an
applied magnetic field, as shown in figure~\ref{figC8C9kG} for $B$=0.111~T.
The recorded absorption signals are similar to the computed ones
(dotted lines), but two significant differences appear. First, a small line is
visible for instance in the upper graph around $\varepsilon/h$=30~GHz, where
no $\sigma_{+}$ transition should be observed. This is due to an imperfect
circular polarisation of the probe beam in the cell, resulting in part from
stress-induced circular dichroism in the cell windows (measured to be of order
0.2\%). The amount of light with the wrong polarisation is determined for each
recording, and each signal is corrected by substracting the appropriate
fraction of the signal with the other circular polarisation (0.79\% and 0.18\%
corrections for $\sigma_{+}$ and $\sigma_{-}$ recordings in
figure~\ref{figC8C9kG})\footnote{Imperfect light polarisation has no effect in
very low field where Zeeman energy splittings are negligible and transition
intensities of each component (C$_{1}$ to C$_{9}$ and D$_{0}$ to D$_{2}$) do
not depend on light polarisation.}. Second, the corrected signals may then
tentatively be fit by the sum of three Gaussian profiles centred on the
frequencies independently determined in a saturated absorption experiment (see
figure~\ref{figexp2kG}). But, as is the case in zero field, this procedure
does not provide satisfactory results for the amplitudes and linewidths of the
split C$_{9}$ lines (corresponding to the transitions originating from any of
the lowest levels of the $2^{3}\mathrm{S}$ state, A$_{1}$ to A$_{4}$, to the
highest levels of the $2^{3}\mathrm{P}$ state, B$_{17}$ and B$_{18}$). In
figure~\ref{figC8C9kG}, traces b display the differences between the corrected
signals and the computed sums of three Gaussian profiles with a common
linewidth and with amplitudes proportional to the nominal transition
probabilities in this field. Two global amplitudes and the linewidth ($\Delta
$=1.221~GHz) are adjusted to fit the two C$_{8}$ transitions, and the
resulting differences are best accounted for assuming a proportion $R$=0.4\%
of $^{4}\mathrm{He}$ in the gas (traces c in figure~\ref{figC8C9kG}). In order
to better reproduce the experimental recordings, the amplitudes of the split
C$_{9}$ lines are finally allowed to vary. Fits with excellent reduced
$\chi^{2}$ and no visible feature in the residues are obtained and the
obtained ratios of intensities are given in table~\ref{tabreskG} for this
typical recording . \begin{table}[tbhtbh]
\centerline{
\begin{tabular}
[c]{|c|c|c|c|c|c|}\hline
\multicolumn{2}{|c}{polarisation} & \multicolumn{2}{|c|}{$\sigma_{-}$} &
\multicolumn{2}{|c|}{$\sigma_{+}$}\\\hline
\multicolumn{2}{|c}{computed} & \multicolumn{2}{|c|}{$T_{6,18}$} &
\multicolumn{2}{|c|}{$T_{5,17}$}\\
\multicolumn{2}{|c}{C$_{8}$ intensity} & \multicolumn{2}{|c|}{0.3757} &
\multicolumn{2}{|c|}{0.18803}\\\hline
\multicolumn{2}{|c|}{computed} & $T_{4,17}$ & $T_{3,18}$ & $T_{2,17}$ &
$T_{1,18}$\\
\multicolumn{2}{|c|}{C$_{9}$ intensity} & 0.31311 & 0.04731 & 0.16111 &
0.25107\\\hline
C$_{8}$/C$_{9}$ & comp. & 0.83340 & 0.12592 & 0.85683 & 1.33527\\\cline
{2-6}ratio & meas. & 0.85198 & 0.15275 & 0.86360 & 1.30055\\\hline
\multicolumn{2}{|c|}{deviation}%
& +2.2 \% & +21 \% & +0.8 \% & $-$3.6 \%\\\hline
\end{tabular}
} \caption{Computed line intensities $T_{ij}$ for $B$=0.1111~T and their
ratios are compared to the experimental line amplitude ratios of figure
\ref{figC8C9kG}.}%
\label{tabreskG}%
\end{table}The same kind of discrepancy as in zero field is obtained for the
strong components of the transitions; it happens to be much worse for the
weakest line in some recordings, including that of figure~\ref{figC8C9kG}, for
reasons which are not understood.

\begin{figure}[h]
\centerline{ \psfig{file=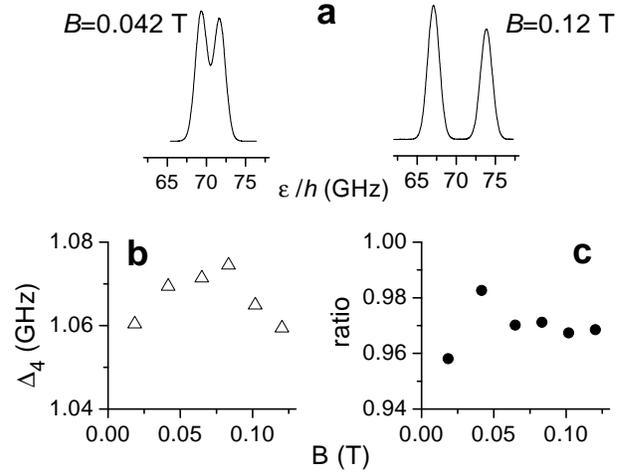, width=8 cm, clip= } }
\caption{Absorption measurements on the D$_{0}$ line of $^{4}\mathrm{He}$
performed in fields up to $B$=0.12~T with a lin$\bot$ polarised probe beam in
a $^{3}\mathrm{He}$-$^{4}\mathrm{He}$ mixture at 0.53 mbar (isotopic ratio
$R$=3). \textbf{a}~: Recordings of absorption signals are given for two values
of the field. The lower plots display the results of Gaussian fits to the
signals. \textbf{b}~: Doppler width $\Delta_{4}$, \textbf{c}~: ratios of
measured to computed relative line amplitudes $\left(  A_{\sigma+}/A_{\sigma
-}\right)  /\left(  T_{3,9}^{(4)}/T_{1,9}^{(4)}\right)  .$}%
\label{figpoidsD0kG}%
\end{figure}

In an attempt to test whether the remaining line intensity disagreements with
the computed values may result from imperfect corrections for the presence of
$^{4}\mathrm{He}$, or from difficulties due to fitting several overlapping
lines, we have also probed the well resolved $\mathrm{D}_{0}$ line of
$^{4}\mathrm{He}$ in a gas mixture in the 0-0.12~T field range. Using a
circularly polarised probe beam, the two transitions connecting levels Y$_{1}$
and Y$_{3}$ to Z$_{9}$ ($\sigma_{+}$ and $\sigma_{-}$ polarisations
respectively) are split in frequency by 56~GHz/T (see
section~\ref{SECnumresults}). The ratio of the corresponding line intensities
is computed to be approximately given in this field range by:%
\begin{equation}
T_{3,9}^{(4)}/T_{1,9}^{(4)}\approx1+1.88B+2.02B^{2},\label{compratio}%
\end{equation}
where $B$ is the field intensity in Tesla. The results of a series of
absorption measurements for a transverse probe beam (with linear perpendicular
polarisation) are displayed in figure~\ref{figpoidsD0kG}. The absorption signals are fit by the sum of three Gaussian
profiles centred on the transition frequencies (measured in a saturated
absorption experiment). This allows for a small amount of $\pi$ polarisation
in the probe beam, which experimentally arises from imperfect alignment of the
direction of the polarisation with respect to $B.$ In spite of this drawback,
the transverse probe beam configuration is preferred to a longitudinal one,
for which polarisation defects may directly affect the ratio of line amplitudes.

The common linewidth $\Delta_{4}$=1.067 $\pm$0.006~GHz
(figure~\ref{figpoidsD0kG}b, corresponding to a temperature of 321~K) is found
to be independent of the field. The ratio $A_{\sigma+}/A_{\sigma-}$ of the
measured line amplitudes for the $\sigma_{+}$ and $\sigma_{-}$ polarisation
components is found to increase with $B$ as expected from
equation~\ref{compratio}, with however a slightly lower value ($\sim$3\%,
figure~\ref{figpoidsD0kG}c).

To summarise the results of this section, the difference of measured line
intensities in our experiments with respect to the computed ones is usually
small (of order $\pm$3\%) after correcting for the presence of some
$^{4}\mathrm{He}$ in our $^{3}\mathrm{He}$ cells and for the observed light
polarisation imperfections. This difference does not arise from
signal-to-noise limitations and its origin is not known. This may set a
practical limit to the accuracy with which an absolute measurement of
population ratios can be made using this experimental technique. Similar
discrepancies are also observed at higher fields, situations for which the
additional systematic effects described in the following sections must first
be discussed.

\subsubsection{\label{SECgradients}Effects of magnetic field gradients}

\begin{figure}[h]
\centerline{ \psfig{file=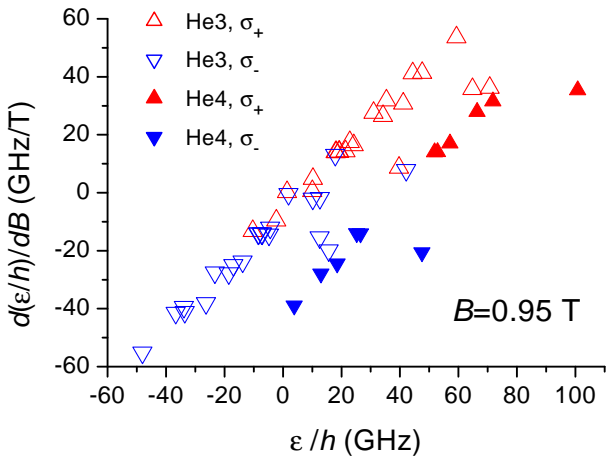, width=8 cm, clip= } }
\caption{Computed field-derivatives of the transitions energies $\epsilon
_{ij}/h$ are plotted for all $^{3}\mathrm{He}$ and $^{4}\mathrm{He}$ circular
polarisation transitions as a function of $\epsilon_{ij}/h$ for $B$=0.95~T.
Inhomogeneous line broadenings proportional to these values are induced by a
field gradient.}%
\label{figspreadzee}%
\end{figure}

As previously mentioned (section \ref{SECexpsetup2}), absorption experiments
above 0.25~T have been performed in the fringe field of a 1.5~T magnet. The
exact field map of this magnet was not measured, but the field decreases from
1.33~T to 0.95~T in only 25~cm, so that field gradients along the axis may
exceed 15~mT/cm at some locations. In contrast, field variations in transverse
planes are much smaller in the vicinity of the field axis. As a result,
a\ negligible broadening of absorption lines is induced by the field gradient
for perpendicular probe beams ($\sigma$ or $\pi$ polarisation for a linear
polarisation perpendicular or parallel to $B$). For probe beams propagating
along the field axis ($\sigma$ polarisation for a linear polarisation,
$\sigma_{+}$ or $\sigma_{-}$ polarisation for a circular polarisation), the
range of fields $\delta B$ applied to the probed atoms is the product of the
field gradient by the cell length. This field spread $\delta B$ induces
spreads $\delta\epsilon_{ij}/h$ of optical transition frequencies which are
proportional to $\delta B$ and to the derivatives $d\epsilon_{ij}/dB.$ Each
spread $\delta\epsilon_{ij}/h$ thus depends on the probed transition as
illustrated in figure~\ref{figspreadzee} for $B$=0.95~T. 

\begin{figure}[h]
\centerline{ \psfig{file=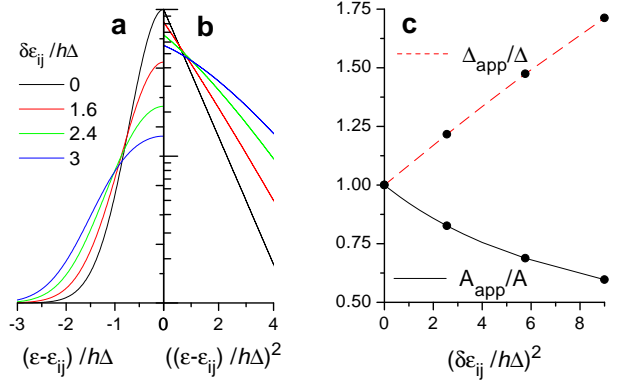, width=8 cm, clip= } }
\caption{\textbf{a}~: Computed absorption (half-)profiles in a magnetic field
gradient resulting from the convolution of a Gaussian Doppler profile (width
$\Delta$) and of a uniform spread of the resonance frequency (over a width
$\delta\epsilon_{ij}/h$). \textbf{b}~: The same profiles are plotted in
semi-logarithmic scale as a function of the square of the detuning. A Gaussian
profile appears as a straight line. \textbf{c}~: Apparent linewidths
$\Delta_{app}$ and amplitudes $A_{app}$ of these computed profiles are plotted
as a function of the square of the reduced resonance frequency spreads. The
solid symbols correspond to the profiles displayed in \textbf{a} and
\textbf{b}.}%
\label{figconvolsplit}%
\end{figure}

The resulting absorption profiles are given by the convolution of the Doppler
velocity distribution and of the frequency spread functions, computed from the
density distribution of metastable atoms, the field gradient at the cell
location and the derivatives $d\epsilon_{ij}/dB.$ Assuming for simplicity a
uniform density of atoms in the $2^{3}\mathrm{S}$ state, computed broadened
absorption profiles only depend on the dimensionless ratio $\delta
\epsilon_{ij}/h\Delta$ of the transition frequency spread to the Doppler
width. Figures \ref{figconvolsplit}a and \ref{figconvolsplit}b display
examples of computed profiles (absorption as a function of this reduced
detuning). For increasing frequency spreads, profiles become broader and
their maximum amplitude decreases (the area is strictly independent of
$\delta\epsilon_{ij}/h\Delta$). Deviation from a true Gaussian profile is more
clearly seen in figure~\ref{figconvolsplit}b as a deviation from a straight
line. To test the procedure used in the analysis of experimental data, one may
fit a Gaussian function to these computed profiles. The apparent linewidths
and amplitudes thus obtained, $\Delta_{app}$ and $A_{app},$ are plotted in
figure~\ref{figconvolsplit}c. For small frequency spreads, $\Delta_{app}$ and
$1/A_{app}$ are almost proportional to $\delta\epsilon_{ij}^{2}.$ However,
significant deviations appear at large $\delta\epsilon_{ij}$, and the product
$\Delta_{app}A_{app}$ is not strictly constant (it increases by 2~\% for
$\delta\epsilon_{ij}/h\Delta\sim$3). This means that whenever frequency
spreads cannot be avoided (e.g. using a transverse probe beam), fitting
procedures to Gaussian profiles introduce systematic errors. Computation of
line areas would\ provide the true line intensities, but this procedure cannot
be used to separately obtain the intensities of ill-resolved transitions.

\begin{figure}[h]
\centerline{ \psfig{file=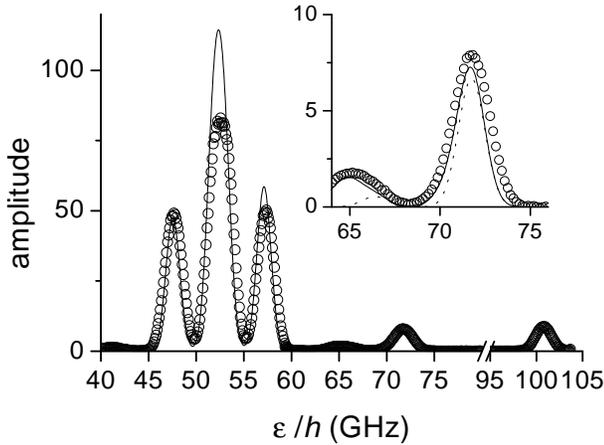, width=8 cm, clip= } } \caption{Part
of an absorption spectrum recorded at $B$=0.95~T in a $^{3}\mathrm{He}$%
-$^{4}\mathrm{He}$ mixture at 2.13 mbar (isotopic ratio $R$=3) using a
longitudinal probe beam (open symbols). The solid curve is the spectrum
computed for this $R,$ assuming thermal Doppler widths for the different
lines. Its amplitude is scaled to reproduce that of the 47 GHz line.
Experimental broadening of the absorption lines is directly found on the 101
GHz line, and suggested by the residual absorption between the main lines
(non-zero signal at 50 and 55 GHz), which is 2-3 times larger than computed.
The insert is an expanded portion of the spectrum, in which several
$^{3}\mathrm{He}$ and $^{4}\mathrm{He}$ lines overlap (the computed absorption
of pure $^{4}\mathrm{He}$ is displayed as a dotted line). Experimental
broadening of the absorption lines also clearly appears for this part of the
spectrum.}%
\label{fig1Tgrads}%
\end{figure}

Figure~\ref{fig1Tgrads} displays part of an absorption measurement performed
at $B$=0.95~T with a longitudinal $\sigma$-polarised probe beam.
It includes the two split D$_{0}$ transitions connecting levels
Y$_{1}$ and Y$_{3}$ to Z$_{9}$ : the $\sigma_{+}$ $^{4}\mathrm{He}$ line with
highest $\epsilon/h$ (101~GHz), and $\sigma_{-}$ $^{4}\mathrm{He}$ line with
highest $\epsilon/h$ (47~GHz), respectively. The recorded absorption signal
(symbols) roughly correspond to the profile computed assuming the nominal
Doppler widths $\Delta$ and neglecting frequency spreads (solid line). Clear
amplitude differences are however observed, in particular for the most intense
line. They can be attributed to OP effects, and will be discussed in detail in
the next section (\ref{SECOP}). Moreover, significant linewidth differences
appear, for instance on the isolated most shifted $^{4}\mathrm{He}$ line
(fitted apparent linewidth $\Delta_{app}$=1.42 GHz, 30\% larger than
$\Delta_{4}$). Most of the other lines actually result from the overlap of
several atomic transitions, and a straightforward Gaussian fit is not
relevant. For example, the computed total absorption around $\epsilon
/h$=72~GHz (insert in figure~\ref{fig1Tgrads}, solid line) results from the
two $^{4}\mathrm{He}$ transitions at 66.7 and 72.1\ GHz (dotted line) and the
two $^{3}\mathrm{He}$ transitions \ at 65.2 and 71.1~GHz.

\begin{figure}[h]
\centerline{ \psfig{file=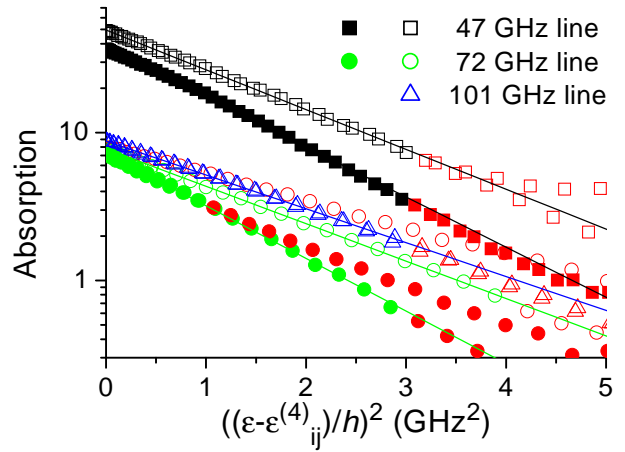, width=8 cm, clip= } }
\caption{Semi-logarithmic plot of absorption data recorded at $B$=0.95~T in a
$^{3}\mathrm{He}$-$^{4}\mathrm{He}$ mixture at 2.13 mbar (isotopic ratio
$R$=3). They are presented for 3 different $^{4}\mathrm{He}$ lines as a
function of the square of the frequency mistuning $\left(  \epsilon
-\epsilon_{ij}^{(4)}\right)  /h.$ Open symbols : data of
figure~\ref{fig1Tgrads}, longitudinal probe beam. Solid symbols~: data for a
transverse probe beam. The straight lines represent Gaussian absorption
profiles, and the values of their slopes are used to extract the apparent
linewidths $\Delta_{app}.$ The smallest linewidths, consistent with the
$^{4}\mathrm{He}$ Doppler width, are obtained for a transverse probe
($\Delta_{4}$=1.13\ and 1.105~GHz for the 47 and 72~GHz lines). Larger
apparent linewidths are obtained for a longitudinal probe ($\Delta_{app}%
$=1.27, 1.31 and 1.42~GHz for the 47, 72 and 101~GHz lines).}%
\label{figecartsgrad}%
\end{figure}

To better analyse the effect of the probe beam direction on the shape and
widths of different lines, we plot in figure~\ref{figecartsgrad} the
absorption data using the same method as in figure~\ref{figconvolsplit}b.
The open symbols correspond to three of the lines in
figure~\ref{fig1Tgrads}, with a probe beam along the direction field, and the
solid symbols to lines from a similar recording with a perpendicular probe
beam. Apart from the asymmetries on the 71~GHz lines which result from the
additional $^{3}\mathrm{He}$ line on the low-frequency side of the absorption
profile (which is thus not retained in the analysis), satisfactory linear fits
reveal Gaussian-like profiles and provide the values of the corresponding
linewidths. The experiments performed with a transverse probe provide data
with the steepest slopes and hence the smallest linewidths, while those using
a longitudinal probe lead to larger apparent linewidths as expected . Using
the convolution results of figure~\ref{figconvolsplit}c, the frequency spreads
$\delta\epsilon_{ij}/h$ for these lines are computed to be 1.4, 1.65 and
1.85~GHz respectively. From the derivatives $d\epsilon_{ij}/dB$ for these
transitions (figure~\ref{figspreadzee}), one computes that a field spread
$\delta B$=52~mT would produce frequency spreads of 1.08, 1.65 and 1.85 GHz,
corresponding to the observed broadenings for the 72 and 101~GHz lines. The
additional apparent broadening for the 47~GHz~experimental line may be
attributed to OP effects for this rather intense line (see the next
section~\ref{SECOP}). For the estimated value of the field gradient (15
mT/cm), this 52~mT field spread would be obtained for an effective cell length
of 3.5 cm. This is quite satisfactory since the density of atoms in the
metastable $2^{3}\mathrm{S}$ state vanishes at the cell walls, so that most of
the probed atoms lie within a distance shorter than the actual gap between the
end windows of our 5~cm-long cells (external length).

To summarise the results obtained in this section, the expected effects of
strong field gradients on absorption lines have indeed been observed. They
introduce a systematic broadening and amplitude decrease of absorption signals
which can be significant (up to 30~\% changes in our experiments), and must be
considered in quantitative analyses. For isolated lines, area measurements are
not affected by the transition frequency spreads. However, most lines usually
overlap in a recorded absorption spectrum (especially in an isotopic mixture),
and the quantitative analysis of such spectra is in practice too inaccurate if
the different linewidths are not known. It is thus important to only use a
transverse probe beam configuration whenever accurate absorption measurements
are needed and a field gradient is present. The same argument would also apply
to saturated absorption measurements, for which even limited frequency spreads
(e.g. of order 100~MHz) would strongly decrease the signal amplitudes.

\subsubsection{\label{SECOP}Effects of optical pumping}

\begin{figure}[h]
\centerline{ \psfig{file=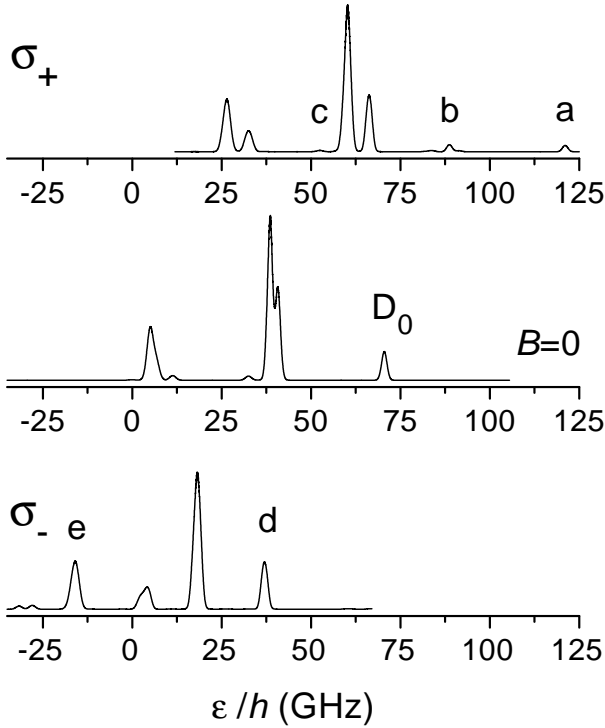, width=8 cm, clip= } }
\caption{Recordings of complete absorption spectra at $B$=0 and 1.5~T for
circular polarisations in a $^{3}\mathrm{He}$-$^{4}\mathrm{He}$ mixture at
2.13 mbar (isotopic ratio $R$=3). The lines labelled a to e are especially
considered in the analysis of OP effects. The $^{4}\mathrm{He}$ transitions
with the highest energy (involving the Z$_{9}$ level) considered in
figure~\ref{figpoidsD0kG} for moderate fields are labelled a and d at 1.5~T,
and correspond to the D$_{0}$ line at zero field. The assignment of the lines
at zero field can be made using figure~\ref{figraies}.}%
\label{figexp15kG}%
\end{figure}

A comparison of absorption recordings at $B$=0 and 1.5~T is presented in
figure~\ref{figexp15kG}. As expected in such a high magnetic field, the absorption spectra
are deeply modified and strongly depend on the probe light polarisation. Line
shifts of order of the fine structure splittings, and large changes in the
intensities of the absorption lines are observed, for instance on the
well-resolved transitions to the excited Z$_{9}$ level (labelled a, D$_{0}$
and d in figure~\ref{figexp15kG}, which we shall generically call D$_{0}$
transitions). For these lines, the computed transition intensities and the
corresponding weights in the total spectra are given in table~\ref{tabres15kG}%
. \begin{table}[htbh]
\centerline{
\begin{tabular}
[c]{|c|c|c|c|}\hline
line & d \ (D$_{0}$, $\sigma_{-}$) & D$_{0}$ ($B$=0) & a \ (D$_{0}$,
$\sigma_{+}$)\\\hline
$T_{i9}$ & 0.7722 & 0.3333 & 0.06459\\\hline
weight (calc.) & 19.31 \%& 8.333 \% & 1.615 \%\\\hline
weight (exp.) & 16.6 \% & 7.9 \% & 1.8 \%\\\hline
\end{tabular}
} \caption{Computed line intensities $T_{i9}$ of the D$_{0}$ line for $B$=0
and 1.5~T. The weights in the spectra (fraction of the line area over the
total area) are computed for an isotopic ratio $R$=3 corresponding to the
experimental conditions. The last line displays the experimental weights for
the measurements shown in figure \ref{figexp15kG}.}%
\label{tabres15kG}%
\end{table}It is not possible to directly compare absorption signals at $B$=0
and 1.5~T because the plasma intensity and distribution in the cell are
strongly modified in high field. However, according to the sum rules of
equation~\ref{sumrule}, the average density of atoms in the $2^{3}\mathrm{S}$
state along the probe beam is deduced by integrating the absorption signal
over the whole spectrum. This is used to scale the absorption amplitudes in
figure~\ref{figexp15kG} and have a meaningful comparison of line weights in
table~\ref{tabres15kG}. While there is a satisfactory agreement with
expectations for all line weights at $B$=0, the ratio of the weight of the
strong (D$_{0},$ $\sigma_{-}$) absorption line d to that of the weak (D$_{0},$
$\sigma_{+}$) line a is substantially smaller than computed.

The measurements displayed in figure~\ref{figexp15kG} have been performed with
a rather intense probe beam, so that OP effects may play a significant role.
As for all experiments performed in low field, we can assume that OP creates
no nuclear polarisation $M$ for the ground state $^{3}\mathrm{He}$ atoms. It
is not possible to enforce rapid nuclear relaxation as is done in low field
(section \ref{SECzerofield}) by imposing a strong relative field inhomogeneity
at 1.5~T. However the angular momentum deposited in the cell by the absorption
of a fraction of the probe beam could only lead to a very slow growth rate of
$M$ for our experimental conditions\footnote{It can be estimated from the
number of $^{3}\mathrm{He}$ nuclei in the cell (10$^{18}$ for a partial
pressure of 0.53~mbar) and the absorbed photon flux from the light beam (e.g.
4$\times$10$^{14}$ photons/s for 0.1~mW, corresponding to a 10~\% absorption
of a 1~mW probe beam): the growth rate of $M$ is of order 0.05~\%/s. This
result is consistent with pumping rates measured for optimal OP conditions at
low field~\cite{Stoltz}).}. As a result, only a very small nuclear
polarisation $M$ can be reached in a frequency sweep during the time spent on
an absorption line ($\sim$5~s). Negligible changes of the Zeeman populations
and hence of absorption signal intensities are induced by this process. In
contrast, the OP effects \emph{in the} $2^{3}\mathrm{S}$ \emph{metastable
state} described in section~\ref{SECOPtheo} may have a strong influence on the
distribution of the Zeeman populations, and thus significantly decrease the
absorption (see figure~\ref{figPO4}).

\begin{figure}[h]
\centerline{ \psfig{file=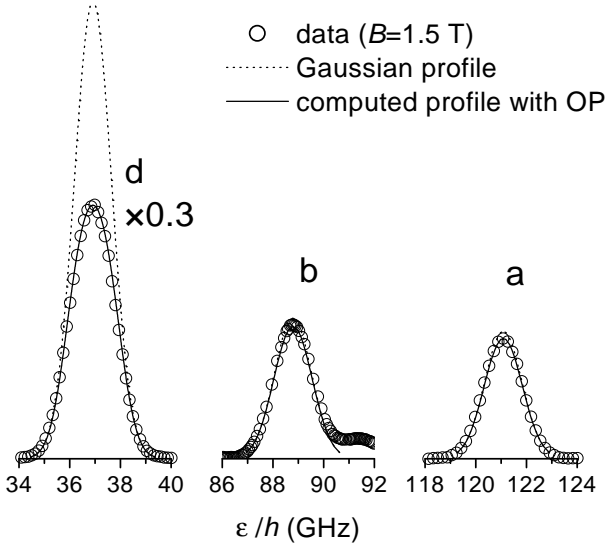, width=8 cm, clip= } }
\caption{Expanded plot of selected absorption profiles from experimental
recordings at 1.5 T (open symbols). The labelling of the lines is that of
figure~\ref{figexp15kG}. The vertical scale has been reduced ($\times$0.3) for
the strongest line d (the actual ratio of the maximum absorptions of lines d
and a is 7.0). Dotted lines~: computed Gaussian profiles for these
transitions. Solid lines~: computed profiles taking into account OP effects
with the parameters given in the text (hardly distinguishable from the dotted
Gaussian profiles for lines a and b).}%
\label{figOP15kG}%
\end{figure}

To discuss and analyse in detail these OP effects at 1.5~T, we show in
figure~\ref{figOP15kG} (symbols) parts of the experimental spectra extracted
from figure~\ref{figexp15kG}, together with different computed profiles which
are discussed below. It is important to note that in spite of the double-pass
configuration in the cell used to enhance absorption signals, no narrow
saturated absorption dips appear in the data. This is different from what is
observed at low field when a counterpropagating pump beam induces population
changes in a given velocity class, and when velocity changing collisions only
partly redistribute these changes over the whole Doppler profile (see
section~\ref{SECzerofield} and figures \ref{figB0He3} - \ref{figD0Bnul}). We
shall thus assume that in the conditions of figure~\ref{figexp15kG} no
correlation is introduced by the probe beam between atomic velocity and Zeeman
population changes, i.e. that any out-of-equilibrium orientation produced by
OP due to the probe is uniformly redistributed among all atoms in the Doppler
profile. The pumping time $\tau_{p}$ introduced in equations \ref{rateyOP} and
\ref{rateaOP} can thus be computed from the photon absorption rates defined in
equations \ref{tauij4} and \ref{tauij} using an averaging procedure similar to
that of \cite{Nacher85}. The averaging is performed over the atomic velocity
projection $v$ along the probe beam, which introduces the Doppler shift
$\omega v/c$ for the atomic transition angular frequencies $\omega$%
=$\omega_{ij}^{(4)}$ and $\omega_{ij}$ in equations \ref{tauij4} and
\ref{tauij}. For $^{3}\mathrm{He}$ transitions, one obtains\footnote{This
differs by a factor 1/$\sqrt{2}$ from the (incorrect) corresponding result in
reference~\cite{Nacher85}.}:
\begin{equation}
\frac{1}{\tau_{p}}=\frac{4\pi\alpha f}{m_{e}\omega\Gamma^{\prime}}%
~~\frac{T_{ij}I_{las}}{\bar{v}\sqrt{\pi}}\int_{-\infty}^{\infty}\frac{\left(
\Gamma^{\prime}/2\right)  ^{2}e^{-(v/\bar{v})^{2}}dv~}{\left(  \Gamma^{\prime
}/2\right)  ^{2}+\left(  \omega-\omega_{ij}-\omega_{ij}v/c\right)  ^{2}}%
\end{equation}
where the mean velocity $\bar{v}=2\pi c\Delta_{3}/\omega_{ij}$ is related to
the Doppler width defined in equation~\ref{eqdoppler}. For a single-frequency
laser and a small enough damping rate of the optical transition coherence
($\Gamma^{\prime}/2\ll2\pi\Delta_{3}$), the integral is easily computed:%
\begin{equation}
\frac{1}{\tau_{p}}\simeq\frac{\sqrt{\pi}\alpha f}{m_{e}\omega\Delta_{3}%
}~T_{ij}I_{las}~e^{-(\left(  \omega-\omega_{ij}\right)  /2\pi\Delta_{3})^{2}%
}.\label{eqtaup}%
\end{equation}
A similar result is obtained for $^{4}\mathrm{He}$ transitions by substitution
of $\Delta_{4},$ $T_{ij}^{(4)}$ and $\omega_{ij}^{(4)}$ in
equation~\ref{eqtaup}. The numerical value of the\ first factor appearing in
the pumping rate expression is~for $^{4}\mathrm{He}$:%
\begin{equation}
\frac{\sqrt{\pi}\alpha f}{m_{e}\omega\Delta_{4}}~=4.269\times10^{3}%
~\mathrm{s}^{-1}/(\mathrm{W/m}^{2}),
\end{equation}
and it is 15 \% smaller for $^{3}\mathrm{He}$ due to the larger value of the
Doppler width.

For a given laser intensity, the pumping rate is thus expected to be
proportional to the transition intensity $T_{ij}$ and, for each absorption
line, to decrease with increasing detuning from transition frequencies of
atoms at rest. Relying on the OP\ model of section~\ref{SECOPtheo}, Zeeman
populations can be numerically computed as a function of the pumping rate. The
absorption signals, which are proportional to the probed populations, to
$T_{ij}$ and to the Gaussian factor in equation~\ref{eqtaup}, can then in
principle be computed using a single global scaling parameter (the number
density of atoms in the $2^{3}\mathrm{S}$ state).

\begin{figure}[h]
\centerline{ \psfig{file=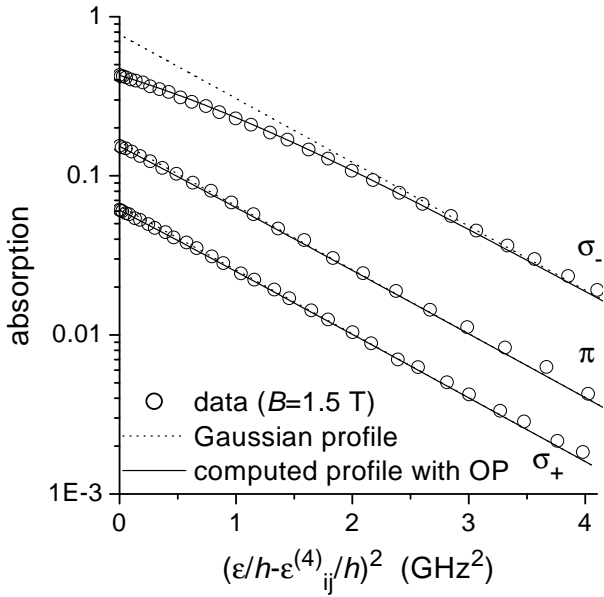, width=8 cm, clip= } }
\caption{Semi-logarithmic plot of absorption data (open symbols) recorded at
$B$=1.5~T in a $^{3}\mathrm{He}$-$^{4}\mathrm{He}$ mixture at 2.13 mbar
(isotopic ratio $R$=3). They are presented for 3 different $^{4}\mathrm{He}$
lines (D$_{0}$, 3 light polarisations) as a function of the square of the
frequency detuning $\epsilon/h-\epsilon_{ij}^{(4)}/h.$ The dotted straight
lines represent gaussian absorption profiles, the solid lines are the computed
profiles with OP effects (see text).}%
\label{figOPlog15kG}%
\end{figure}

In our experiments, the laser intensity $I_{las}$ of the probe beam at the
cell location could not be precisely measured, and instead was deduced from
the comparison of the measured ratio of the maximum absorptions of the
$\sigma_{-}$ and $\sigma_{+}$ D$_{0}$ lines, labelled d and a in figures
\ref{figexp15kG} and \ref{figOP15kG}, to the computed ratio which would be
obtained in the absence of OP effects, i.e. for $y_{1}$=$y_{3}$=1/3
($y_{3}T_{3,9}/y_{1}T_{1,9}$=7.0 instead of $T_{3,9}/T_{1,9}$=11.96). The
value $\left(  \tau_{e}/\tau_{p}\right)  /T_{ij}$=3.28$\times$10$^{-3}$ is
numerically obtained from this comparison of OP effects on these lines at
resonance (the two values for $T_{ij}$ are given in table~\ref{tabres15kG}).
The exchange rate $1/\tau_{e}$ is computed using equation~\ref{unsurtaue} and
the value of the exchange rate $1/\tau_{e}^{0}$ in a pure $^{3}\mathrm{He}$
gas at the operating pressure(3.75$\times$10$^{6}$~s$^{-1}$/mbar
\cite{JDR71}):
\begin{equation}
1/\tau_{e}=1/\tau_{e}^{0}\left(  1+R\left(  \mu-1\right)  \right)  .
\end{equation}
For the conditions of the experiment (isotopic ratio $R$=3, total pressure
2.13 mbar) $\tau_{e}$=1.04$\times$10$^{-7}$~s, hence $1/(\tau_{p}T_{ij})$
=3.15$\times$10$^{4}$~s and $I_{las}$=7.4~W/m$^{2}$ or 0.74~mW/cm$^{2}$, which
is quite consistent with its independently estimated value. For this value of
$I_{las},$ OP effects are computed for all lines as a function of the detuning
from resonance, and the resulting absorption profiles precisely agree with the
experimental data (figure~\ref{figOP15kG}, solid lines, compared to the
Doppler profiles, dotted lines). This agreement over the whole profiles is
better seen in figure~\ref{figOPlog15kG}. There is a slight systematic deviation for all lines at large
detuning, where absorption is larger than computed. This may be due to a
temperature slightly above room temperature (as is systematically observed in
other experiments), and to the difference between the true atomic (Voigt)
profile and a Gaussian one. The agreement is otherwise very good and fully
supports the collisional redistribution assumption used to derive
equation~\ref{eqtaup}. These measurements can also be viewed as indirect
checks of the variation of the population of the pumped sublevel with the
pumping rate, such as plotted in figures \ref{figPO4} and \ref{figPO4states},
up to a 45~\% population reduction for $y_{3}$. The influence of $I_{las}$ was
indeed also directly measured by using a linear $\sigma$ polarisation instead
of a circular one, thus reducing the effective pump power by a factor 2 on
isolated lines. All measured OP effects are reduced according to calculations,
with for instance a 29~\% maximum population reduction for $y_{3}$ on the
$\sigma_{-}$ D$_{0}$ line.

OP effects for other transitions also agree with expectations. For instance,
probing the population $y_{2}$ with a $\pi$ polarisation
(figure~\ref{figOPlog15kG}) induces relatively small OP effects in spite of a
significant transition intensity. This is consistent with the reduced
sensitivity of the Y$_{2}$ state to OP which was noticed in section
\ref{SECOPtheo} (figure~\ref{figPO4states}). Most $^{3}\mathrm{He}$
transitions are difficult to analyse due to the proximity of several
overlapping lines, but the lines c and e in figure~\ref{figexp15kG} can be
considered to evaluate OP effects. Line c is an isolated line probing the
A$_{5}$ state, so weak that no significant OP effect is expected (a 0.2~\%
maximum population reduction for $a_{5}$), and observed. In contrast, line e
consists of 4 lines (two with $\epsilon/h\sim$-16.1 GHz, two with
$\epsilon/h\sim$-14.9 GHz), each with a transition intensity $\sim$1.
Experimentally, the ratio of the maximum absorptions of lines e and c is 40.2,
4~\% lower than expected from transition intensities, thus suggesting a 4~\%
relative decrease of the probed populations. However these transitions concern
4 different sublevels in the $2^{3}\mathrm{S}$ state, this is an average
decrease for the populations in these sublevels. Furthermore, the simple OP
model of section \ref{SECOPtheo} cannot be directly applied since it assumes
that a single level is pumped. Still, if the A$_{5}$ state alone was pumped
with the probe intensity used in the experiment, a computed 3.2~\% maximum
population reduction would be obtained for $a_{5}$, in reasonable agreement
with the small observed OP effects.

Let us finally mention that OP effects have been observed to have all the
expected features at lower magnetic field. For instance, for the same laser
intensity $I_{las}$ as at 1.5~T, the maximum population reduction for $y_{3}$
on the (D$_{0}$, $\sigma_{-}$) line is 38~\% at 1.3~T, 26~\% at 0.95~T and
8~\% at 0.4~T. The much weaker effects expected for other transitions could
not clearly be observed, especially in the presence of strong field gradients
which also affect the apparent line intensities and shapes as discussed in the
previous section.

To summarise the results of this section, the computed effects of optical
pumping in the $2^{3}\mathrm{S}$ state have been observed and quantitatively
verified in some detail. This supports the validity of the assumptions made on
the effect of metastability exchange collisions on the Zeeman populations and
of the simple OP model considered in section~\ref{ME}. It also confirms that
orientation is fully redistributed by collisions over the whole velocity
profile in high field. This can be understood if one considers that ME
collisions usually strongly modify the velocities of the colliding atoms, but
that due to hyperfine decoupling little angular momentum is transferred to the
nuclear spin. An intuitive measure of this decoupling can be deduced from the
high efficiency of OP processes at small pumping rate: a 45~\% population
decrease can be obtained at 1.5~T for $\tau_{e}/\tau_{p}$=1/400, which means
that relaxation of electronic orientation typically requires 400 exchange
collisions (or $\sim$100 collisions with an unpolarised ground state
$^{3}\mathrm{He}$ atom). One consequence is that performing saturated
absorption measurements similar to those of section~\ref{SECabsat} to probe
$^{4}\mathrm{He}$ transitions becomes increasingly difficult in fields above
0.5-1~T, and certainly requires the presence of a larger amount of
$^{3}\mathrm{He}$ than in our experiments (an isotopic ratio $R\ll$1).
Finally, a practical result of this study of OP effects is that the
attenuation level of a probe beam required to perform an unperturbed
measurement of relative populations is well determined. It strongly depends on
the probed isotope and on the chosen transition, and on the gas composition
and pressure, but a laser intensity $I_{las}$ typically 100 times lower (i.e.
10 $\mu$W/cm$^{2}$) than the highest one we have used must be chosen to have a
1~\% accuracy on all population measurements up to 1.5\ T.

\subsection{\label{SECoptmeas}Optical measurement of nuclear polarisation}

In this section we address the measurement of the nuclear polarisation $M$ of
the \emph{ground state} of $^{3}\mathrm{He}$, such as can be efficiently
prepared by performing OP of the $2^{3}\mathrm{S}$ state, both in pure
$^{3}\mathrm{He}$ and in isotopic mixtures.

NMR is a valuable technique which can be used to directly measure the nuclear
magnetisation resulting from the polarisation. It is however usually not well
suited to systematic studies of OP dynamics and efficiency: it has a reduced
sensitivity for a low pressure gas, it induces polarisation losses at each
measurement and thus cannot continuously measure $M$, and it is affected by
the noise-generating plasma discharge used for OP. In addition, it requires a
high absolute homogeneity of the magnetic field $B$, and a retuning or
redesign of the NMR spectrometer at each change of the field value.

In contrast, optical measurement techniques have a very high sensitivity to
changes in the electronic states of the excited states of helium in the
plasma, even in a low pressure gas. The standard optical detection technique
\cite{Bigelow92,Pavlovic70} mentioned in the introduction, in which the
circular polarisation of a chosen helium spectral line emitted by the plasma
is measured and the nuclear polarisation is inferred is simple to implement
and provides an accurate non-destructive measurement of $M$. It relies on
hyperfine coupling to transfer angular momentum from nuclear to electronic
spins in the excited state which emits the monitored spectral line. The
decoupling effect of an applied magnetic field unfortunately reduces the
efficiency of this angular momentum transfer, which is also sensitive to
depolarising collisions. The usefulness of this technique is thus restricted
to low fields ($\lesssim$10~mT), low gas pressures ($\lesssim$5~mbar) and
limited $^{4}\mathrm{He}$ isotopic ratios ($R\lesssim$1) to avoid a
significant sensitivity loss ($\div$2 for each of the quoted limits).
Different optical methods, which rely on absorption measurements on the
$2^{3}\mathrm{S}$ - $2^{3}\mathrm{P}$ transition, have been successfully used
to quantitatively determine the nuclear polarisation of $^{3}\mathrm{He}%
$~\cite{Colegrove63,Bigelow92,Greenhow64,Daniels71}. It is a method of this
kind, designed to operate for arbitrary magnetic fields and for all isotopic
mixtures, which we describe and discuss in the rest of this article.

\subsubsection{Principle}

Optical absorption measurement methods provide information both on the total
number density of atoms in the $2^{3}\mathrm{S}$ state and on the relative
populations of the probed sublevels. In usual situations, the population
distribution in the $2^{3}\mathrm{S}$ state is strongly coupled by ME
collisions to that in the ground state, characterised by $M$. These
populations would exactly be ruled by a spin temperature distribution (see
section~\ref{SECspinT}) in the absence of OP or relaxation processes. For
simplicity, we shall assume that these additional processes have a negligible
effect compared to that of ME. The validity of this assumption will be
discussed in the final section~\ref{SECdiscuss}.

When two absorption measurements directly probe two populations of atoms in
the $2^{3}\mathrm{S}$ state (for instance at low field when the line C$_{8}$
or D$_{0}$ is probed with $\sigma_{+}$ and $\sigma_{-}$ circular
polarisations), the derivation of $M$ is a straightforward procedure. When
transitions simultaneously probe several sublevels (e.g. in low field with
$\sigma$ polarisation on any line, with any polarisation on line C$_{9}$,
etc...), the measurements of two independent combinations of populations can
still be used to infer the nuclear polarisation $M$, but specific calculations
are then required~\cite{Bigelow92}. Here, we are mostly interested in
measurements at high enough magnetic fields for the Zeeman shifts to remove
all level degeneracies ($B\gtrsim$50~mT). It is thus very easy to selectively
probe a single level by an adequate choice of the transition frequency and of
the polarisation of the light.

With the spin temperature 1/$\beta$ defined in section~\ref{SECspinT}, the
nuclear polarisation is given by:%
\begin{equation}
M=\left(  e^{\beta}-1\right)  /\left(  e^{\beta}+1\right)  . \label{Mvsbeta}%
\end{equation}
Assuming that the two measured absorption signals, $S$ and $S^{\prime}$, are
simply proportional to the number densities of atoms in the probed sublevels
(averaged in the volume of the probe beam), we can relate them to the total
number density and to the relative populations. When $^{3}\mathrm{He}$ atoms
are probed, one obtains:%
\begin{equation}
S=Kn_{3}a\quad\mathrm{and}\quad S^{\prime}=K^{\prime}n_{3}a^{\prime},
\label{Svsa}%
\end{equation}
where only the total number density $n_{3}$ and the relative populations $a$
and $a^{\prime}$ of the two probed sublevels vary with $M$. The $M$%
-independent coefficients $K$ and $K^{\prime}$ depend on the probed
transitions; they are proportional to the transition intensities and thus
depend on the chosen transitions and they vary with the magnetic field. The
ratio $S^{\prime}/S$ only depends on the ratio of transition intensities and
of relative populations (see equation \ref{aivsbeta}):%
\begin{equation}
S^{\prime}/S=K^{\prime}/Ke^{\beta(m_{F}^{\prime}-m_{F})}, \label{rapS}%
\end{equation}
where $m_{F}$ and $m_{F}^{\prime}$ are the total angular momentum projections
in the two probed states. To eliminate systematic errors resulting from an
inaccurate determination of the ratio $K^{\prime}/K$, it is often convenient
to also measure absorption signals, $S_{0}$ and $S_{0}^{\prime}$, in the
unpolarised system ($M$=0, hence $S_{0}^{\prime}/S_{0}=K^{\prime}/K$). The
spin temperature can thus be directly deduced from ratios of measured
absorption signals:%
\begin{equation}
e^{\beta}=\left(  S^{\prime}S_{0}/SS_{0}^{\prime}\right)  ^{1/(m_{F}^{\prime
}-m_{F})}, \label{MvsratS}%
\end{equation}
provided that $m_{F}^{\prime}\neq m_{F}.$ The nuclear polarisation can then be
obtained using equation \ref{Mvsbeta}. The total number density $n_{3}$ in the
$2^{3}\mathrm{S}$ state can in turn be computed from equations \ref{aivsbeta},
\ref{Svsa} and \ref{rapS}. In the following we only discuss its relative
variation with $M$:%
\begin{equation}
\frac{n_{3}(M)}{n_{3}(0)}=\frac{S}{S_{0}}\frac{e^{2\beta}+2e^{\beta
}+2e^{-\beta}+e^{-2\beta}}{6e^{\beta m_{F}}}. \label{varn3}%
\end{equation}

Similar results are indeed obtained when $^{4}\mathrm{He}$ atoms are probed to
measure two relative populations, $y$ and $y^{\prime}$, among the three
sublevels in the $2^{3}\mathrm{S}$ state. In this case, the angular momentum
projections in equations \ref{rapS} and \ref{MvsratS} must be replaced by
$m_{S}$ and $m_{S}^{\prime}$, and equation \ref{varn3} is replaced by:%
\begin{equation}
\frac{n_{4}(M)}{n_{4}(0)}=\frac{S}{S_{0}}\frac{e^{\beta}+1+e^{-\beta}%
}{3e^{\beta m_{S}}}. \label{varn4}%
\end{equation}

\subsubsection{Examples of operation}

As a demonstration of the performance of an absorption-based optical
measurement of $M$ in non-standard conditions, we report here on experimental
measurements performed at 0.0925~T. This field value is high enough for the
Zeeman splittings to be at least of the order of the Doppler line widths, so
that generic problems arising from high-field operation are present. The main
difference with low-field techniques is that the two absorption measurements
required to record the signals $S$ and $S^{\prime}$ must be performed using
different probe laser frequencies (a few line crossings at some field values,
e.g. around 0.16 T - see figure~\ref{figexp2kG} - may be used to avoid this
constraint, but this will not be considered in the following). Before
describing experimental schemes to obtain such dual-frequency measurements, we
present examples of continuous frequency scans of the probe laser in
steady-state OP situations.

\begin{figure}[h]
\centerline{ \psfig{file=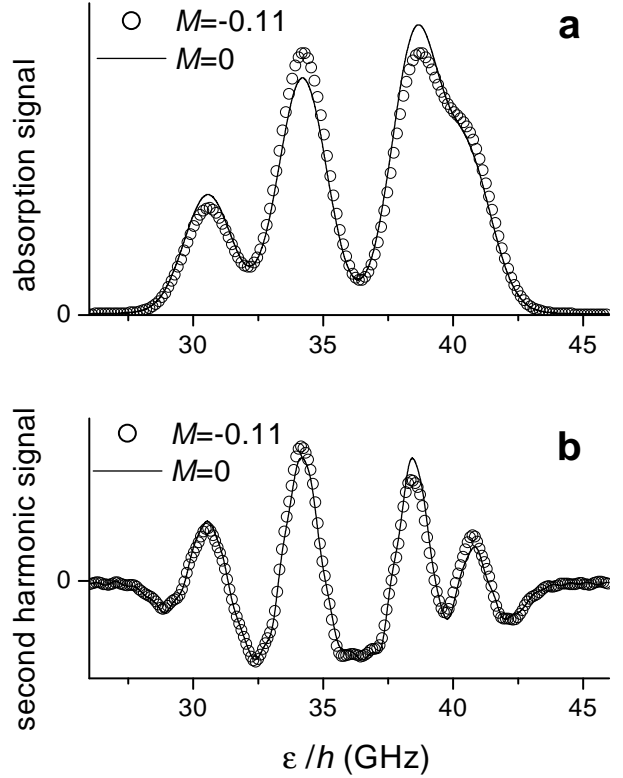, width=8 cm, clip= } }
\caption{\textbf{a}~: Experimental recordings of absorption signals obtained
at 0.0925~T in pure $^{3}\mathrm{He}$ (pressure 8.5~mbar) for $M$=0 (solid
line) or in a steady-state OP situation (open symbols). The OP is performed
using a laser diode (different from the one used to probe absorption) tuned on
the maximum absorption of the ($\sigma_{-},$ C$_{9}$) transition. The incident
pump power is 30~mW, and the transmitted beam is back reflected once through
the cell. Lock-in detection is performed at the modulation frequency
$f_{\mathrm{RF}}$ of the RF discharge. \textbf{b}~: Experimental recordings of
signals obtained with a frequency-modulated probe beam and second-harmonic
phase-sensitive detection (see text) in the same conditions.}%
\label{figC8C9OPscan}%
\end{figure}

Figure\ \ref{figC8C9OPscan}a displays recordings of absorption signals on the
C$_{8}$ and C$_{9}$ $^{3}\mathrm{He}$ lines for a $\pi$ polarisation of the
transverse probe. They are obtained with the same technique as for line intensity
measurements (section \ref{SECexpsetup2}), using a modulation (at a frequency
$f_{\mathrm{RF}}\sim$100~Hz) of the RF discharge excitation of the plasma and
a phase-sensitive analysis at $f_{\mathrm{RF}}$ of the transmitted probe
intensity. The recorded spectra are quite similar to those of
figure~\ref{figC8C9kG}, but four lines are obtained for this $\pi$
polarisation. A fit of the spectra by a sum of four Gaussian lines is used to
obtain the relative line intensities given in table~\ref{tababspiM}.
\begin{table}[htbh]
\centerline{
\begin{tabular}
[c]{|c|cccc|}\hline
$\epsilon/h$ (GHz) & 30.61 & 34.24 & 38.60 & 40.60\\
level, $m_{F}$ & A$_{6}$, + & A$_{5}$, $-$ & A$_{3}$, + & A$_{2}$, $-$\\
\hline
$S_{0}/T_{ij}$ & 1.003 & 0.991 & 0.989 & 1.018\\
$S/T_{ij}$ & 0.890 & 1.098 & 0.884 &  1.127\\
$S/S_{0}$ & 0.888 & 1.109 & 0.894 & 1.108\\
\hline
combination & $a_{6}/a_{5}$ & $a_{6}/a_{2}$ & $a_{3}/a_{5}$ & $a_{3}/a_{2}$\\
\hline
$M$ for $S_{0}$ & 0.006 & $-$0.001 & $-$0.007 & $-$0.014\\
$M$ for $S$ & $-$0.105 & $-$0.108 & $-$0.117 & $-$0.121\\
$M$ for $S/S_{0}$ & $-$0.111 & $-$0.107 & $-$0.110 & $-$0.107\\\hline
\end{tabular}
} \caption{Table of normalised relative line intensities $S_{0}/T_{ij}$ and
$S/T_{ij}$ obtained from fits of the two recordings of
figure~\ref{figC8C9OPscan}a. The nuclear polarisation $M$ is computed from
ratios of intensities for the four relevant combinations of lines
($m_{F}^{\prime}\neq m_{F}$). From the ratios $S/S_{0}$ (last line), the
nuclear polarisation is deduced to be $M$=$-$0.109$\pm$0.002. This negative
polarisations is obtained due to the use of $\sigma_{-}$ OP light.}%
\label{tababspiM}%
\end{table}As for the absorption measurements discussed in section
\ref{SECisotopic}, several systematic effects actually limit the accuracy of
the measurements of relative line intensities (e.g. imperfect light
polarisation or $^{3}\mathrm{He}$ purity). A $\pm$1.5~\% deviation from the
computed intensity ratios is observed for the $M$=0 recording in
figure~\ref{figC8C9kG} (first line in table~\ref{tababspiM}). This accounts
for the observed scatter in the determinations of $M$ deduced from the ratios
of signal amplitudes (``$M$ for $S$'' values in table~\ref{tababspiM}), and
justifies the use of the absorption measurements at null polarisation and of
equation~\ref{MvsratS} to actually compute $M$ (last line in
table~\ref{tababspiM}).

Replacing a continuous frequency scan by operation at two fixed frequencies
requires accurate independent settings of the laser frequencies to probe the
centres of the absorption lines. With a different modulation scheme, in which
the laser frequency is modulated over a small fraction of the Doppler width at
a frequency $f_{\mathrm{probe}}\sim$1~kHz, a phase-sensitive analysis at
$f_{\mathrm{probe}}$ of the transmitted probe intensity provides an error
signal proportional to the derivative of the absorption signal. This error
signal allows to lock the laser frequency on any absorption line. A second
harmonic phase-sensitive analysis (at $2f_{\mathrm{probe}}$) provides a signal
proportional to the second derivative of the absorption signal. This is
displayed in figure\ \ref{figC8C9OPscan}b for the same conditions as those of
the absorption measurement in figure\ \ref{figC8C9OPscan}a. The four
transitions appear to be better resolved, and the amplitude variations can
similarly be used to deduce the polarisation $M$. The respective advantages of
the two modulation schemes will be discussed in the next section.

\begin{figure}[h]
\centerline{ \psfig{file=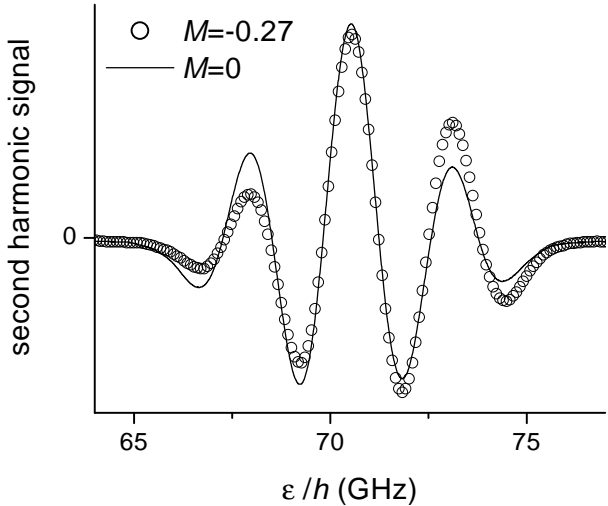, width=8 cm, clip= } }
\caption{Experimental recordings of absorption signals 0.0925~T in an isotopic
mixture ($R$=3, total pressure 8.5~mbar) for $M$=0 (solid line) or in a
steady-state OP situation (open symbols). $\sigma$ and $\pi$ polarisations are
equally present in the transverse probe beam. The OP is performed in the same
way as for the recordings in figure~\ref{figC8C9OPscan} but on the
($\sigma_{-},$ D$_{0}$) transition.}%
\label{figD0OPscan}%
\end{figure}

In isotopic mixtures the same transitions could still be probed, but the
overlap of the intense $^{4}\mathrm{He}$ lines precludes any reliable analysis
of the C$_{9}$ transition. Instead we present recordings of absorption spectra
on the D$_{0}$ lines in figure\ \ref{figD0OPscan} for a transverse probe beam
with equal weights on $\sigma$ and $\pi$ polarisations. While the same laser power is used for OP, a much larger nuclear
polarisation ($M$=0.27) is deduced from this measurement in an isotopic
mixture. This higher efficiency is similar to that systematically observed in
low field OP experiments~\cite{Stoltz}, and is related to the more efficient
absorption of the pumping light by the ($\sigma_{-}$,D$_{0}$) line.

\begin{figure}[h]
\centerline{ \psfig{file=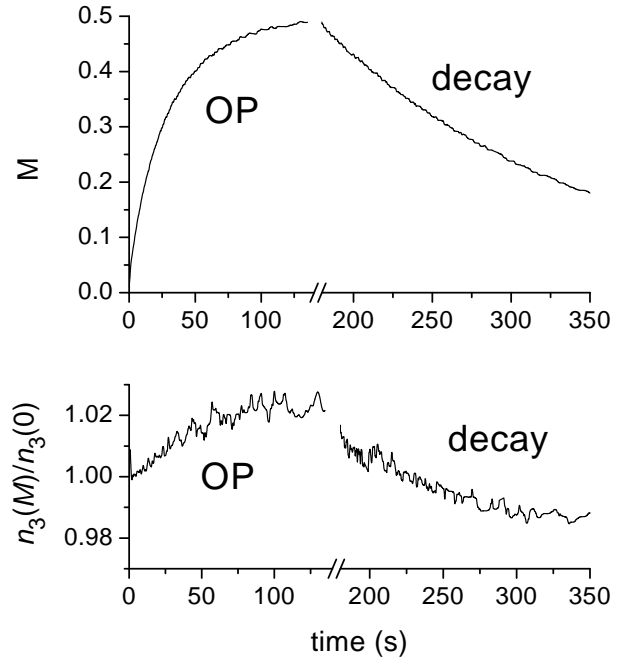, width=8 cm, clip= } } \caption{Nuclear
polarisation $M$ and relative changes in total metastable density $n_{3}$
deduced from successive measurements of absorption signals by states A$_{5}$
and A$_{3}$ ($B$=0.0925~T, $^{3}\mathrm{He}$ pressure 1.07~mbar). The OP is
performed on the ($\sigma_{+},$ C$_{9}$) transition with 150 mW of incident
pumping power starting from an unpolarised gas (this sets the initial time
$t$=0). The pumping beam is blocked after 180~s and $M$ exponentially decays
back to 0 ($T_{1}$=156~s). Here polarisation is positive due to the choice of
$\sigma_{+}$ OP light.}%
\label{figOP1kG}%
\end{figure}

Frequency scans over 20~GHz as displayed in figures \ref{figC8C9OPscan} and
\ref{figD0OPscan} are performed in $\sim$100~s and can be used only to
characterise steady-state situations. A simple method to track the dynamics of
buildup or decay of $M$ in an OP experiment consists in actually repeating
identical experiments with the probe laser frequency tuned to different
transitions. The result of the analysis of two successively performed
absorption recordings is given in figure~\ref{figOP1kG}. The probed lines in these measurements are the two most intense
lines in figure~\ref{figC8C9OPscan}a. The recorded signals are obtained with a
standard modulation of the RF discharge excitation and a phase-sensitive
analysis at $f_{\mathrm{RF}}$ of the transmitted probe intensity. A higher
nuclear polarisation is obtained here for two reasons: the gas pressure is
lower, and a fibre amplifier is used to increase the pumping beam
power~\cite{Chernikov}. \ 

Small changes of the number density $n_{3}$ are derived from the recordings.
They result in part from a slow drift in the RF\ discharge conditions, since
in the analysis $n_{3}(0)$ is obtained from the absorptions measured for the
$M$=0 situations at the beginning of the recordings. This drift accounts for
the $\sim$1~\% difference observed over the total duration of the recordings (%
$>$%
700~s). Another part ($\sim$2~\%) of the change of $n_{3}$ in
figure~\ref{figOP1kG} is clearly correlated to the value of the nuclear
polarisation. It is known that the electronic orientation in the
$2^{3}\mathrm{S}$ state has an influence on the rate of ionising collisions in
the plasma:%
\begin{equation}%
\begin{array}
[c]{ccccccccc}%
\mathrm{He}^{\mathrm{\ast}} & + & \mathrm{He}^{\mathrm{\ast}} & \rightarrow &
\mathrm{He} & + & \mathrm{He}^{\mathrm{+}} & + & \mathrm{e}^{\mathrm{-}}%
\end{array}
\label{Penning}%
\end{equation}
(which would ultimately be forbidden in a fully polarised
system~\cite{Schearer70}). Such an effect probably explains the systematically
observed variations of $n_{3}$ with $M$ (here a small increase). However these
variations (even their sign) strongly depend on the RF discharge intensity
\cite{Bigelow92} and are thus difficult to quantitatively predict. This makes
the simultaneous measurements of $M$ and $n_{3}$ of special interest for an
accurate characterisation of OP kinetics.

\subsubsection{\label{SECdiscuss}Discussion}

The frequency scans in figures \ref{figC8C9OPscan}a and \ref{figC8C9OPscan}b
illustrate the difference arising from the chosen absorption measurement
method: modulation of the RF discharge intensity and detection at
$f_{\mathrm{RF}}$, or modulation of the probe frequency and detection at
$2f_{\mathrm{probe}}$. The signal-to-noise ratio of the phase-sensitive
detection at $f_{\mathrm{RF}}$ is observed to be higher when both amplitudes
of the modulations are optimally set. Detection at $f_{\mathrm{RF}}$ may hence
be preferred, but this choice is not mandatory (a fair enough signal-to-noise
ratio can be obtained, e.g. in figure~\ref{figD0OPscan}).

Whatever the chosen detection scheme, a modulation of the probe frequency may
be applied to obtain a frequency error signal from a phase-sensitive detection
at $f_{\mathrm{probe}}$. This allows to precisely adjust the probe frequency
on the centre of isolated absorption lines, but is affected by the presence of
any neighbouring lines (the effect depends on their distance and relative
intensity). Whenever ill-resolved lines are probed to deduce relative
populations of individual states, a careful analysis must be performed and
appropriate corrections must be applied to the measured signals. These
corrections are usually smaller for signals detected at $2f_{\mathrm{probe}}$
when the distance to the nearest line is smaller than the Doppler width
$\Delta.$ Conversely, smaller corrections are required for a detection at
$f_{\mathrm{RF}}$ if this distance is larger than 1.3$\Delta,$ as is the case
for the C$_{9}$ doublet in Figure~\ref{figC8C9OPscan} (the exact crossover
distance depends on the relative amplitude of the perturbing line). This may
seem surprising since ill-resolved lines have narrower features with a
detection at $2f_{\mathrm{probe}}$. However it is only the central part of the
second derivative of a Gaussian profile which appears to be narrower than the
profile itself. Side lobes actually extend beyond the Doppler width, and as a
result there may be a significant crosstalk between apparently resolved lines.
Indeed, the choice of a detection scheme impacts both on the data analysis and
on the accuracy of the results only when measurements at two fixed frequencies
are performed: for a complete frequency scan of a set of lines, integration of
signals obtained by a detection at $2f_{\mathrm{probe}}$ can be used to
exactly reconstruct the usual absorption spectra.

We have tested several methods to obtain time-resolved results with
dual-frequency operation of the probe laser. The one used for the recordings
in figure~\ref{figOP1kG}, which consists in repeating once every measurement,
is very simple to implement but is time consuming. However, since the measured
values of the atomic density $n_{3}$ and $n_{4}$ usually vary very little with
the polarisation, one may alternatively neglect this variation and only
perform one absorption measurement. Using equation \ref{varn3} or \ref{varn4}
with its left-hand site set to 1, the spin-temperature (and hence $M$) can
then be obtained from this single absorption measurement. For the measurement
of figure~\ref{figOP1kG}, this approximation would have resulted in a maximum
relative error of 2~\% on $M$ and 1~\% on the decay rate $T_{1}$.

For situations where very accurate measurements are required, or where the
plasma conditions and hence $n_{3}$ and $n_{4}$ may substantially change in
the course of an experiment, alternating operation of the probe laser on two
frequencies can be obtained by two techniques. A temperature change of the DBR
diode can be used to induce the required frequency change (see section
\ref{SECexpsetup}), with a typical response time is of order 0.1~s. Following
a frequency change of a few GHz, the frequency is actually sufficiently stable
to provide a meaningful absorption signal after $\sim$1~s of dead time. This
means that the period of the probe frequency changes cannot be shorter that a
few seconds, and that the technique is not suitable to measure fast pumping
transients of the kind obtained using powerful pumping lasers \cite{hypint}. A
change of the current in the DBR diode can also be used to induce the required
frequency change. This is best done at a high enough frequency for the
temperature of the device to remain constant (%
$>$%
1~kHz), and very fast OP transients can be monitored with this technique.
However, the current changes also drive significant power changes of the probe
beam. The full elimination of this effect in the absorption measurements
requires a careful signal acquisition and analysis, and the absolute accuracy
of the results is somewhat reduced compared to the previous techniques.

We finally come back to the initial assumption made in this section, namely
that relaxation and OP processes do not affect the ME-ruled distribution of
populations in the $2^{3}\mathrm{S}$ state, and discuss their perturbing
influence on the optical absorption measurements.

Relaxation of the spin variables in the $2^{3}\mathrm{S}$ state\footnote{The
main origin of this relaxation lies in the limited lifetime of the atoms in
the $2^{3}\mathrm{S}$ state in the cell, due to diffusion to the wall (faster
at low pressure), ionising collisions (equation~\ref{Penning}, more frequent
at high density of $\mathrm{He}^{\ast}$ atoms), 3-body collisions (formation
of $\mathrm{He}_{2}^{\ast}$ molecules~\cite{NousMolec}, more likely at high
pressure).} would drive the spin distribution to its true thermal (Boltzmann)
equilibrium in the applied field in the absence of ME collisions. Even for
$B$=1.5~T and for electronic spins, this equilibrium corresponds to a small
polarisation ($\leq$1~\%) compared to $M$, so that the effect of relaxation is
to reduce the orientation in the probed $2^{3}\mathrm{S}$ state. This leads to
under-estimating the true value of the ground state nuclear polarisation with
absorption-based optical measurements.

Conversely, OP with an intense pump laser results in an overpolarisation of
the pumped $2^{3}\mathrm{S}$ atoms with respect to the spin temperature
population distribution (this is the driving process which actually transfers
angular momentum to the ground state reservoir of atoms through
ME\ collisions). We have indeed demonstrated in section \ref{SECOP} that even
a moderate laser intensity can induce strong deviations from equilibrium of
the population distributions in high magnetic field. However this
overpolarisation is a local effect which can simply be measured, and
eliminated by displacing the probe beam so as not to overlap the pump beam.
Figure~\ref{figOP1kG} clearly shows that blocking the pumping beam to let the
nuclear polarisation decay does not change the measured relative populations
nor the inferred value for $M$.

Experimentally, the effects of relaxation and OP which we have observed at
$B$=0.0925~T are similar to those observed in low field~\cite{Bigelow92}, and
do not significantly affect our optical measurement technique. This remains to
be quantitatively assessed in higher fields and at higher pressures, for which
hyperfine decoupling and additional relaxation processes might decrease the
efficacy of ME collisions to impose a single spin temperature for all
populations. Still, based on our systematic study of OP effects up to 1.5~T,
we believe that an absorption measurement technique can provide reliable
results, especially if one probes the atomic states least sensitive to OP
effects, and thus most efficiently coupled to the ground state polarisation
(the $^{3}\mathrm{He}$ $2^{3}\mathrm{S}$ states $\mathrm{A}_{+}^{l}$ and
$\mathrm{A}_{-}^{h}$, as shown in figure~\ref{figPO3}).

\section{Prospects}

From a theoretical point of view, we have developed and tested convenient
tools to numerically obtain the eigenstates of the $2^{3}\mathrm{S}$ and
$2^{3}\mathrm{P}$ levels in $^{3}\mathrm{He}$ and $^{4}\mathrm{He}$ and
explore the characteristic features of the 1083 nm transition in an arbitrary
magnetic field. We have also extended the quantitative treatment of ME
collisions by deriving rate equations for populations of all sublevels of the
$2^{3}\mathrm{S}$ state which are valid for isotopic mixtures in an arbitrary
magnetic field. The use of an elementary OP model for the case $M$=0 led to
quantitative predictions which have been accurately verified experimentally. A
more elaborate OP model, similar to the one introduced for pure $^{3}%
\mathrm{He}$ in low field~\cite{Nacher85}, can now be developed. This work is
in progress, based on the rate equations describing the effects of ME
collisions and absorption / emission of light which we have obtained and on an
adequate phenomenological treatment of relaxation phenomena.

Experimentally, we have demonstrated that the use of a DBR laser diode and of
simple techniques based on standard or Doppler-free absorption measurement can
provide a wealth of spectroscopic information on the 1083 nm transition in
helium up to high magnetic fields.

Saturated absorption experiments (or velocity-selective OP measurements) have
been performed in this work only as a means to perform Doppler-free
measurement of transition frequencies, to check for the soundness of our
experimental techniques. However the complex signals which we have obtained,
combining narrow lines and broad pedestals, also provide quantitative
information on collisional redistribution of orientation between atoms of
different velocity classes in the $2^{3}\mathrm{S}$ state. These collisions
are a key process when OP is performed at high pumping intensity and
saturation effects are important \cite{hypint}. A detailed study is currently
being made using independent pump and probe laser frequencies to fully
characterise the collisional transfer of orientation and the saturation
effects. This work will also provide useful guidance for designing improved
laser sources for high power OP.

A detailed study of several systematic effects on absorption measurements has
allowed us to assess the accuracy of relative population measurements in
various situations. It could be expected that hyperfine decoupling would
hamper angular momentum transfer by ME collisions at very high field. We have
demonstrated that ME collisions actually have a very different efficacy for
coupling populations in the $2^{3}\mathrm{S}$ state to the nuclear
polarisation in the ground state depending on the value of the magnetic field,
the helium isotope and the implied $2^{3}\mathrm{S}$ state sublevel. Based on
these techniques and results, and using the optical measurement methods
described in the last section of this article, a systematic study of OP of
helium in non-standard conditions (high magnetic field, high pressure) can now
be performed. One of the motivations of this study is the observation,
mentioned in the introduction, that higher nuclear polarisations may be
obtained by applying a stronger field (0.1~T) than usual. We believe that a
systematic study will help to elucidate the nature of relaxation processes in
the plasma, and may lead to reduce their effect in situations of practical
interest for applications requiring hyperpolarised $^{3}\mathrm{He}$ .

\begin{acknowledgement}
We wish to thank the C.I.E.R.M. in the Kremlin-Bic\^{e}tre hospital
for hosting the experiment. One of us (T.D) acknowledges support
from the Ecole Normale Sup\'erieure. PICS and Polonium programmes
have permitted repeated travel and continued collaboration between
Paris and Krakow.
\end{acknowledgement}

\section*{Appendix: tables and matrices}

The tables and matrices used in the body of the text are collected in this Appendix.

\paragraph{Fine-structure term parameters}

are listed in tables \ref{sfinemat} and~\ref{sfineval}: \begin{table}[h]
\centerline{
\begin{tabular}
[c]{c|ccc|c|}\cline{2-5}& $2^{3}\mathrm{P}_{\mathrm{0}}$ & $2^{3}\mathrm
{P}_{\mathrm{1}}$ &
$2^{3}\mathrm{P}_{\mathrm{2}}$ & $2^{1}\mathrm{P}$\\\cline{1-5}\multicolumn
{1}{|c|}{$2^{3}\mathrm{P}_{\mathrm{0}}$} & $E_{0}$ & 0 & 0 & \\
\multicolumn{1}{|c|}{$2^{3}\mathrm{P}_{\mathrm{1}}$} & 0 & $E_{1}$ & 0 & \\
\multicolumn{1}{|c|}{$2^{3}\mathrm{P}_{\mathrm{2}}$} & 0 & 0 & $E_{2}$ &
\\\hline
\multicolumn{1}{|c|}{$2^{3}\mathrm{P}_{\mathrm{0}}$} & $E_{0}$ & 0 & 0 & 0\\
\multicolumn{1}{|c|}{$2^{3}\mathrm{P}_{\mathrm{1}}$} & 0 & $E_{1}^{\prime}$ &
0 & $E_{M}$\\
\multicolumn{1}{|c|}{$2^{3}\mathrm{P}_{\mathrm{2}}$} & 0 & 0 & $E_{2}$ &
0\\\hline
\multicolumn{1}{|c|}{$2^{1}\mathrm{P}$} & 0 & $E_{M}$ & 0 & $\Delta$\\\hline
\end{tabular}
} \caption{$m_{J}$-independent matrix elements of the fine-structure term
$H_{fs}$ in the $\left\{  \left|  J;m_{J}\right\rangle \right\}  $ basis, used
to compute the $2^{3}\mathrm{P}$ state structure. $E_{0},$ $E_{1},$ $E_{2}$
are the true energies of the $2^{3}\mathrm{P}_{0,1,2}$ levels in zero $B$
field. The last 4 lines of the table provide the expression of the effective
Hamiltonian explicitly considering the $2^{1}\mathrm{P}$---$2^{3}\mathrm{P}$
mixing, with additional parameters $\Delta,$ $E_{M}$ and $E_{1}^{\prime}.$}%
\label{sfinemat}%
\end{table}\begin{table}[hh]
\centerline{
\begin{tabular}
[c]{|c|c|c|c|}\hline
& $^{4}\mathrm{He}$ (MHz) & $^{3}\mathrm{He}$ (MHz) &
reference\\\hline
$E_{2}$ & 0 & 0 & \\
$E_{1}$ & 2291.175  & 2291.926 & \cite{Storry00,Castillega00}\\
$E_{0}$ & 29616.950 & 29616.675  & \cite{Minardi99}\\
$\Delta$ & 6.1423$\times$10$^{7}$ & 6.1431$\times$10$^{7}$ &  \cite{Hinds85}\\
$E_{M}$ & -17035 & -17037 &  \cite{Hinds85}\\
$E_{1}^{\prime}$ & 2295.899 & 2296.651 & see caption\\\hline
\end{tabular}
} \caption{Values of the energies for $^{4}\mathrm{He}$ and $^{3}\mathrm{He}$
are given with reference to that of the $2^{3}\mathrm{P}_{2}$ level of each
isotope. We use the most recent and accurate values whenever possible, and
mass effects are evaluated as corrections~\cite{Hinds85,Hijikata88}. $E_{1}$
and $E_{1}^{\prime}$ are almost exactly 1 MHz larger, and $E_{M}$ 1000 times
larger, than the values of reference~\cite{Hinds85}, probably due to misprints
in the article. $E_{1}^{\prime}$ is not a free parameter since the eigenvalue
$E_{1}$ must be obtained. It is given by $E_{1}^{\prime}=E_{1}+E_{M}%
^{2}/\left(  \Delta-E_{1}\right)  .$}%
\label{sfineval}%
\end{table}

\paragraph{The Zeeman Hamiltonian,}

including relativistic corrections of $O\left(  \alpha^{2}\right)  ,$ consists
of a linear term $H_{1}$ and a quadratic term $H_{2}.$ The linear term
contains a correction with respect to the simple expression in
equations~\ref{Hz4} and \ref{Hz3}:%
\begin{equation}
H_{1}=H_{Z}+\mu_{B}g_{x}\left(  SO^{(2)}\right)  \cdot B, \tag{A1}%
\label{Hzlin}%
\end{equation}
in which the tensor operator $O^{(2)}$ involves $L$
\cite{Lhuillier76,Yan94,Lewis70}. Due to this additional term, $H_{1}$ is more
easily expressed in the coupled $\left|  J,m_{J}\right\rangle $ basis, as is
the quadratic term $H_{2}.$ $H_{1}$+$H_{2}$ has non-zero elements only between
sublevels of equal $m_{J},$ which involve five different $g$-factors in the
$2^{3}\mathrm{P}$ state, and two in the $2^{3}\mathrm{S}$ state, all listed in
table III of \cite{Yan94}. In the present work, this full expression is used
only to evaluate the error resulting from neglected terms involving $g_{x}$
and $H_{2}$. The values listed in table \ref{lande} which are used for
$g_{S}^{\prime}$ and $g_{L}^{\prime}$ are taken from \cite{Yan94}, limiting
the number of significant digits in order to be consistent with having
neglected $g_{x}^{\prime}.$ The mass-dependence of $g_{L}^{\prime}$ is also
computed from results of \cite{Yan94}, separating out the mass-dependent term:%
\begin{equation}
g_{L}=1-\frac{m}{M_{i}}\left\{  1+\frac{2F_{1}}{\sqrt{6}}\left(  \frac{M_{i}%
}{M_{4}}-1\right)  \right\}  +\delta g_{L}, \tag{A2}%
\end{equation}
in which $M_{i}$ stands for the mass of either isotope ($i$=3,4) and $\delta
g_{L}$ is the correction for $^{4}\mathrm{He}$ given in the reference.
\begin{table}[h]
\centerline{
\begin{tabular}
[c]{|c|ccc|}\hline
& $g_{S}^{\prime}$ & $g_{L}^{\prime}\left(  ^{4}\mathrm{He}\right)  $ &
$g_{L}^{\prime}\left(  ^{3}\mathrm{He}\right)  $\\\hline
$2^{3}\mathrm{S}$ & 2.002237319 & 0 & 0\\
$2^{3}\mathrm{P}$ & 2.002238838 & 0.999873626 & 0.999827935 \\\hline
\end{tabular}
}\caption{Values of the $g$-factors used in the simplified form of the Zeeman
Hamiltonian given in equations \ref{Hz4} and \ref{Hz3}. The nuclear term in
equation~\ref{Hz3} involves $g_{I}$=2.31748$\times$10$^{-3}.$ The value of the
Bohr magneton in frequency units is $\mu_{B}$=13996.242 MHz/T \cite{DrakeHB}.}%
\label{lande}%
\end{table}

\paragraph{The transition matrix elements $T_{m_{S}j}^{(4)}$ and $T_{ij}$}

are obtained using the eigenstates of the $2^{3}\mathrm{S}$ and $2^{3}%
\mathrm{P}$ states and the selection rule operator $T(e_{\lambda})$:
\begin{align}
T_{m_{S}j}^{(4)}  &  =\left|  \left\langle m_{S}\left|  T(e_{\lambda})\right|
Z_{j}\right\rangle \right|  ^{2}\tag{A3}\\
T_{ij}  &  =\left|  \left\langle A_{i}\left|  T(e_{\lambda})\right|
B_{j}\right\rangle \right|  ^{2}, \tag{A4}%
\end{align}
with matrix elements of $T(e_{\lambda})$ given by:
\begin{equation}
\left\langle m_{S}\left|  T(e_{\lambda})\right|  m_{S}^{\prime},m_{L}%
\right\rangle =\delta(m_{S},m_{S}^{\prime})\times\Sigma(e_{\lambda},m_{L})
\tag{A5}%
\end{equation}
for $^{4}\mathrm{He}$, and by:
\begin{multline}
\left\langle m_{S}:m_{I}\left|  T(e_{\lambda})\right|  m_{S}^{\prime}%
,m_{L}:m_{I}^{\prime}\right\rangle =\nonumber\\
\delta(m_{S},m_{S}^{\prime})\delta(m_{I},m_{I}^{\prime})\times\Sigma
(e_{\lambda},m_{L}), \tag{A6}%
\end{multline}
for $^{3}\mathrm{He}$, with $\Sigma(e_{\lambda},m_{L})$ as given in
table~\ref{srule}. \begin{table}[h]
\centerline{
\begin{tabular}
[c]{|c|ccc|}\hline
& $m_{L}$=1 & $m_{L}$=0 & $m_{L}$=-1\\\hline
$\sigma_{+}$ & 1 & 0 & 0\\
$\pi$ & 0 & 1 & 0\\
$\sigma_{-}$ & 0 & 0 & 1\\\hline
\end{tabular}
} \caption{Selection rules allow transitions only if angular momentum is
conserved, depending on the polarisation state of the light, $e_{\lambda
}=\sigma_{+},$ $\sigma_{-}$ or $\pi,$ and on $m_{L}.$}%
\label{srule}%
\end{table}

\paragraph{The matrix $F_{6}$}

which represents the contact interaction term of the hyperfine Hamiltonian
$H_{hfs}$ (equation~\ref{Hhfs}) in the decoupled $\left\{  \left|  m_{S}%
:m_{I}\right\rangle \right\}  $ basis and the values of the hyperfine coupling
constants $A_{\mathrm{S}}$ and $A_{\mathrm{P}}$ for the $2^{3}\mathrm{S}$ and
$2^{3}\mathrm{P}$ states respectively are given in table~\ref{shfine}:

\begin{table}[h]
\centering%
\begin{tabular}
[c]{c|cccccc|}\cline{2-7}%
& \multicolumn{6}{|c|}{$m_{S}:m_{I}$}\\\hline
\multicolumn{1}{|c|}{$m_{S}$:$m_{I}$} & $1$:$+$ & $0$:$+$ & $-1$:$+$ & $1$:$-$%
& $0$:$-$ & $-1:-$\\\hline
\multicolumn{1}{|c|}{$1$:$+$} & 1/2 &  &  &  &  & \\
\multicolumn{1}{|c|}{$0$:$+$} &  &  &  & 1/$\sqrt{2}$ &  & \\
\multicolumn{1}{|c|}{$-1$:$+$} &  &  & -1/2 &  & 1/$\sqrt{2}$ & \\
\multicolumn{1}{|c|}{$1$:$-$} &  & 1/$\sqrt{2}$ &  & -1/2 &  & \\
\multicolumn{1}{|c|}{$0$:$-$} &  &  & 1/$\sqrt{2}$ &  &  & \\
\multicolumn{1}{|c|}{$-1$:$-$} &  &  &  &  &  & 1/2\\\hline
\end{tabular}
\caption{Matrix elements of $F_{6}/A_{\mathrm{S}}$ (equation~\ref{H6}). Values
of the hyperfine coupling constants of equation \ref{Hhfs} are $A_{\mathrm{S}%
}$= $-4493.134$~MHz \cite{Rosner70} and $A_{\mathrm{P}}$=$-4283.026$~MHz,
0.019\% lower than the value the corresponding parameter $C$ of
reference~\cite{Prestage85} to account for singlet admixture effects (see
text).}%
\label{shfine}%
\end{table}

\paragraph{Computed line positions and intensities}

for $B$=0 which are represented in figure~\ref{figraies} are listed in
table~\ref{tabB0}. The results for $^{3}\mathrm{He}$ slightly differ from
previously published data \cite{Nacher85} due to the use of a more accurate
Hamiltonian.\begin{table}[h]
\centering%
\begin{tabular}
[c]{c|c|c|}\cline{2-3}%
& $\epsilon/h$ (GHz) & Intensity\\\hline
\multicolumn{1}{|c|}{C$_{1}$} & 0 & 0.03566\\
\multicolumn{1}{|c|}{C$_{2}$} & 4.51159 & 0.37482\\
\multicolumn{1}{|c|}{C$_{3}$} & 4.95920 & 2\\
\multicolumn{1}{|c|}{C$_{4}$} & 5.18076 & 1.29767\\
\multicolumn{1}{|c|}{C$_{5}$} & 6.73970 & 1.29767\\
\multicolumn{1}{|c|}{C$_{6}$} & 11.25129 & 0.29185\\
\multicolumn{1}{|c|}{C$_{7}$} & 11.92046 & 0.03566\\
\multicolumn{1}{|c|}{C$_{8}$} & 32.60452 & 0.29185\\
\multicolumn{1}{|c|}{C$_{9}$} & 39.34422 & 0.37482\\\hline
\multicolumn{1}{|c|}{D$_{2}$} & 38.53362 & 5/3\\
\multicolumn{1}{|c|}{D$_{1}$} & 40.82479 & 1\\
\multicolumn{1}{|c|}{D$_{0}$} & 70.44174 & 1/3\\\hline
\end{tabular}
\caption{Optical transition energies and intensities for $B$=0. The energies
$\epsilon$ are referenced to that of the C$_{1}$ transition, the intensities
are the partial sums of the transition matrix elements $T_{ij}$ for each line,
and obey the total sum rule of equation~\ref{sumrule}.}%
\label{tabB0}%
\end{table}

\paragraph{The matrix elements $\left\langle J,m_{J}\left|  P_{4}\right|
m_{L},m_{S}\right\rangle $}

of the transformation operator $P_{4}$ from the $\left(  L,S\right)  $ to the
$\left(  J\right)  $ representation for the $2^{3}\mathrm{P}$ state of
$^{4}\mathrm{He}$ are given in table \ref{P4}~. The additional term
$H_{hfs}^{cor}$ in equation~\ref{Hhfs} is given in table~\ref{shfinecor}. It
only involves 2 parameters $d$ and $e$, similar to parameters $D/2$ and $E/5$
defined by the Yale group~\cite{Hinds85} but with values adjusted to
effectively take couplings to singlet states into account with this simple
model. \begin{table*}[ht]
\centering%
\begin{tabular}
[c]{c|ccccccccc|}\cline{2-10}%
& \multicolumn{9}{|c|}{$m_{L},m_{S}$}\\\hline
\multicolumn{1}{|c|}{$J$;$m_{J}$} & 1,1 & 0,1 & $-$1,1 & 1,0 & 0,0 & $-$1,0 &
1,$-$1 & 0,$-$1 & $-$1,$-$1\\\hline
\multicolumn{1}{|c|}{2;2} & 1 &  &  &  &  &  &  &  & \\
\multicolumn{1}{|c|}{2;1} &  & 1/$\sqrt{2}$ &  & 1/$\sqrt{2}$ &  &  &  &  & \\
\multicolumn{1}{|c|}{2;0} &  &  & 1/$\sqrt{6}$ &  & 2/$\sqrt{6}$ &  & 1/$\sqrt{6}$ &  & \\
\multicolumn{1}{|c|}{2;$-$1} &  &  &  &  &  & 1/$\sqrt{2}$ &  & 1/$\sqrt{2}$ & \\
\multicolumn{1}{|c|}{2;$-$2} &  &  &  &  &  &  &  &  & 1\\
\multicolumn{1}{|c|}{1;1} &  & $-$1/$\sqrt{2}$ &  & 1/$\sqrt{2}$ &  &  &  &  & \\
\multicolumn{1}{|c|}{1;0} &  &  & $-$1/$\sqrt{2}$ &  &  &  & 1/$\sqrt{2}$ &  & \\
\multicolumn{1}{|c|}{1;$-$1} &  &  &  &  &  & $-$1/$\sqrt{2}$ &  & 1/$\sqrt{2}$ & \\
\multicolumn{1}{|c|}{0;0} &  &  & 1/$\sqrt{3}$ &  & $-$1/$\sqrt{3}$ &  & 1/$\sqrt{3}$ &  &
\\\hline
\end{tabular}
\caption{Matrix elements of the transformation operator $P_{4}$ (in
equation~\ref{H9}). The inverse transformation is given by the
transposed matrix : $P_{4}^{-1}\left(  i,j\right)  $=$P_{4}\left(  j,i\right)
.$}%
\label{P4}%
\end{table*}

\begin{table*}[ht]
\centering%
\begin{tabular}
[c]{c|ccccccccc|}\cline{2-10}%
& \multicolumn{9}{|c|}{$m_{L},m_{S}:+$}\\\hline
\multicolumn{1}{|c|}{$m_{L}$,$m_{S}$:$m_{I}$} & 1,1 & 0,1 & $-$1,1 & 1,0 & 0,0 & $-$1,0 & 1,$-$1 &
0,$-$1 & $-$1,$-$1\\\hline
\multicolumn{1}{|c|}{1,1:+} & $d$+2$e$ &  &  &  &  &  &  &  & \\
\multicolumn{1}{|c|}{0,1:+} &  & $-$4$e$ &  & 3$e$ &  &  &  &  & \\
\multicolumn{1}{|c|}{$-$1,1:+} &  &  & $-$$d$+2$e$ &  & $-$3$e$ &  &  &  & \\
\multicolumn{1}{|c|}{1,0:+} &  & 3$e$ &  & $d$ &  &  &  &  & \\
\multicolumn{1}{|c|}{0,0:+} &  &  & $-$3$e$ &  & 0 &  & 3$e$ &  & \\
\multicolumn{1}{|c|}{$-$1,0:+} &  &  &  &  &  & $-$$d$ &  & $-$3$e$ & \\
\multicolumn{1}{|c|}{1,$-$1:+} &  &  &  &  & 3$e$ &  & $d$$-$2$e$ &  & \\
\multicolumn{1}{|c|}{0,$-$1:+} &  &  &  &  &  & $-$3$e$ &  & 4$e$ & \\
\multicolumn{1}{|c|}{$-$1,$-$1:+} &  &  &  &  &  &  &  &  & $-$$d$$-$2$e$\\\hline
\multicolumn{1}{|c|}{1,1:$-$} &  & $\sqrt{2}$($d$+3$e$) &  & $-$$\sqrt{2}e$ &  &  &  &
& \\
\multicolumn{1}{|c|}{0,1:$-$} &  &  & $\sqrt{2}$($d$$-$3$e$) &  & 2$\sqrt{2}e$ &  &  &
& \\
\multicolumn{1}{|c|}{$-$1,1:$-$} &  &  &  &  &  & $-$$\sqrt{2}e$ &  &  & \\
\multicolumn{1}{|c|}{1,0:$-$} &  &  & 6$\sqrt{2}e$ &  & $\sqrt{2}d$ &  & $-$$\sqrt
{2}e$ &  & \\
\multicolumn{1}{|c|}{0,0:$-$} &  &  &  &  &  & $\sqrt{2}d$ &  & 2$\sqrt{2}e$ & \\
\multicolumn{1}{|c|}{$-$1,0:$-$} &  &  &  &  &  &  &  &  & $-$$\sqrt{2}e$\\
\multicolumn{1}{|c|}{1,$-$1:$-$} &  &  &  &  &  & 6$\sqrt{2}e$ &  & $\sqrt{2}$($d$%
$-$3$e$) & \\
\multicolumn{1}{|c|}{0,$-$1:$-$} &  &  &  &  &  &  &  &  & $\sqrt{2}$($d$+3$e$)\\
\multicolumn{1}{|c|}{$-$1,$-$1:$-$} &  &  &  &  &  &  &  &  & \\\hline
\end{tabular}
\caption{Matrix elements of $H_{hfs}^{cor}$ in equation~\ref{Hhfs}. Half of the matrix 
is explicitely given, the missing elements being obtained using the matrix symmetries 
with respect to the two diagonals (the same matrix element is obtained when all six signs
of angular momentum projections are changed). The values of
the parameters are $d$=$-$14.507~MHz and $e$= 1.4861~MHz, slightly larger than the values
of the corresponding parameters $D/2$ and $E/5$ of reference~\cite{Prestage85}
(see text).}%
\label{shfinecor}%
\end{table*}

\begin{figure}[h]
\centerline{ \psfig{file=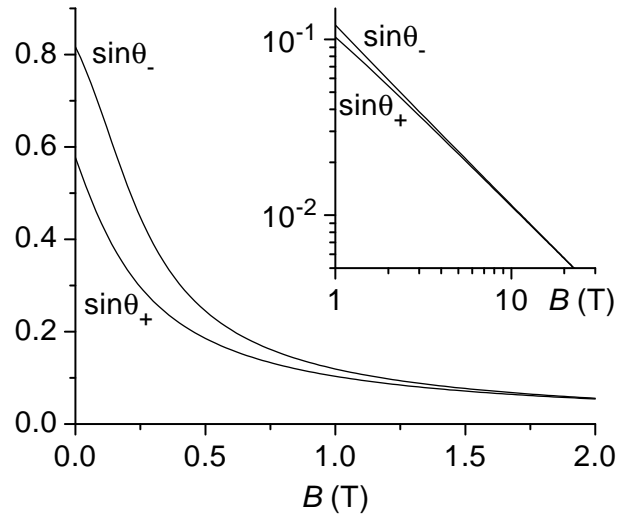, width=8 cm, clip= } }\caption{The
level-mixing parameters $\sin\theta_{+}$ and $\sin\theta_{-}$ are plotted as a
function of the magnetic field $B.$ Insert : asymptotic 1/$B$ decays at high
field, with almost equal values of order 1.1$\times$10$^{-2}$ at 10 Tesla.}%
\label{figthetas}%
\end{figure}

\paragraph{Given the total Hamiltonian in the $2^{3}\mathrm{S}$ state,}

which couples only states $\left|  m_{S}:m_{I}\right\rangle $ of given
$m_{F}=m_{S}+m_{I},$ a simple analytical expression of the eigenstates and of
the energies can be derived. To simplify the following expressions, we
introduce the reduced magnetic field $b$ (note that $A_{\mathrm{S}}<0$) and
the nuclear correction $\epsilon$:%
\begin{align}
b &  =-g_{S}^{^{\prime}}\mu_{B}B/A_{\mathrm{S}}\tag{A7}\\
\epsilon &  =g_{I}/2g_{S}^{^{\prime}}.\tag{A8}%
\end{align}
The $m_{F}=\pm3/2$ states, $\left|  -1:-\right\rangle $ and $\left|
1:+\right\rangle ,$ are eigenvalues with energies $\left(  1/2\pm\beta\left(
1+\epsilon\right)  \right)  A_{\mathrm{S}}.$ In the $m_{F}=+1/2$ subspace, the
matrix of the reduced Hamiltonian $H_{6}/A_{\mathrm{S}}$ on the basis
$\left\{  \left|  0:+\right\rangle ,\left|  1:-\right\rangle \right\}  $ is:%
\begin{equation}
\left(
\begin{array}
[c]{cc}%
-b\epsilon & 1/\sqrt{2}\\
1/\sqrt{2}\  & \ -b\left(  1-\epsilon\right)  -1/2
\end{array}
\right)  \tag{A9}\label{matundemi}%
\end{equation}
(the matrix for $m_{F}=-1/2$ on the basis $\left\{  \left|  0:-\right\rangle
,\left|  -1:+\right\rangle \right\}  $ is obtained replacing $\beta$ by
-$\beta$ in equation~\ref{matundemi}). The level energies are:%
\begin{multline}
W_{+}^{h,l}=-\frac{A_{\mathrm{S}}}{4}\left(  1+2b\right.  \nonumber\\
\pm\sqrt{9+4b+4b^{2}-8b\epsilon-16b^{2}\epsilon\left(  1-\epsilon\right)
})\tag{A10}\label{Wplushl}%
\end{multline}%
\begin{multline}
W_{-}^{h,l}=-\frac{A_{\mathrm{S}}}{4}\left(  1-2b\right.  \nonumber\\
\pm\sqrt{9-4b+4b^{2}+8b\epsilon-16b^{2}\epsilon\left(  1-\epsilon\right)
})\allowbreak,\tag{A11}\label{Wmoinshl}%
\end{multline}
in which the upper indices $h\ $and $l$ refer to the higher and lower energies
respectively, and the lower index is the sign of $m_{F}.$ With different
notations, these expressions would be identical to those given in
reference~\cite{Prestage85} but for a misprint (a missing factor 1/2 in the
last line of their equation 3). Here, both $W_{+}^{h}$ and $W_{-}^{h}$ are
obtained for positive sign in the second lines of equations \ref{Wplushl} and
\ref{Wmoinshl}. The field dependence of all energies is shown in
figure~\ref{figniv} in the main text. The corresponding eigenstates can be
simply written introducing the field-dependent angles $\theta_{+}$ and
$\theta_{-}$:
\begin{align}
\tan\theta_{+} &  =-\sqrt{2}\left(  W_{+}^{h}/A_{\mathrm{S}}+1/2+b\left(
1-\epsilon\right)  \right)  \tag{A12}\\
\tan\theta_{-} &  =-\sqrt{2}\left(  W_{-}^{h}/A_{\mathrm{S}}-b\epsilon\right)
\tag{A13}\\
\mathrm{A}_{+}^{h} &  =-\sin\theta_{+}\left|  0:+\right\rangle +\cos\theta
_{+}\left|  1:-\right\rangle \tag{A14}\label{Ahplus}\\
\mathrm{A}_{+}^{l} &  =\cos\theta_{+}\left|  0:+\right\rangle +\sin\theta
_{+}\left|  1:-\right\rangle \tag{A15}\\
\mathrm{A}_{-}^{h} &  =-\sin\theta_{-}\left|  -1:+\right\rangle +\cos
\theta_{-}\left|  0:-\right\rangle \tag{A16}\\
\mathrm{A}_{-}^{l} &  =\cos\theta_{-}\left|  -1:+\right\rangle +\sin\theta
_{-}\left|  0:-\right\rangle .\tag{A17}\label{Almoins}%
\end{align}
The variation of the angles $\theta_{+}$ and $\theta_{-}$ in shown in
figure~\ref{figthetas}. The large amount of state mixing due to hyperfine coupling at low
field ($\sin^{2}\theta_{+}$=1/3, $\sin^{2}\theta_{-}$=2/3 for $B$=0) is
strongly reduced above 1T ($\sin^{2}\theta_{\pm}\sim0.012/B^{2}$). Our state
naming convention in most of this article ($\mathrm{A}_{1}$ to $\mathrm{A}%
_{6}$ by increasing energies) introduces three field intervals due to two
level crossings. The first crossing occurs at 0.1619 T between $\left|
1:+\right\rangle $ and $\mathrm{A}_{-}^{h},$ the second one at 4.716 T between
$\mathrm{A}_{-}^{h}$ and $\mathrm{A}_{+}^{l}.$ Altogether, three states are
affected by these level crossings, as summarised in table~\ref{tabnomsAi}%
.\begin{table}[hh]
\centering%
\begin{tabular}
[c]{|c|ccc|c|}\hline
$B\text{ (T)}$ & $<\text{0.1619}$ & \multicolumn{1}{|c}{0$\text{.1619--4.716}%
$} & \multicolumn{1}{|c|}{$>\text{4.716}$} & $\text{any }B$\\\hline\hline
$\mathrm{A}_{1}$ &  &  &  & $\left|  -1:-\right\rangle $\\
$\mathrm{A}_{2}$ &  &  &  & $\mathrm{A}_{-}^{l}$\\\hline
$\mathrm{A}_{3}$ & $\mathrm{A}_{+}^{l}$ & $\mathrm{A}_{+}^{l}$ &
$\mathrm{A}_{-}^{h}$ & \\
$\mathrm{A}_{4}$ & $\left|  1:+\right\rangle $ & $\mathrm{A}_{-}^{h}$ &
$\mathrm{A}_{+}^{l}$ & \\
$\mathrm{A}_{5}$ & $\mathrm{A}_{-}^{h}$ & $\left|  1:+\right\rangle $ &
$\left|  1:+\right\rangle $ & \\\hline
$\mathrm{A}_{6}$ &  &  &  & $\mathrm{A}_{+}^{h}$\\\hline
\end{tabular}
\caption{Table of correspondence between the state names used in this Appendix
and the energy-ordered naming convention $\mathrm{A}_{1}$ to $\mathrm{A}_{6}%
$.}%
\label{tabnomsAi}%
\end{table}

\paragraph{ME matrices:}

we now provide the elements of the $B$-dependent ME matrices $L,$ $G^{4},$
$E^{3},$ $F^{3},$ $E^{4}$ and $F^{4}$ which appear in rate
equations\ \ref{rateM} to \ref{ratea}. The six relative populations $a_{i}$ of
the text are here designated by $a_{--},$ $a_{-}^{l},$ $a_{+}^{l},$ $a_{++},$
$a_{-}^{h},$ $a_{+}^{h}.$ This indexing convention is the same for the state
names and the corresponding populations, so that table~\ref{tabnomsAi} can be
used to transform the indexing, or to determine the permutation of rows and
columns required for $B>$0.1619~T between the matrices provided in tables
\ref{tabL} to \ref{tabF3} and those appearing in equations\ \ref{rateM} to
\ref{ratea}. For more compact expressions, we use the following notations:%
\begin{equation}
c_{\pm}=\cos^{2}\theta_{\pm},~s_{\pm}=\sin^{2}\theta_{\pm}.\tag{A18}%
\end{equation}
These matrices are not fully independent. In particular, the steady-state
solutions of equations~\ref{raterho4} and~\ref{raterho3}, given in
equations~\ref{equilrho4} and~\ref{equilrho3}, are independent on the isotopic
ratio $R.$ Substitution of the steady state solutions $y_{i}=%
{\textstyle\sum_{k}}
G_{ik}^{4}a_{k}$ of equation~\ref{ratey} in equation~\ref{ratea} can be used
to show that $E^{3}=E^{4}\cdot G^{4}$ and $F^{3}=F^{4}\cdot G^{4},$ two
field-independent relations which can be directly verified on the matrices
given in tables \ref{tabG4} to \ref{tabF3}.\begin{table}[h]
\centering%
\begin{tabular}
[c]{|cccccc|}\hline
$a_{--}$ & $a_{-}^{l}$ & $a_{+}^{l}$ & $a_{++}$ & $a_{-}^{h}$ & $a_{+}^{h}%
$\\\hline
-1 & $c_{-}$-$s_{-}$ & $c_{+}$-$s_{+}$ & 1 & $s_{-}$-$c_{-}$ & $s_{+}$-$c_{+}%
$\\\hline
\end{tabular}
\caption{Matrix elements of operator $L$ (equation~\ref{rateM}).}%
\label{tabL}%
\end{table}\begin{table}[hh]
\centering%
\begin{tabular}
[c]{c|cccccc|}\cline{2-7}%
& $a_{--}$ & $a_{-}^{l}$ & $a_{+}^{l}$ & $a_{++}$ & $a_{-}^{h}$ & $a_{+}^{h}%
$\\\hline
\multicolumn{1}{|c|}{$y_{1}$} & 1 & $c_{-}$ & 0 & 0 & $s_{-}$ & 0\\
\multicolumn{1}{|c|}{$y_{2}$} & 0 & $s_{-}$ & $c_{+}$ & 0 & $c_{-}$ & $s_{+}%
$\\
\multicolumn{1}{|c|}{$y_{3}$} & 0 & 0 & $s_{+}$ & 1 & 0 & $c_{+}$\\\hline
\end{tabular}
\caption{Matrix elements of $G^{4}$ (equation~\ref{ratey}).}%
\label{tabG4}%
\end{table}\begin{table}[hhh]
\centering%
\begin{tabular}
[c]{c|ccc|ccc|}\cline{2-7}%
&  & 2$E^{4}$ &  &  & 2$F^{4}$ & \\\cline{2-7}%
& $y_{1}$ & $y_{2}$ & $y_{3}$ & $y_{1}$ & $y_{2}$ & $y_{3}$\\\hline
\multicolumn{1}{|c|}{$a_{--}$} & 1 & 0 & 0 & -1 & 0 & 0\\
\multicolumn{1}{|c|}{$a_{-}^{l}$} & $c_{-}$ & $s_{-}$ & 0 & $c_{-}$ & -$s_{-}$%
& 0\\
\multicolumn{1}{|c|}{$a_{+}^{l}$} & 0 & $c_{+}$ & $s_{+}$ & 0 & $c_{+}$ &
-$s_{+}$\\
\multicolumn{1}{|c|}{$a_{++}$} & 0 & 0 & 1 & 0 & 0 & 1\\
\multicolumn{1}{|c|}{$a_{-}^{h}$} & $s_{-}$ & $c_{-}$ & 0 & $s_{-}$ & -$c_{-}$%
& 0\\
\multicolumn{1}{|c|}{$a_{+}^{h}$} & 0 & $s_{+}$ & $c_{+}$ & 0 & $s_{+}$ &
-$c_{+}$\\\hline
\end{tabular}
\caption{Matrix elements of 2$E^{4}$ and 2$F^{4}$ (twice those of $E^{4}$ and
$F^{4}$ in equation~\ref{ratea}).}%
\label{tabEF4}%
\end{table}\begin{table}[hhhh]
\centering%
\begin{tabular}
[c]{c|cccccc|}\cline{2-7}%
& $a_{--}$ & $a_{-}^{l}$ & $a_{+}^{l}$ & $a_{++}$ & $a_{-}^{h}$ & $a_{+}^{h}%
$\\\hline
\multicolumn{1}{|c|}{$a_{--}$} & 1 & $c_{-}$ & 0 & 0 & $s_{-}$ & 0\\
\multicolumn{1}{|c|}{$a_{-}^{l}$} & $c_{-}$ & $c_{-}^{2}$+$s_{-}^{2}$ &
$c_{+}s_{-}$ & 0 & 2$c_{-}s_{-}$ & $s_{+}s_{-}$\\
\multicolumn{1}{|c|}{$a_{+}^{l}$} & 0 & $c_{+}s_{-}$ & $c_{+}^{2}$+$s_{+}^{2}$%
& $s_{+}$ & $c_{+}c_{-}$ & 2$c_{+}s_{+}$\\
\multicolumn{1}{|c|}{$a_{++}$} & 0 & 0 & $s_{+}$ & 1 & 0 & $c_{+}$\\
\multicolumn{1}{|c|}{$a_{-}^{h}$} & $s_{-}$ & 2$c_{-}s_{-}$ & $c_{+}c_{-}$ &
0 & $c_{-}^{2}$+$s_{-}^{2}$ & $c_{-}s_{+}$\\
\multicolumn{1}{|c|}{$a_{+}^{h}$} & 0 & $s_{+}s_{-}$ & 2$c_{+}s_{+}$ & $c_{+}$%
& $c_{-}s_{+}$ & $c_{+}^{2}$+$s_{+}^{2}$\\\hline
\end{tabular}
\caption{Matrix elements of 2$E^{3}$ (twice those of $E^{3}$ in
equation~\ref{ratea}).}%
\label{tabE3}%
\end{table}\begin{table}[hhhhh]
\centering%
\begin{tabular}
[c]{c|cccccc|}\cline{2-7}%
& $a_{--}$ & $a_{-}^{l}$ & $a_{+}^{l}$ & $a_{++}$ & $a_{-}^{h}$ & $a_{+}^{h}%
$\\\hline
\multicolumn{1}{|c|}{$a_{--}$} & -1 & -$c_{-}$ & 0 & 0 & -$s_{-}$ & 0\\
\multicolumn{1}{|c|}{$a_{-}^{l}$} & $c_{-}$ & $c_{-}^{2}$-$s_{-}^{2}$ &
-$c_{+}s_{-}$ & 0 & 0 & -$s_{+}s_{-}$\\
\multicolumn{1}{|c|}{$a_{+}^{l}$} & 0 & $c_{+}s_{-}$ & $c_{+}^{2}$-$s_{+}^{2}$%
& -$s_{+}$ & $c_{+}c_{-}$ & 0\\
\multicolumn{1}{|c|}{$a_{++}$} & 0 & 0 & $s_{+}$ & 1 & 0 & $c_{+}$\\
\multicolumn{1}{|c|}{$a_{-}^{h}$} & $s_{-}$ & 0 & -$c_{+}c_{-}$ & 0 &
$s_{-}^{2}$-$c_{-}^{2}$ & -$c_{-}s_{+}$\\
\multicolumn{1}{|c|}{$a_{+}^{h}$} & 0 & $s_{+}s_{-}$ & 0 & -$c_{+}$ &
$c_{-}s_{+}$ & $s_{+}^{2}$-$c_{+}^{2}$\\\hline
\end{tabular}
\caption{Matrix elements of 2$F^{3}$ (twice those of $F^{3}$ in
equation~\ref{ratea}).}%
\label{tabF3}%
\end{table}


\begin{thebibliography}{99}
\bibitem{Rohe99}D. Rohe \textit{et al., }Phys. Rev. Lett. \textbf{83} (1999) 4257.

\bibitem {Becker98}J. Becker \textit{et al.}, Nuc. Instr. \& Meth. A
\textbf{402} (1998) 327.

\bibitem {Jones00}G.L.\ Jones \textit{et al.}, Nuc. Instr. \& Meth. A
\textbf{440} (2000) 772.

\bibitem {Tastevin00}G.\ Tastevin, Physica Scripta \textbf{T86} (2000) 46.

\bibitem {Chupp01}T.\ Chupp and S.\ Swanson, Adv. At. Mol. Opt. Phys.
\textbf{45} (2001) 51.

\bibitem {Colegrove63}F.D.\ Colegrove, L.D.\ Schearer and G.K.\ Walters, Phys.
Rev. \textbf{132} (1963) 2561.

\bibitem {Nacher85}P.-J.\ Nacher and M.\ Leduc, J.\ Phys.\ Paris \textbf{46}
(1985) 2057.

\bibitem {Bigelow92}N.P. Bigelow, P.J. Nacher and M. Leduc, J. Phys. II France
\textbf{2} (1992) 2159.

\bibitem {Leduc83}M Leduc, S.B. Crampton, P.J. Nacher and F.\ Lalo\"{e},
Nuclear Sci. App. \textbf{1} (1983) 1.

\bibitem {Becker94}J. Becker \textit{et al}., Nucl. Instr. and Meth. A
\textbf{346} (1994) 45.

\bibitem {Nacher99}P.-J. Nacher, G. Tastevin, X. Maitre, X. Dollat, B. Lemaire
and J. Olejnik, Eur. Radiol. \textbf{9} (1999) B18.

\bibitem {Gentile00}T.\ R.\ Gentile \textit{et al}., Magn. Reson.\ Med.
\textbf{43} (2000) 290.

\bibitem {MRIorsay}L. Darrasse \textit{et al.} in: Proc. Int. Soc. for Mag.
Res. in Medicine, Sydney\textit{ (}1998) 449.

\bibitem {Flowers90}J.L.\ Flowers, B.W.\ Petley and M.G.\ Richards,
J.\ Phys.\ B \textbf{23} (1990) 1359.

\bibitem {Flowers97}J.L.\ Flowers, C.J.\ Bickley, P.W.\ Josephs-Franks and
B.W.\ Petley, IEEE Trans.\ Instrum.\ Meas. \textbf{46} (1997) 104.

\bibitem {Courtade00}E.\ Courtade, F.\ Marion, P.J.\ Nacher, G.\ Tastevin,
T.\ Dohnalik and K.\ Kiersnowski, Hyperfine Interactions \textbf{127} (2000) 451.

\bibitem {Pavlovic70}M.\ Pavlovic and F.\ Lalo\"{e}, J.\ Phys.\ France
\textbf{31} (1970) 173.

\bibitem {Fitzsimmons68}W.A.\ Fitzsimmons, N.F.\ Lane and G.K.\ Walters,
Phys.\ Rev. A \textbf{19} (1968) 193.

\bibitem {Chapman76}R.\ Chapman and M.\ Bloom, Canad. J.\ Phys. \textbf{54}
(1976) 861.

\bibitem {BarbéTh}R.\ Barb\'{e}, PhD\ Thesis (1977) Paris.

\bibitem {NacherTh}P.-J. Nacher, PhD Thesis (1985) Paris.

\bibitem {CourtadeTh}E.\ Courtade, PhD\ Thesis (2001) Paris.

\bibitem {NousMolec}E. Courtade, P.-J. Nacher, C. Dedonder, C. Jouvet and T.
Dohnalik, \textit{in preparation}.

\bibitem {Philips82}W.D.\ Philips and H.\ Metcalf, Phys.\ Rev.\ Lett.
\textbf{48} (1982) 596.

\bibitem {Orsay}A.\ Robert \textit{et al.}, Science \textbf{292} (2001) 461.

\bibitem {Paris}F.\ Pereira Dos Santos et al., Phys.\ Rev.\ Lett. \textbf{86}
(2001) 3459.

\bibitem {Storry00}C.H.\ Storry, \ M.C.\ George and E.A.\ Hessels,
Phys.\ Rev.\ Lett. \textbf{84} (2000) 3274.

\bibitem {Castillega00}J.\ Castillega, D.\ Livingstone, A.\ Sanders and
D.\ Shiner, Phys.\ Rev.\ Lett. \textbf{84} (2000) 4321.

\bibitem {Shiner95}D.\ Shiner, R.\ Dixson and V.\ Vedantham,
Phys.\ Rev.\ Lett. \textbf{74} (1995) 3553.

\bibitem {Lhuillier76}C.\ Lhuillier, J.-P.\ Faroux and N.\ Billy, J.\ Physique
\textbf{37} (1976) 335.

\bibitem {Yang86}D.-H. Yang, P. McNicholl and H. Metcalf, Phys.\ Rev. A
\textbf{33} (1986) 1725.

\bibitem {Yan94}Z.-C.\ Yan and G.W.F.\ Drake, Phys.\ Rev A \textbf{50} (1994) R1980.

\bibitem {Kponou81}A.\ Kponou \textit{et al}., Phys.\ Rev.\ A \textbf{24}
(1981) 264 ; W.\ Frieze \textit{et al}, \textit{ibid}., 279.

\bibitem {Prestage85}J.D.\ Prestage, C.E.\ Johnson, E.A.\ Hinds and
F.M.J.\ Pichanick, Phys. Rev. A.\textbf{32} (1985) 2712.

\bibitem {Gonzalo97}I.\ Gonzalo and E.\ Santos, Phys. Rev. A \textbf{56}
(1997) 3576.

\bibitem {Hinds85}E.A.\ Hinds, J.D.\ Prestage and F.M.J.\ Pichanick,
Phys.\ Rev.\ A \textbf{32} (1985) 2615.

\bibitem {Hijikata88}K.\ Hijikata and K.\ Ohtsuki, J.\ Phys.\ Soc.\ Jap.
\textbf{57} (1988) 4141.

\bibitem {reff}\textit{Atomic transition probabilities}, W.L.\ Wiese,
M.W.\ Smith and B.M.\ Glennon, National Standard Reference Data Series NBS 4 (1966).

\bibitem {DrakeHB}\textit{Atomic, molecular, and optical physics handbook}
(AIP\ Press, G.W.F.\ Drake ed., 1996).

\bibitem {Bloch85}D.\ Bloch, G. Trenec and M.\ Leduc, J.\ Phys.\ B:
At.\ Mol.\ Phys. \textbf{18} (1985) 1093.

\bibitem {Numrec}See for instance W.H.\ Press \textit{et al}.,
\textit{Numerical recipes in Fortran }(Cambridge University Press, 1992) 460.

\bibitem {medemander}Source files or executable programs may be provided on
request by one of the authors (P.-J. N.).

\bibitem {Lewis70}S.A.\ Lewis, F.M.J.\ Pichanick and V.W.\ Hughes,
Phys.\ Rev.\ A \textbf{1} (1970) 86.

\bibitem {Partridge66}R.B.\ Partridge and G.W.\ Series, Proc.\ Phys.\ Soc.
\textbf{88} (1966) 983.

\bibitem {JDR73}J.\ Dupont-Roc, M.\ Leduc and F.\ Laloe, J.\ Physique
\textbf{34} (1973) 961 and 977.

\bibitem {Pinard80}M.\ Pinard and F.\ Lalo\"{e}, J.\ Physique \textbf{41}
(1980) 799.

\bibitem {Anderson59}L.W.\ Anderson, F.M.\ Pipkin and J.C.\ Baird, Phys. Rev.
\textbf{116} (1959) 87.

\bibitem {Happer72}W.\ Happer, Rev.\ Mod.\ Phys. \textbf{44} (1972) 169.

\bibitem {JDR71}J.\ Dupont-Roc, M.\ Leduc and F.\ Lalo\"{e},
Phys.\ Rev.\ Lett. \textbf{27} (1971) 467.

\bibitem {Larat91}C.\ Larat, PhD Thesis (1991) Paris.

\bibitem {Minardi99}F.\ Minardi \textit{et al}., Phys.\ Rev.\ Lett.
\textbf{82} (1999) 1112.

\bibitem {Shiner94}D.\ Shiner, R.\ Dixson and P.\ Zhao, Phys.\ Rev.\ Lett.
\textbf{72} (1994) 1802.

\bibitem {Prevedelli}M.\ Prevedelli, P.\ Cancio, G.\ Giusfredi, F.S.\ Pavone
and M.\ Inguscio, Optics Commun. \textbf{125} (1996) 231.

\bibitem {CIERM}This magnet is part of the GE Signa MRI\ system of the Centre
Inter Etablissements de R\'{e}sonance Magn\'{e}tique in the
Kremlin-Bic\^{e}tre hospital. We wish to thank J.\ Bittoun and his team for
providing the opportunity to perform these \ high-field measurements.

\bibitem {Stoltz}E.~Stoltz, M.~Meyerhoff, N.~Bigelow, M.~Leduc, P.-J.~Nacher
and G.~Tastevin, Appl. Phys.\ \textbf{B63} (1996) 629.

\bibitem {Greenhow64}R.C.~Greenhow, Phys. Rev. \textbf{126} (1964) A660.

\bibitem {Daniels71}J.M.~Daniels and R.S.~Timsit, Can. J.\ Phys. \textbf{49}
(1971) 539; R.S.~Timsit and J.M.~Daniels, Can. J.\ Phys. \textbf{49} (1971) 545.

\bibitem {Chernikov}S.V.~Chernikov, J.R.~Taylor, N.S.~Platonov,
V.P.~Gaponstev, P.-J.~Nacher, G.~Tastevin, M.~Leduc and M.J.~Barlow,
Electronics Lett. \textbf{33} (1997) 787.

\bibitem {Schearer70}L.D.~Schearer and L.A.~Riseberg, Phys. Lett. \textbf{33A}
(1970) 325.

\bibitem {hypint}M.~Leduc, P.J.~Nacher, G.~Tastevin and E.~Courtade, Hyperfine
Interactions \textbf{127} (2000) 443.

\bibitem {Rosner70}S.D.\ Rosner and F.M.\ Pipkin, Phys.\ Rev.\ A \textbf{1}
(1970) 571.
\end{thebibliography}
\end{document}